\pdfoutput=1
\RequirePackage{ifpdf}
\ifpdf 
\documentclass[pdftex]{sigma}
\else
\documentclass{sigma}
\fi

\numberwithin{equation}{section}

\usepackage{bbm}

\newcommand{\bma}{\left(\begin{array}}
\newcommand{\ema}{\end{array}\right)}
\newcommand{\be}{\begin{equation}}
\newcommand{\ee}{\end{equation}}
\newcommand{\ba}{\begin{eqnarray}}
\newcommand{\ea}{\end{eqnarray}}
\newcommand{\bs}{\begin{subequations}}
\newcommand{\es}{\end{subequations}}
\newcommand{\bc}{\begin{center}}
\newcommand{\ec}{\end{center}}

\newcommand{\Pl}{{\text{Pl}}}
\def\dpl{\delta_{\rm Pl}}
\def\bu{\bar\mu}
\def\rme{e}
\def\rmd{d}
\def\rmi{i}

\def\dpl{\delta_{\rm Pl}}
\def\a{\alpha}

\def\g{\gamma}
\def\la{\lambda}
\def\k{\kappa}
\def\e{\epsilon}

\def\s{\sigma}

\def\cN{{\cal N}}

\def\p{\partial}
\begin{document}

\allowdisplaybreaks

\renewcommand{\thefootnote}{$\star$}

\renewcommand{\PaperNumber}{016}

\FirstPageHeading

\ShortArticleName{Introduction to Loop Quantum Cosmology}

\ArticleName{Introduction to Loop Quantum Cosmology\footnote{This
paper is a contribution to the Special Issue ``Loop Quantum Gravity and Cosmology''. The full collection is available at \href{http://www.emis.de/journals/SIGMA/LQGC.html}{http://www.emis.de/journals/SIGMA/LQGC.html}}}

\Author{Kinjal BANERJEE~$^\dag$, Gianluca CALCAGNI~$^\ddag$ and Mercedes MART\'IN-BENITO~$^\ddag$}

\AuthorNameForHeading{K.\ Banerjee, G.\ Calcagni and M.\ Mart\'in-Benito}

\Address{$^\dag$~Department of Physics, Beijing Normal University,  Beijing 100875, China}
\EmailD{\href{mailto:kinjalb@gmail.com}{kinjalb@gmail.com}}

\Address{$^\ddag$~Max Planck Institute for Gravitational Physics (Albert Einstein Institute),\\
\hphantom{$^\ddag$}~Am M\"uhlenberg 1, D-14476 Golm, Germany}
\EmailD{\href{mailto:calcagni@aei.mpg.de}{calcagni@aei.mpg.de}, \href{mailto:mercedes@aei.mpg.de}{mercedes@aei.mpg.de}}

\ArticleDates{Received September 30, 2011, in f\/inal form March 13, 2012; Published online March 25, 2012}

\Abstract{This is an introduction to loop quantum cosmology (LQC) reviewing mini- and midisuperspace models as well as homogeneous and inhomogeneous ef\/fective dynamics.}


\Classification{83C45; 83C75; 83F05}

\renewcommand{\thefootnote}{\arabic{footnote}}
\setcounter{footnote}{0}

\tableofcontents

\section*{Introduction}
\addcontentsline{toc}{section}{Introduction}

General relativity (GR) and quantum mechanics are two of the best verif\/ied theories of modern physics. While general
relativity has been spectacularly successful in explaining the universe at astronomical and cosmological scales, quantum mechanics gives an equally coherent physical picture on small scales. However, one of the biggest unfulf\/illed challenges in physics remains
to incorporate the two theories in the same framework. Ordinary quantum f\/ield theories, which have managed to describe
the three other fundamental forces (electromagnetic, weak and strong), have failed for general relativity because it
is not perturbatively renormalizable.

Loop quantum gravity (LQG) \cite{lqg3,lqg4,lqg2,lqg1} is an attempt to construct a mathematically ri\-go\-rous, non-perturbative, background independent
formulation of quantum general relativity. GR is reformulated in terms of Ashtekar--Barbero variables, namely the densitized triad and the Ashtekar
connection. The basic classical variables are taken to be the holonomies of the connection and the f\/luxes of the triads and these are then promoted
to basic quantum operators. The quantization is not the standard Schr\"odinger quantization but an unitarily inequivalent choice known as
loop/polymer quantization. The kinematic structure of LQG has been well developed. A~robust feature of LQG, not imposed but emergent, is the underlying discreteness of space.

With the aim of obtaining physical implications from LQG, in the last years the application of loop quantization techniques to cosmological models has undergone a notable development. This f\/ield of research is known under the name of loop quantum cosmology (LQC). The models analyzed in LQC are
mini- and midisuperspace models. These models have Killing vectors which reduce the degrees of freedom of full GR. In the case of minisuperspaces, the reduced theories have no f\/ield-theory degrees of freedom remaining. Although there are f\/ield-theory degrees of freedom in the midisuperspace models, their number is smaller than in the full theory. Therefore, these are simplif\/ied systems which provide toy models suitable for studying some aspects of the full quantum gravity theory. Moreover, classical solutions are well known (in fact, we are aware of very few systems which have closed-form solutions of Einstein equations with no Killing vectors) and it is relatively easy to study the ef\/fects of the quantization.

LQC cannot be considered the cosmological sector of LQG because the symmetry reduction is carried out before quantizing, and the results so obtained may not be the same if the reduction is done after quantization. However by adapting the techniques used in the full theory to the symmetry-reduced cosmological models we may hope to capture some of the crucial features of the full theory, as well as to obtain hints about how to tackle them. Indeed, one of the generic characteristics of LQC is the avoidance of the classical singularity. In the present absence of recognized experimental and observational signatures of quantum gravity, this novel and robust result has been increasing the hope that LQG may indeed be the correct theory of quantum gravity.

In this article we will review the progress made in the various cosmological models studied in LQC in the last few years. A recent review \cite{AsS}
emphasizes aspects that are only brief\/ly mentioned here, such as the ``simplif\/ied'' or ``solvable'' LQC framework, the details of ef\/fective dynamics for FRW models with non-zero curvature and/or cosmological constant, and inf\/lationary perturbation theory in LQC. On the other hand, here we focus more on midisuperspaces and discuss lattice ref\/inement parametrizations at some length. Before starting, we shall brief\/ly recall the main features in the kinematic structures of LQG. Similar ingredients are used in the kinematic structure of the LQC models to be discussed later.

\section{Loop quantization}

\subsection{Ashtekar--Barbero formalism}

In the Hamiltonian formulation, the four-dimensional spacetime metric is described by a three-metric $q_{ab}$ induced in
the spatial sections $\Sigma$ that foliate the spacetime manifold, the lapse function~$N$ and the shift vector $N^{a}$ \cite{adm1,adm2}\footnote{Latin indices from the beginning of the alphabet, $a,b,\dots$, denote spatial indices.}. Both the lapse $N$ and the
shift vector $N^{a}$ are Lagrange multipliers accompanying the constraints that encoded the general covariance of general relativity. These constraints
are, respectively, the scalar or Hamiltonian constraint and the dif\/feomorphisms constraint (which is a three-vector). Therefore, the physically
relevant information is encoded in the spatial three-metric and in its canonically conjugate momentum, or equivalently, in the extrinsic curvature
$K_{ab}=\mathcal L_n q_{ab}/2$, where $n$ is the unit normal to~$\Sigma$ and~$\mathcal L_n$ is the Lie derivative along~$n$~\cite{wald}.

LQG is based in a formulation of general relativity as a gauge theory \cite{lqg0c,lqg0ab, lqg0aa,lqg0ba,lqg0bb}, in which the phase space is described
by a $su(2)$ gauge connection, the Ashtekar--Barbero connection $A^i_a$, and its canonically conjugate momentum,
the densitized triad\footnote{Latin indices from the middle of the alphabet, $i,j,\dots$ are $SU(2)$ indices and label new degrees of freedom introduced when passing to the
triad formulation.} $E^a_i$, that plays the role of an ``electric f\/ield''. To def\/ine these objects, f\/irst one introduces the co-triad
$e^i_a$, def\/ined as $q_{ab}=e^i_ae^j_b\delta_{ij}$, where $\delta_{ij}$ stands for the Kronecker delta in three dimensions, and then one def\/ines the
triad, $e^a_i$, as its inverse $e^a_ie^j_b=\delta_i^j\delta^a_b$. The densitized triad then reads $E^a_i=\sqrt{q}e^a_i$,  where $q$ stands for the
determinant of the spatial three-metric. In turn, the Ashtekar--Barbero connection reads \cite{bar} $A^i_a=\Gamma^i_a+\gamma K^i_a$, where $\gamma$ is an
arbitrary real and non-vanishing para\-me\-ter, called the Immirzi parameter \cite{gior1,gior2}, $K^i_a=K_{ab}e^b_j\delta^{ij}$ is the extrinsic curvature in
triadic form, and $\Gamma^i_a$ is the spin connection compatible with the densitized triad. Namely, it verif\/ies
$\nabla_b E^a_i+\epsilon_{ijk}\Gamma^j_b E^{ak}=0$, where $\epsilon_{ijk}$ is the totally antisymmetric symbol and $\nabla_b$ is the usual spatial
covariant derivative~\cite{wald}. The canonical pair $(A,E)$ has the following Poisson bracket:
\begin{gather*}
 \{A^a_i(x),E^j_b(y)\}=8\pi G\gamma\delta^a_b\delta^j_i\delta(x-y) , 
\end{gather*}
where $G$ is Newton constant and $\delta(x-y)$ denotes the three-dimensional Dirac delta distribution
on the hypersurface $\Sigma$.

Since the internal Euclidean metric $\delta_{ij}$ is invariant under $SU(2)$ rotations, the internal $SU(2)$ degrees of freedom are gauge.
Therefore, in this formulation of general relativity, besides the dif\/feomorphisms constraint $\mathcal{C}_a$ and the scalar (or Hamiltonian)
constraint $\mathcal{C}$, there is a gauge (or Gauss) constraint $ \mathcal{G}_i$ f\/ixing the rotation freedom that we have just introduced.
In the variables $(A^a_i,E^j_b)$, those constraints have the following expression (in vacuum)\footnote{In the presence of matter coupled to the geometry, there is a matter term contributing to each constraint.} \cite{lqg1},
\begin{gather*}
 \mathcal{G}_i =\partial_aE^a_i+\epsilon_{ijk}\Gamma^j_a E^{ak}=0,\\
\mathcal{C}_a =F_{ab}^iE^b_i=0,\\
\mathcal{C} =\frac1{\sqrt{|\det(E)|}}\epsilon_{ijk}\left[F^i_{ab}-(1+\gamma^2)\epsilon^i_{mn}
K^m_aK^n_b\right]E^{aj} E^{bk}=0,
\end{gather*}
where $F^i_{ab}$ is the curvature tensor of the Ashtekar--Barbero connection,
\begin{gather*}
 F^i_{ab}=\partial_a A^i_b-\partial_b A^i_a+\epsilon_{ijk}A^j_aA^k_b.
\end{gather*}

\subsubsection{Holonomy-f\/lux algebra}

The next step is to def\/ine the holonomies and f\/luxes which will later be promoted to basic quantum variables.

The conf\/iguration variables chosen are the holonomies of $A^i_a$. They are more convenient than the connection itself thanks to their properties under gauge
transformation. The holonomy of the connection $A$ along the edge $e$ is given by
\begin{gather*}
 h_e(A)=\mathcal P e^{\int_e dx^a A^i_a(x)\tau_i},
\end{gather*}
where $\mathcal P$ denotes path ordering and $\tau_i$ are the generators of
$SU(2)$, such that $[\tau_i,\tau_j]=\epsilon_{ijk}\tau^k$.

The momentum conjugate to the holonomy is given by the f\/lux of
$E^a_i$ over surfaces $S$ and smeared with a $su(2)$-valued function $f^i$:
\begin{gather*}
 E(S,f)=\int_Sf^i E^a_i\epsilon_{abc}dx^bdx^c.
\end{gather*}

The description of the phase space in terms of holonomies and f\/luxes is not only suitable for its transformation properties,
but also because these objects are dif\/feomorphism invariant and their def\/inition is background independent.
Moreover, their Poisson bracket is divergence-free
\begin{gather*}
 \{E(S,f),h_e(A)\}=2\pi G\gamma \epsilon(e,S)f^i\tau_i h_e(A),
\end{gather*}
where $\epsilon(e,S)$ represents the regularization of the Dirac delta: it vanishes if~$e$ does not intersect~$S$, as well as if
$e\subset S$, and $|\epsilon(e,S)|=1$ if $e$ and $S$ intersect in one point, the sign depending on the relative orientation between
$e$ and~$S$~\cite{lqg3}.

\subsection{Kinematic Hilbert space}

In LQG, the holonomy-f\/lux algebra is represented over a kinematical Hilbert space that is
dif\/ferent from the more familiar
Schr\"odinger-type Hilbert space. It is given by the
completion of the space of cylindrical functions (def\/ined on the space of generalized connections)
with respect to the so-called Ashtekar--Lewandowski measure
\cite{ALmeasure1,ALmeasure3,ALmeasure2,baez}.
We give a very brief description of this kinematical Hilbert space below, while the details can be
found in \cite{lqg3,lqg4,lqg2,lqg1} (and references therein).

A generalized connection $h_e(A)\equiv\bar A_e$ is an assignment of $\bar A \in$ $SU(2)$ to any analytic path $e \subset \Sigma$.
A graph $\Gamma$ is a collection of analytic paths $e\subset \Sigma$ meeting at most at their endpoints.
We will consider only closed graphs. The point at which two edges meet is called a vertex.
Let n be the number of edges in $\Gamma$. A function cylindrical with respect to $\Gamma$ is given by
\begin{equation*}
\psi_\Gamma(\bar A) := f_\Gamma \big(\bar A_{e_1},\dots, \bar A_{e_n}\big) ,
\end{equation*}
where $f_\Gamma$ is a smooth function on $SU(2)^n$. The space of states cylindrical with respect to $\Gamma$ are denoted by Cyl$_\Gamma$.
The space of all functions cylindrical with respect to some $\Gamma \in \Sigma$ is denoted by Cyl and is given by
\begin{equation*}
\mbox{Cyl} = \bigcup_\Gamma \mbox{Cyl}_\Gamma.
\end{equation*}

Given a cylindrical function $\psi_\Gamma(\bar A) \in {\rm Cyl}$, the Ashtekar--Lewandowski measure, denoted by~$\mu_0$, is def\/ined by
\begin{equation*}
\int_{\overline{{\mathcal A}}}d\mu_0[\psi_\Gamma(\bar A)] :=
\int_{SU(2)^n} \prod_{e \subset \Gamma} d h^e   f_\Gamma \big(\bar A_{e_1},\dots, \bar A_{e_n}\big) ,
\qquad \forall\, \psi_\Gamma(\bar A) ,
\end{equation*}
where d$h$ is the normalized Haar measure on $SU(2)$.
Using this measure we can def\/ine an inner product on Cyl:
\begin{gather*}
\langle \psi_\Gamma, \psi_\Gamma '\rangle  :=  \langle f_\Gamma\big(\bar A_{e_1},\dots, \bar
A_{e_n}\big),g_{\Gamma'}\big(\bar A_{e_1},\dots, \bar A_{e_m}\big)\rangle\nonumber\\
\phantom{\langle \psi_\Gamma, \psi_\Gamma '\rangle}{}  =  \int_{SU(2)^n} \prod_{e \subset \Xi_{\Gamma \Gamma'}}  d h^e\overline{f_\Gamma\big(\bar A_{e_1},\dots,
\bar A_{e_n}\big)}g_{\Gamma'}\big(\bar A_{e_1},\dots, \bar A_{e_m}\big),
\end{gather*}
where ${\Xi}_{\Gamma \Gamma'}$ is any graph such that $\Gamma \subset \Xi_{\Gamma \Gamma'}$ and $\Gamma'
\subset \Xi_{\Gamma \Gamma'}$. Then, the kinematical Hilbert space of LQG is the Cauchy completion of
$\mbox{Cyl}$ in the Ashtekar--Lewandowski norm: ${\mathcal H}_{\rm kin}=L^2(\overline{{\mathcal A}},d\mu_0)$.

A basis on this Hilbert space is provided by \emph{spin network} states, which are
constructed as follows. Given a graph $\Gamma$, each edge $e$ is colored by a non-trivial irreducible representation
$\pi_{j_e}$ of $SU(2)$. Spin network states are cylindrical functions with respect to this colored graph. They are denoted by
$T_s := T_{\Gamma,\vec{j}}(\bar A)$ where $\vec{j} = \{ j_e \}$.
Then, every cylindrical function can be expanded in the basis of spin network states.

On $\mbox{Cyl}_\Gamma$ the operators representing the corresponding holonomies act by multiplication, while the operator representing the f\/lux is given by
\begin{gather*}
\hat{E}_\Gamma(S,f)=i2\pi G\hbar\sum_{e\subset\Gamma}\epsilon(e,S)\text{Tr}\left(f^i\tau_i\bar{A}_e\frac{\partial}{\partial \bar{A}_e}\right).
\end{gather*}

To obtain the quantum version of the more general operators, they have to be f\/irst rewritten in terms of the basic holonomy-f\/lux operators.
Note that the quantum conf\/iguration space is not the space of smooth connections but rather the space of holonomies (or generalized connections).
Since the Ashtekar--Lewandowski measure is discontinuous in the connection, there is no well-def\/ined operator for the connection on
${\mathcal H}_{\rm kin}$. Consequently, the curvature must be def\/ined in terms of holonomies before it can be promoted to a quantum operator.
The strategy in the full theory is to def\/ine any general quantum operator via regularization as follows (see~\cite{lqg3, lqg1} for details):
\begin{itemize}\itemsep=0pt
\item the spatial manifold $\Sigma$ is triangulated into elementary tetrahedra;
\item the integral over $\Sigma$ is replaced by a Riemann sum over the cells;
\item for each cell, we def\/ine a regularized expression in terms of the basic operators, such that we get the correct classical expression in the limit
the cell is shrunk to zero;
\item this is promoted to a quantum operator provided it is densely def\/ined on ${\mathcal H}_{\rm kin}$.
\end{itemize}
In the subsequent sections we shall see how the same strategy is applied for def\/ining the quantum operators in LQC. One signif\/icant dif\/ference is that
in the full theory the f\/inal expressions are independent of the regularization, while in the symmetry-reduced models the regularization
(i.e., the size of the cells) cannot be removed and has to be treated as an ambiguity. However, we can f\/ix the form of the ambiguity by taking hints
from the full theory.

One of the most interesting features of LQG is that the spectra of the operators
representing geometrical quantities like area and volume are discrete. Discrete eigenvalues imply that the underlying spatial manifold is also
discrete at least when we are close to the quantum gravity scale. This is a feature of the quantization scheme and it also plays and important role in
the singularity avoidance in LQC minisuperspace models.

This is the kinematical structure of LQG. However we are interested in physical states, i.e.\ states which are annihilated by the all the constraints.
To obtain the physical Hilbert space we now need to solve the quantum constraints. The Gauss constraint is
easy to solve and we can obtain gauge invariant Hilbert space spanned by the gauge invariant spin networks. The inf\/initesimal
dif\/feomorphism constraint cannot be expressed as a self-adjoint operator on~${\mathcal H}_{\rm kin}$. However we can consider f\/inite dif\/feomorphisms
and the solutions to the f\/inite dif\/feomorphism constraint are obtained via \textit{group averaging}.
It turns out that these solutions do not lie in~${\mathcal H}_{\rm kin}$ but in~Cyl$^\star$, the algebraic dual of Cyl.

In the construction of the Hamiltonian constraint operator we face a number of problems (see~\cite{pftreview} and
references therein for details). Although a well-def\/ined Hamiltonian constraint ope\-ra\-tor can be constructed which satisf\/ies an on-shell anomaly-free quantum constraint algebra, the quantization procedure suf\/fers from a number of ambiguities: in the choice of the regulators, in the transcription in terms of basic quantum variables, and in the choice of curvature appro\-xi\-mants. Also the domain of the Hamiltonian constraint operator is not known. Ef\/forts have been made to reduce the ambiguities by studying the of\/f-shell closure of the constraint algebra and by trying to f\/ind the correct semiclassical limit, but no signif\/icant progress has been made so far. So, although we have a well-def\/ined full quantum theory of gravity at the kinematical level, the physical Hilbert-space construction is beset by a number of open problems and is not yet complete.

LQC tries to study some of the features of Loop quantization while avoiding the problems of the full theory. As we shall see later, the programme of LQC tries to closely follow the same steps, as far as possible, in the much simpler case of cosmological models with no (or at most one) f\/ield-theory degrees of freedom. In minisuperspace models it is possible to go beyond the kinematics and construct the physical Hilbert space. Another useful procedure developed to study the ef\/fect of the underlying discreteness is the use of ef\/fective equations to study homogeneous cosmologies and perturbations therein. This has opened up a large number of systems to semiclassical analyses. It is hoped that lessons learned from LQC can give hints about how to tackle the issues being faced in LQG.

\section{Plan of the review}

Signif\/icant progress has been made in the study of a number of cosmologies in LQC. Here, we shall give an overall account of various facets of LQC, outlining technical aspects, reviewing the results achieved and indicating the directions of further research. The rest of the paper is divided into three parts.

In Part~\ref{part1} we discuss LQC minisuperspace models. The simplest cases of minisuperspace are Friedmann--Robertson--Walker (FRW) models, which are homogeneous and isotropic. The kinematical quantization programme followed for these models will be discussed in detail, using the example of f\/lat FRW. We also describe the results obtained in the physical Hilbert space including the dynamical singularity resolution and the bounce.
Open and closed FRW models, with and without a cosmological constant, are brief\/ly discussed. The next level of complication, Bianchi models, consists in removing the assumption of isotropy. In this case, our illustrative example will be the Bianchi~I model but we also indicate the work done so far for Bianchi~II and Bianchi~IX cases.

Then, Part~\ref{part2} focusses on the LQC of midisuperspace models which are neither homogeneous nor isotropic. We describe the only case whose loop quantization has been studied in some detail, the linearly polarized Gowdy $T^3$ model. Two contrasting approaches have been taken in the study of this model. In the f\/irst approach, the degrees of freedom have been separated into homogeneous and inhomogeneous sectors. The homogeneous sector is quantized using
the tools developed in LQC, while the inhomogeneous sector is Fock quantized. In the second approach, the model is studied as a whole mimicking the steps of LQG. We describe and compare both procedures.

Finally, in Part~\ref{part3} we discuss the programme of ef\/fective dynamics developed in LQC. In contrast to the previous two parts, this approach aims to incorporate the ef\/fects of the discrete geometry as corrections to the classical equations. In this way it may be possible to link LQC to phenomenological evidence.

In the end we summarize the current directions of ongoing research. This review is intended as an introduction of the main results achieved in the f\/ield in the past few years, especially in the Hamiltonian formalism, and it does not cover more recent work being done in the area of cosmological perturbations, phenomenology, and spin-foam cosmology. We will comment about these and other lines of research in Sections~\ref{latti} and~\ref{concl}.

\part{Minisuperspaces in loop quantum cosmology}\label{part1}

LQC \cite{lqc2a,lqc2b,lqc1,lqc3} adapts the techniques developed in loop quantum gravity~\cite{lqg3,lqg2, lqg1} to
the quantization of simpler models than the full theory, as minisuperspace models. Minisuperspace models are solutions of Einstein's equations with a high degree
of symmetry, so much so that there are no f\/ield theory degrees of freedom remaining. They lead to homogeneous cosmological solutions
all of which suf\/fer from a singularity where the classical equations of motion break down.
Since, after quantization, these are essentially quantum
mechanical systems, they serve as good toy models for testing the predictions of~LQG.

In LQC, we start from the classically symmetry-reduced phase space and then try to apply the steps followed in LQG to
these systems. Owing to simplif\/ications due to classical symmetry reduction, many technical complications typical of LQG can be avoided, and the quantization
programme can be carried out beyond what has been achieved so far in the full theory. The fact that there is a well-def\/ined full theory which tells us
that the underlying spatial geometry is discrete is a crucial ingredient in the formulation of LQC. A signif\/icant achievement of LQC is the development of
a well-def\/ined quantum theory for cosmological models where the classical singularity is absent. This resolution of the classical singularity is a
robust feature of LQC as it is seen in all the minisuperspace models studied so far, as well as under various choices made in addressing the ambiguities arising
in quantization. In this part we shall review the LQC of various known minisuperspace cosmological scenarios.

\section{Friedmann--Robertson--Walker models}\label{chap:1-flatFRW}

LQC started with the pioneering works by Bojowald~\cite{boj2, boj1a,boj1b,boj1c,boj1d}, that showed the f\/irst attempts of implementing the methods
of LQG to the quantization of the simplest cosmological model: the f\/lat  Friedmann--Robertson--Walker (FRW) model
(homogeneous and isotropic with f\/lat spatial sections), whose geometry is described by a single degree of freedom, the scale factor.
This system, even if very simple, is physically interesting since, at large scales, our universe is approximately homogeneous and isotropic.
In addition, cosmological observations are compatible with a spatially f\/lat geometry.

After the early papers by Bojowald, the kinematic structure of LQC was revised and more rigorously established~\cite{abl}, which made it
possible to complete the quantization of the model in presence of a homogeneous massless scalar f\/ield minimally coupled to the geometry,
as well as to study the resulting quantum evolution \cite{acs, aps1,aps2,aps3}.
Classically, this model represents expanding universes with an initial \emph{big bang} singularity, where certain physical observables, such as the
matter density, diverge. Remarkably, the quantum dynamics resolves the singularity replacing it with a \emph{quantum bounce}, while for semiclassical
states it agrees with the classical dynamics far away form the singularity.
Therefore, even though this is the simplest cosmological model, its loop quantization, also called polymeric quantization, already leads to relevant
results, the most important one being the avoidance of the singularity.

Using the example of the f\/lat FRW model coupled to a massless scalar, we shall discuss in detail the basics and the mathematical structure
of LQC, adopting the so-called improved dynamics prescription \cite{aps3}.

\subsection{Classical phase space description}

\subsubsection{Ashtekar--Barbero formalism}

The classical phase space in the presence of homogeneity is much simpler than the general situation described in the introduction.
In homogeneous cosmology, the gauge and dif\/feomorphisms constraints are trivially satisf\/ied, the
Hamiltonian constraint being the only survivor in the model. Moreover, for f\/lat FRW the spin connection vanishes. In this case, the
geometry part of the scalar constraint in its integral version is\footnote{The lapse function $N$ goes out of the integral due to the homogeneity.} $C_\text{grav}(N)=
NC_\text{grav}$, with
\begin{gather}\label{eq:lig-escalar-lqg-hom}
C_\text{grav}=\int_{\Sigma} d^3x\, \mathcal C=-\frac1{\gamma^2}\int_{\Sigma} d^3x
\frac{\epsilon_{ijk}F^i_{ab}E^{aj}E^{bk}}{\sqrt{|\det(E)|}}.
\end{gather}

Since f\/lat FRW spatial sections $\Sigma$ are non-compact, and the variables that describe it are spatially homogeneous, integrals such as~\eqref{eq:lig-escalar-lqg-hom} diverge. To avoid that, one usually restricts the analysis to a f\/inite cell~$\mathcal V$.
Owing to homogeneity, the study of this cell reproduces what happens in the whole universe. When imposing also isotropy,
the connection and the triad can be described (in a convenient gauge) by a single parameter~$c$ and~$p$, respectively, in the form~\cite{abl}
\begin{gather*}
 A^i_a=c  V_o^{-1/3} \,{}^oe^i_a, \qquad E^a_i=p  V_o^{-2/3} \sqrt{{}^oq}\,{}^oe^a_i.
\end{gather*}
Here we have introduced a f\/iducial co-triad ${}^oe^i_a$ that we will choose to be diagonal,
${}^oe^i_a=\delta^i_a$, and the determinant $\sqrt{{}^oq}$ of the corresponding f\/iducial metric. The
results do not depend on the f\/iducial choice.
With the above def\/initions, the symplectic structure is def\/ined via,
\begin{gather*}
 \{c,p\}=\frac{8\pi G\gamma}{3}.
\end{gather*}
The variable~$p$ is related to the scale factor $a$ commonly employed in geometrodynamics through
the expression
$a(t)=\sqrt{|p(t)|} V_o^{-1/3}$. Note that~$p$ is positive (negative) if physical and f\/iducial triads have the same (opposite) orientation.

On the other hand, a (homogeneous) massless scalar f\/ield~$\phi$, together with its momentum~$P_\phi$, provide the canonical pair describing the matter content, with Poisson bracket
$\{\phi,P_\phi\}=1$.
Then, the total Hamiltonian constraint contains a matter contribution beside the
geometry one, given in equation~\eqref{eq:lig-escalar-lqg-hom}, and reads
\begin{gather}\label{eq:ligFRW}
 C=C_\text{grav}+C_\text{mat}=-\frac6{\gamma^2}c^2\sqrt{|p|}+8\pi G\frac{P_\phi^2}{V}=0,
\end{gather}
where $V=|p|^{3/2}$ is the physical volume of the cell $\mathcal V$.

\subsubsection{Holonomy-f\/lux algebra}

When def\/ining holonomies and f\/luxes in LQC, and in the particular case of isotropic FRW models, owing to the homogeneity it is
suf\/f\/icient to consider straight edges oriented along the f\/iducial directions, and with oriented length equal to $\mu V_o^{1/3}$,
where $\mu$ is an arbitrary real number. Therefore, the holonomy along one such edge, in the $i$-th direction, is given by
\begin{gather*}
 h_i^\mu(c)=e^{\mu c \tau_i}=\cos\left(\frac{\mu c}{2}\right)\mathbbm{1}+2\sin\left(\frac{\mu
c}{2}\right)\tau_i.
\end{gather*}
Then, the gravitational part of the conf\/iguration algebra is the algebra generated by the matrix elements of the holonomies, namely,
the algebra of quasi-periodic functions of $c$, that are the complex exponentials
\begin{gather*}
 \mathcal N_\mu(c)=e^{\frac{i}{2}\mu c}.
\end{gather*}
In analogy with the terminology employed in LQG \cite{lqg3, lqg1}, the vector space of these quasi-periodic functions is called the space of
cylindrical functions def\/ined over symmetric connections, and it is denoted by  $\text{Cyl}_\text{S}$.

In turn, the f\/lux is given by
\begin{gather*}
 E(S,f)=p V_o^{-2/3}A_{S,f},
\end{gather*}
where $A_{S,f}$ is the f\/iducial area of $S$ times an orientation factor (that depends on~$f$). Then, the f\/lux is essentially described by~$p$.

In summary, in isotropic and homogeneous LQC the phase space is described by the va\-riab\-les~$\mathcal N_\mu(c)$ and~$p$, whose Poisson bracket is
\begin{gather*}
 \{\mathcal N_\mu(c),p\}=i\frac{4\pi G\gamma}{3}\mu\mathcal N_\mu(c).
\end{gather*}

\subsection{Kinematical structure}
\label{1sec:kin}

Mimicking the quantization implemented in LQG, in LQC we adopt a representation of the algebra generated by the phase space variables
$\mathcal N_\mu(c)$ and $p$ that is not continuous in the connection, and therefore there is no
operator representing $c$ \cite{abl}. More concretely, the quantum conf\/iguration space is the Bohr
compactif\/ication of the real line, $\mathbb{R}_\text{Bohr}$, and the corresponding Haar measure that
characterizes the kinematical Hilbert space is the so-called Bohr measure~\cite{Vel}.
It is simpler to work in momentum representation. In fact, such Hilbert space is isomorphic to the space of functions of $\mu\in\mathbb{R}$
that are square summable with respect to the discrete measure~\cite{Vel}, known as polymeric space.
In other words, employing the kets $|\mu\rangle$ to denote the quantum states~$\mathcal N_\mu(c)$, whose linear span is the space
$\text{Cyl}_\text{S}$ (dense in $\mathbb{R}_\text{Bohr}$), the kinematical Hilbert space is the completion of
$\text{Cyl}_\text{S}$ with respect to the inner product $\langle \mu|\mu'\rangle=\delta_{\mu\mu'}$.
We will denote this Hilbert space by $\mathcal H_{\text{grav}}$. Note that $\mathcal
H_{\text{grav}}$ is non-separable, since the states $|\mu\rangle$ form a non-countable orthogonal basis.

Obviously, the action of $\hat{\mathcal N}_{\mu}$ on the basis states is
\begin{gather*}
 \hat{\mathcal N}_{\mu'} |\mu\rangle= |\mu+\mu'\rangle.
\end{gather*}
On the other hand, the Dirac rule
$[\hat{\mathcal N}_\mu,\hat p]=i\hbar \widehat{\{\mathcal N_\mu(c),p\}}$
implies that
\begin{gather*}
 \hat{p}|\mu\rangle=p(\mu)|\mu\rangle, \qquad p(\mu)=\frac{4\pi l_\text{Pl}^2\gamma}{3} \mu,
\end{gather*}
where $l_\text{Pl}=\sqrt{G\hbar}$ is the Planck length. As we see, the spectrum of this operator is
discrete, as a~consequence of the representation not being continuous in~$\mu$. Due to this lack of
continuity, the Stone--von Neumann theorem about the uniqueness of the representation in quantum
mecha\-nics~\cite{SvN1,SvN2} is not applicable in this context. Therefore, the loop quantization of
this model is inequivalent to the standard Wheeler--DeWitt (WDW) quantization~\cite{witt, whe},
where operators have a typical Schr\"odinger-like representation. In fact, while the WDW
quantization fails in solving the problem of the big bang singularity, the loop quantization is
singularity free~\cite{aps2,aps3}, as we will see later.

For the matter f\/ield, we adopt a standard Schr\"odinger-like representation, with
$\hat{\phi}$ acting by multiplication and $\hat{P}_\phi=-i\hbar\partial_\phi$ as derivative, being both operators def\/ined
on the Hilbert space $L^2(\mathbb{R},d\phi)$. As domain, we take the Schwartz space
$\mathcal{S}(\mathbb{R})$ of rapidly decreasing functions, which is dense in
$L^2(\mathbb{R},d\phi)$.
The total kinematical Hilbert space is then $\mathcal H_{\text{kin}}=\mathcal
H_{\text{grav}}\otimes L^2(\mathbb{R},d\phi)$.\footnote{Note that the basic operators
def\/ined above are in the tensor product of both sectors (geometry and matter), acting as the
identity in the sector where they do not have dependence. For instance, the operator
$\hat{p}$ def\/ined on $\text{Cyl}_\text{S}\otimes\mathcal{S}(\mathbb{R})$ really means
$\hat{p}\otimes\mathbbm{1}$. Nonetheless, for the sake of simplicity we will ignore the tensor product by
the identity.}

\subsection{Hamiltonian constraint operator}
\label{1sec:lig-ham-aps}

\subsubsection{Curvature operator and improved dynamics}

Since the connection is not well def\/ined in the quantum theory, the classical expression of the Hamiltonian constraint, given in
equation~\eqref{eq:ligFRW}, cannot be promoted directly to an operator. In order to obtain the quantum analogue of the gravitational part, we follow the
procedure adopted in the full theory. We start from the general expression \eqref{eq:lig-escalar-lqg-hom} and express the curvature tensor in terms
of the holonomies, which do have a well-def\/ined quantum counterpart.

Following LQG, we take a closed square loop with holonomy
\begin{gather*}
 h^\mu_{\square_{ij}}=h_i^\mu h_j^\mu (h_i^\mu)^{-1} (h_j^\mu)^{-1},
\end{gather*}
that encloses a f\/iducial area $A_{\square}=\mu^2 V_o^{2/3}$.
The curvature tensor then reads~\cite{abl}
\begin{align}\label{eq:curvatura-exacta}
 F^i_{ab}=-2 \;\lim_{A_{\square}\rightarrow 0}
\text{tr}\left(\frac{h^\mu_{\square_{jk}}-\delta_{jk}}{A_{\square}}\tau^i\right){}^oe^j_a{}
^oe^k_b.
\end{align}
This limit is classically well def\/ined. However, in the quantum theory we cannot contract the area to zero because that limit does not converge.

Since we have a well def\/ined full theory (unlike WDW quantization), we can appeal to the discretization of geometry coming from it.
In LQG, geometric area has a discrete spectrum with a non-vanishing minimum eigenvalue~$\Delta$~\cite{area2, area1}.
This suggests that we should not take the null area limit, but consider only areas larger than~$\Delta$. Then, we contract the
area of the loop till a minimum value
$A_{\square_{\min}}=\bar\mu^2 V_o^{2/3}$, such that the geometric area corresponding to this f\/iducial area, given by the f\/lux
$E(\square_{\min},f=1)=p\bar\mu^2$, is equal to~$\Delta$. In short, the curvature is def\/ined by the regularized expression
\begin{gather}\label{eq:curvatura}
{F}^i_{ab}=-2\,
\text{tr}\left( \frac{h^{\bar\mu}_{\square_{jk}}-\delta_{jk}}{\bar\mu^2
V_o^{2/3}}\tau^i\right){}^oe^j_a{}^oe^k_b,
\end{gather}
where $\bar\mu$, characterizing the minimum area of the loop, is given by the \emph{Ansatz}
\begin{align}\label{mu}
 \frac1{\bar\mu}=\sqrt{\frac{|p|}{\Delta}}.
\end{align}
This choice of $\bar\mu$ is usually called \emph{improved dynamics} in the LQC literature \cite{aps3}.
Note that the smaller the value of  $\bar{\mu}$ is, or equivalently the bigger the value of $|p|$ is, the better equation~\eqref{eq:curvatura} approximates the classical expression~\eqref{eq:curvatura-exacta}, so that both expressions agree in the regime in which the area of the cell under study is large enough.
Finally, the curvature operator is obtained by promoting equation~\eqref{eq:curvatura} to an operator. Let us remark that there are two kinds of ambiguities in the def\/inition of this operator. On the one hand, the value of the parameter  $\bar{\mu}$, that is f\/ixed by the improved dynamics prescription, as we have just explained. On the other hand, we also have the ambiguity in the $SU(2)$ representation we use for calculating the trace. As usual in LQC \cite{lqc1}, we will compute the holonomies in the fundamental representation of spin $1/2$.

Note that terms of the kind ${\mathcal N}_{\bar\mu}=e^{i\bar\mu c/2}$ contribute to
$h^{\bar\mu}_{\square_{ij}}$. In order to def\/ine the operator $\hat{\mathcal
N}_{\bar\mu}=\widehat{e^{i\bar\mu c/2}}$, it is assumed that this operator generates unit
translations over the af\/f\/ine parameter associated with the vector f\/ield $\bar\mu[p(\mu)]\partial_\mu$~\cite{aps3}.
In other words, we introduce a canonical transformation in the geometry sector of the phase space,
such that it is described by the variable $b=\hbar\bar\mu c/2$ and its canonically conjugate
variable $v(p)=(2\pi\gamma l_\text{Pl}^2\sqrt{\Delta})^{-1}\text{sgn}(p)|p|^{3/2}$ ($\text{sgn}$
denotes the sign), with $\{b,v\}=1$. The variable $v(\mu)=v[p(\mu)]$ indeed verif\/ies
$\partial_v=\bar\mu(\mu)\partial_\mu$. Then, we relabel the basis states of $\mathcal
H_{\text{grav}}$ with this new parameter $v$ that, unlike $\mu$, is adapted to the action of
$\hat{\mathcal N}_{\bar\mu}$. In fact, introducing the operator $\hat{v}$ with action
$\hat{v}|v\rangle=v|v\rangle$, it is straightforward to show that $\hat{\mathcal
N}_{\bar\mu}|v\rangle=|v+1\rangle$, so that the Dirac rule
$[\widehat{e^{ib/\hbar}},\hat{v}]|v\rangle=i\hbar
\widehat{\{e^{ib/\hbar},v\}}|v\rangle$ is satisf\/ied. On the other hand, we obtain
$\hat{p}|v\rangle=(2\pi \gamma l_\text{Pl}^2\sqrt{\Delta})^{2/3}\text{sgn}(v)|v|^{2/3} |v\rangle$.

It is worth mentioning that the parameter $v$ has a geometrical interpretation: its absolute value
is proportional to the physical volume of the cell $\mathcal V$, given by
\begin{gather*}
 \hat{V}=\widehat{|p|}^{3/2}, \qquad \hat{V}|v\rangle= 2\pi \gamma
l_\text{Pl}^2\sqrt{\Delta}|v||v\rangle.
\end{gather*}

The quantization within the prescription~\eqref{mu} meant an important improvement for LQC~\cite{aps3}.
Earlier, it was assumed that the minimum f\/iducial length was just some constant~$\mu_o$ related to~$\Delta$~\cite{abl}. However,
the resulting quantum dynamics was not successful, inasmuch as the quantum ef\/fects of the geometry could be important
at scales where the matter density was not necessarily high. In that case, in the semiclassical regime the physical results deviated signif\/icantly
from the predictions made by general relativity~\cite{aps2}. Improved dynamics solves this problem. Furthermore, it has been proved that it is
the only minisuperspace quantization (among a certain family of possibilities) yielding to a physically admissible model~\cite{cs},
independent of the f\/iducial structures, with a well-def\/ined classical limit in agreement with GR, and giving rise to a scale of
Planck order where quantum ef\/fects are important and solve the singularity problem.

\subsubsection{Representation of the Hamiltonian constraint}

When trying to promote the gravitational part of the scalar constraint~\eqref{eq:lig-escalar-lqg-hom} to an operator, we f\/ind an additional dif\/f\/iculty concerning the
inverse of the volume,
\begin{gather*}
 \frac1{V}=\frac{\sqrt{{}^o q}}{\sqrt{|\det(E)|}V_o}.
\end{gather*}
The volume operator has a discrete spectrum with the eigenvalue zero included, so its inverse
(obtained by using the spectral theorem) is not well def\/ined in zero.
Nonetheless, following LQG \cite{inv-vol-lqg1,inv-vol-lqg2}, from the classical identity
\begin{gather}\label{eq:identidad-lig}
\frac{\epsilon_{ijk}E^{aj}E^{bk}}{\sqrt{|\det(E)|}} =\sum_{k=1}^3\frac{\text{sgn}(p)}{2\pi\gamma
GV_o^{1/3}}\frac1{l}\, {}^o e^k_c \,{}^o \epsilon^{abc}\,
\text{tr}\big(h_k^{l}(c)\big\{[h_k^{l}(c)]^{-1},
V\big\}\tau_i\big),
\end{gather}
we can obtain an operator for the left-hand side of this expression by promoting the functions on the right-hand side to the
corresponding operators, and by making the replacement
$
 \widehat{\{\;\;,\;\;\}}\rightarrow -(i/\hbar)[\hat{\;\;},\hat{\;\;}]
$.
Note that the parameter $l$ labels a quantization ambiguity. In order not to introduce new scales
in the theory, we take for $l$ the value $\bar\mu=\sqrt{{\Delta}/{|p|}}$~\cite{aps3}.

Plugging this result into the Hamiltonian constraint~\eqref{eq:lig-escalar-lqg-hom},
as well as the curvature given in equation~\eqref{eq:curvatura}, we obtain that the geometry (or
gravitational) contribution to the Hamiltonian constraint operator is~\cite{aps3}
\begin{gather}\label{eq:operador-lig-aps}
 \widehat{C}_\text{grav}=
i\frac{3\widehat{\text{sgn}(p)}}{2\pi\gamma^3
l_\text{Pl}^2\Delta^{3/2}}\hat{V}[\widehat{\sin\left(\bar\mu
c\right)}\widehat{\text{sgn}(p)}]^2\big(\hat{\mathcal
N}_{\bar{\mu}}\hat{V}\hat{\mathcal N}_{-\bar{\mu}}- \hat{\mathcal
N}_{-\bar{\mu}}\hat{V}\hat{\mathcal N}_{\bar{\mu}}\big),
\end{gather}
with
\begin{gather*}
\widehat{\sin(\bar\mu c)}=\frac{\hat{\mathcal
N}_{2\bar\mu}-\hat{\mathcal N}_{-2\bar\mu}}{2i}.
\end{gather*}

Let us now deal with the representation of the matter contribution, given in the second term of
equation~\eqref{eq:ligFRW}. To represent the inverse of the volume, we follow the same strategy as before,
now starting with the classical identity
\begin{gather*}
\frac{\text{sgn}(p)}{|p|^{1-s}} =\frac{1}{s4\pi\gamma
G}\frac1{l}\text{tr}\left(\sum_i\tau^ih_i^{l}(c)\big\{[h_i^{l}(c)]^{-1},
|p|^s\big\}\right).
\end{gather*}
As before, we take the trace in the fundamental representation and we choose $l$ equal to $\bar\mu$ in the quantum theory. To f\/ix the ambiguity in the constant $s>0$, we choose for simplicity $s=1/2$. We obtain
\begin{gather}
\widehat{\left[\frac{1}{\sqrt{|p|}}\right]}
 =\frac{3}{4\pi\gamma
l_\text{Pl}^2\sqrt{\Delta}}\widehat{\text{sgn}(p)}\widehat{\sqrt{|p|}}\left(\hat{\mathcal
N}_{-\bar{\mu}}\widehat{\sqrt{|p|}}\hat{\mathcal
N}_{\bar{\mu}}- \hat{\mathcal
N}_{\bar{\mu}}\widehat{\sqrt{|p|}}\hat{\mathcal
N}_{-\bar{\mu}}\right).
\end{gather}
The action of this operator on the basis states is diagonal and given by
\begin{gather*}
 \widehat{\left[\frac1{\sqrt{|p|}}\right]}|v\rangle=b(v)|v\rangle,\qquad
b(v)=\frac{3}{2}\frac1{(2\pi\gamma l_\text{Pl}^2\sqrt{\Delta})^{1/3}}|v|^{1/3}
\big||v+1|^{1/3}-|v-1|^{1/3}\big|.
\end{gather*}
While, for large values of $v$, $b(v)$ is well approximated by the classical value
$1/\sqrt{|p|}$, for small values of $v$ they dif\/fer considerably. In fact, the above operator is
bounded from above and annihilates the zero-volume states.

The matter contribution to the constraint is then given by the operator
\begin{gather*}
\widehat{C}_\text{mat}=-8\pi l_\text{Pl}^2\hbar
\widehat{\left[\frac1{V}\right]}\partial_\phi^2, \qquad
\widehat{\left[\frac1{V}\right]}=\widehat{\left[\frac1{\sqrt{|p|}}\right]}^3.
\end{gather*}

In order for the Hamiltonian constraint operator $\widehat{C}=\widehat{C}_\text{grav}+\widehat{C}_\text{mat}$
to be (essentially) self-adjoint, we need to symmetrize the gravitational term~\eqref{eq:operador-lig-aps}. There is an ambiguity in the chosen symmetric factor ordering and
several possibilities have been studied in the literature~\cite{acs,aps3,klp,mmo,mop,ydm} (see~\cite{mop} for a detailed comparison between them). Due to its suitable properties, here we will
adopt the prescription called sMMO in~\cite{mop}\footnote{The acronym ``MMO'' refers to the
model of \cite{mmo}, by Mart\'in-Benito, Mena Marug\'an, and Olmedo.}, that is a simplif\/ied version
of the prescription of \cite{mmo}. Its two main features are:
\begin{enumerate}\itemsep=0pt
\item [i)] decoupling of the zero-volume state $|v=0\rangle$;
\item [ii)] decoupling of states with opposite orientation of the densitized triad, namely states
$|v<0\rangle$ are decoupled from states $|v>0\rangle$.
\end{enumerate}
As we will see, this will give rise to simple superselection sectors with nice properties.
Re\-markably, the behavior of the resulting eigenstates of the gravitational part of the
constraint already shows the occurrence of a generic quantum bounce dynamically resolving the
singularity. Therefore, this prescription ensures that the quantum bounce mechanism is an intrinsic
feature of the theory, independent of the particular physical state considered\footnote{In~\cite{aps3}, the quantum bounce was shown just for particular semiclassical states. Then, with the
factor ordering adopted in~\cite{acs}, it was shown that the quantum bounce is generic, but the
result is only obtained for a specif\/ic superselection sector. The results of~\cite{mmo} are instead
completely general.}.

Then, following \cite{mmo,mop}, we take
\begin{gather}\label{eq:operador-lig-mmo}
\widehat{C}=\widehat{\left[
\frac{1}{V}\right]}^{1/2}\left(-\frac{6}{\gamma^{2}}
\widehat{\Omega}^2+8\pi
G\hat{P}_{\phi}^2\right)\widehat{\left[
\frac{1}{V}\right]}^{1/2},
\end{gather}
where the operator $\widehat\Omega$ is def\/ined as
\begin{gather}\label{eq:operador-grav-mmo}
\widehat\Omega =\frac1{4i\sqrt{\Delta}}\widehat{|p|}^{3/4}
\left[\big(\hat{\mathcal
N}_{2\bar\mu}-\hat{\mathcal
N}_{-2\bar\mu}\big)\widehat{\text{sgn}(p)}
+\widehat{\text{sgn}(p)}\big(\hat{\mathcal
N}_{2\bar\mu}-\hat{\mathcal
N}_{-2\bar\mu}\big)\right]\widehat{|p|}^{3/4}.
\end{gather}
The action of $\widehat{\text{sgn}(p)}$ on the state $|v=0\rangle$ can be def\/ined
arbitrarily, since the f\/inal action of $\widehat\Omega$ is independent of that choice, provided
that $\widehat\Omega|0\rangle=0$.

Thanks to the splitting of powers of~$p$ on the left and on the right, $\widehat{C}$ annihilates
the subspace of zero-volume states and leaves invariant its orthogonal complement, thus decoupling
the zero-volume states as desired. We can then remove the state~$|0\rangle$ and def\/ine the
operators acting on the geometry sector on the Hilbert space
$\widetilde{\mathcal{H}}_{\textrm{grav}}$ def\/ined as the Cauchy completion (with respect to the
discrete
measure) of the dense domain
\begin{gather*}
 \widetilde{\textrm{Cyl}}_\text{S}=\text{span}\{|v\rangle;\;v\in\mathbb{R}\setminus\{0\}\}.
\end{gather*}
As a consequence, the big bang is resolved already at the kinematical level, in the sense that
the quantum equivalent of the classical singularity (namely, the eigenstate of vanishing physical
volume) has been entirely removed from the kinematical Hilbert space (see also~\cite{boj}).

In view of the operator~\eqref{eq:operador-lig-mmo}, it is more convenient to work with its
densitized version, def\/i\-ned~as
\begin{gather*}
\widehat{{\cal C}}=
\widehat{\left[\frac1{V}\right]}^{-1/{2}}
\widehat{C}\widehat{\left[\frac{1}{V}
\right]}^{-1/{2}}=-\frac{6}{\gamma^{2}}
\widehat{\Omega}^2+8\pi
G\hat{P}_{\phi}^2,
\end{gather*}
since the operators $\widehat{\Omega}^2$ and $\hat{P}_{\phi}^2=-\hbar^2\partial_\phi^2$ become Dirac
observables that commute with the densitized constraint operator $\widehat{{\cal C}}$.
Note that, if we had not decoupled the zero-volume states, zero would be in the discrete
spectrum of $\widehat{[1/V]}$ and the operator $\widehat{\left[1/{V}\right]}^{-1/2}$
(obtained via spectral theorem) would be ill def\/ined. Nonetheless, in
$\widetilde{\mathcal{H}}_{\textrm{grav}}$ (with domain $\widetilde{\text{Cyl}}_\text{S}$) it is
well def\/ined.
Both the densitized and original constraints are equivalent, inasmuch as their solutions are
bijectively related~\cite{mmo}.

\subsection{Analysis of the Hamiltonian constraint operator}
\label{3sec:grav}

With the aim of diagonalizing the Hamiltonian constraint operator $\widehat{{\cal C}}$, let us characterize the spectral properties of the
operators entering its def\/inition. As it is well known, the operator $\hat{P}_{\phi}^2=-\hbar^2\partial_\phi^2$ is
essentially self-adjoint in its domain $\mathcal{S}(\mathbb{R})$, with double degenerate absolutely
continuous spectrum, its generalized eigenfunctions of eigenvalue~$(\hbar\nu)^2$ being the plane waves~$e^{\pm i|\nu|\phi}$. The gravitational operator $\widehat\Omega^2$
is more complicated and we analyze it in detail in the following.

\subsubsection{Superselection sectors}

The action of $\widehat\Omega^2$ on the basis states $|v\rangle$ of the kinematical sector
$\widetilde{\mathcal{H}}_{\textrm{grav}}$ is
\begin{gather*}
\widehat\Omega^2|v\rangle =-f_+(v)f_+(v+2)|v+4\rangle+
\left[f_+^2(v)+f_-^2(v)\right]|v\rangle-f_-(v)f_-(v-2)
|v-4\rangle,
\end{gather*}
where
\begin{gather*}
f_\pm(v)=\frac{\pi\gamma l_{\textrm{Pl}}^2}{2}
\sqrt{|v\pm2|}\sqrt{|v|}s_\pm(v),\qquad
s_\pm(v)=\text{sgn}(v\pm2)+\text{sgn}(v),
\end{gather*}
so that $\widehat\Omega^2$ is a dif\/ference operator of step four.
In addition, note that $f_-(v)f_-(v-2)=0$ if $v\in(0,4]$ and $f_+(v)f_+(v+2)=0$ if $v\in[-4,0)$.
In consequence, the operator~$\widehat\Omega^2$ only relates states $|v\rangle$ with support in a
particular semilattice of step four of the form
\begin{gather*}
{\mathcal L}_{\varepsilon}^\pm=\{v=\pm(\varepsilon+4n),
\,n\in\mathbb{N}\},\qquad \varepsilon\in(0,4].
\end{gather*}
Then, $\widehat\Omega^2$ is well def\/ined in any of the Hilbert subspaces $\mathcal
H^\pm_{\varepsilon}$ obtained as the closure of the respective domains
$\text{Cyl}_{\varepsilon}^{\pm}=\text{lin}\{|v\rangle,\,v\in{\mathcal L}_{\varepsilon}^\pm\}$,
with respect to the discrete inner product. The non-separable kinematical Hilbert space
$\widetilde{\mathcal H}_{\text{grav}}$ can be thus written as a direct sum of separable subspaces
$\widetilde{\mathcal H}_{\text{grav}}=\oplus_{\varepsilon}(\mathcal H^+_{\varepsilon}\oplus\mathcal
H^-_{\varepsilon})$.

The action of the Hamiltonian constraint (and that of the physical observables, as we will see)
preserves the spaces
$\mathcal H^\pm_{\varepsilon}\otimes L^2(\mathbb{R},d\phi)$, which then provide
superselection sectors. Therefore, we can restrict the analysis to any of them, e.g., to
$\mathcal H^+_{\varepsilon}\otimes L^2(\mathbb{R},d\phi)$, for an arbitrary value of
$\varepsilon\in(0,4]$.

The fact that the gravitational part of the Hamiltonian constraint is a dif\/ference
operator is due to the discreteness of the geometry representation, and therefore it is a generic
feature of the theory. Actually, the dif\/ferent factor orderings analyzed within the improved
dynamics prescription (e.g., \cite{acs,aps3,mmo}) display superselection sectors having support in
lattices of step four. The dif\/ference between the superselection sectors considered here \cite{mmo}
and those of \cite{acs, aps3} is that the formers have support contained in a semiaxis of the
real line, whereas the support of the latters is contained in the whole real line.

\subsubsection{Self-adjointness and spectral properties}

Though the gravitational part of the Hamiltonian constraint operator is not a usual dif\/ferential
operator but a dif\/ference operator, there exists a rigorous proof showing that
it is essentially self-adjoint \cite{kale}. Here we sketch that proof for the operator that we are
considering, $\widehat\Omega^2$, but indeed the proof can be extended for the dif\/ferent
orderings explored in the literature (e.g., \cite{acs, aps3})\footnote{$\widehat\Omega^2$ is
analog to the operator $\Theta$ of \cite{aps3}.}.

In \cite{kale} the authors def\/ine certain operator $\widehat{H}'_\text{APS}$,\footnote{The acronym ``APS'' refers to the
model of \cite{aps3} by Ashtekar, Paw{\l}owski and Singh.} which is a dif\/ference operator of step four, and they show that
$\widehat{H}'_\text{APS}$ is unitarily related, through a Fourier transformation, to the Hamiltonian of a
point particle in a one-dimensional P\"oschl--Teller potential, which is a well-known
dif\/ferential operator. In particular, it is essentially self-adjoint, and then so is
$\widehat H'_\text{APS}$ as well.

In our notation, $\widehat{H}'_\text{APS}$ is def\/ined on the Hilbert spaces:
\begin{itemize}\itemsep=0pt
\item $\mathcal H^+_{\varepsilon}\oplus\mathcal H^-_{4-\varepsilon}$, with domain
$\text{Cyl}_{\varepsilon}^{+}\cup
\text{Cyl}_{4-\varepsilon}^{-},\text{ if }\varepsilon\neq4$;
\item $\mathcal H^+_{4}\oplus\mathcal H^-_{4}\oplus\mathcal H_0$, ($\mathcal H_0$ being the
one-dimensional Hilbert space generated by
$|v=0\rangle$), with domain
$\text{Cyl}_{4}^{+}\cup\text{Cyl}_{-4}^{-}\cup\text{lin}\{|0\rangle\},
\text{ if}$ $\varepsilon=4$.
\end{itemize}
Now, one can show that $\widehat\Omega^2$ and $[4/(3 \pi G)]\widehat{H}'_\text{APS}$ (def\/ined on the same Hilbert space)  dif\/fer in a~trace class symmetric operator \cite{mmo,mop}. Then, a theorem by Kato and Rellich \cite{kato} ensures that $\widehat\Omega^2$, def\/ined in the same Hilbert space as $\widehat{H}'_\text{APS}$, is essentially
self-adjoint. From this result, it is not dif\/f\/icult to prove also that the restriction of $\widehat\Omega^2$
to $\mathcal H^+_{\varepsilon}$ (the subspace where we have restricted the analysis) is also essentially self-adjoint \cite{mmo}, just by analyzing its def\/iciency index equation \cite{functional1}.

On the other hand, it was shown in \cite{kale} that the essential and the absolutely
continuous spectra of the operator $H'_\text{APS}$ are both $[0,\infty)$. Once again,
Kato's perturbation theory \cite{kato} allows one to extend these results to the operator
$\widehat\Omega^2$ def\/ined in $\mathcal H^+_{\varepsilon}\oplus
\mathcal H^-_{4-\varepsilon}$. In addition, taking into account the symmetry of
$\widehat\Omega^2$ under a f\/lip of sign in $v$ and assuming the
independence of the spectrum from the label $\varepsilon$, we conclude that the essential and
absolutely continuous spectra of~$\widehat\Omega^2$ def\/ined in $\mathcal H^+_{\varepsilon}$
are $[0,\infty)$ as well. Besides, as we will see in next subsection, the (generalized)
eigenfunctions of $\widehat\Omega^2$ converge for large~$v$ to eigenfunctions of the
WDW counterpart of the operator. This fact, together with the continuity of the spectrum in
geometrodynamics, suf\/f\/ices to conclude that the discrete and singular spectra are empty.

In summary, the operator $\widehat\Omega^2$ def\/ined on~$\mathcal H^+_{\varepsilon}$
is a positive and essentially self-adjoint operator, whose spectrum is absolutely continuous and
given by $\mathbb{R}^+$.

\subsubsection{Generalized eigenfunctions}

Let us denote by $|e^{\varepsilon}_{\lambda}\rangle=\sum_{v\in{\mathcal L}^+_{\varepsilon}}
e^{\varepsilon}_{\lambda}(v)|v\rangle$ the generalized eigenstates of $\widehat\Omega^2$, corresponding to the eigenvalue
(in generalized sense) $\lambda\in[0,\infty)$.
The analysis of the eigenvalue equation
$\widehat\Omega^2|e^{\varepsilon}_{\lambda}\rangle=\lambda|e^{\varepsilon}_{\lambda}
\rangle$ shows that the initial datum $e^{\varepsilon}_{\lambda}(\varepsilon)$ completely determines the rest of
eigenfunction coef\/f\/icients $e^{\varepsilon}_{\lambda}(\varepsilon+4n)$, $n\in \mathbb{N}^+$ \cite{mmo}. Therefore, the spectrum of $\widehat\Omega^2$, besides being positive and absolutely continuous, is also non-degenerate.
We choose a basis of states $|e^{\varepsilon}_{\lambda}\rangle$ normalized to the
Dirac delta such that $\langle e^{\varepsilon}_{\lambda}|e^{\varepsilon}_{\lambda'}\rangle=
\delta(\lambda-\lambda')$.
This condition f\/ixes the complex norm of $e^{\varepsilon}_{\lambda}(\varepsilon)$. The only remaining freedom in the choice of this initial datum is then its phase, that we f\/ix by taking $e^{\varepsilon}_{\lambda}(\varepsilon)$
positive. The generalized eigenfunctions that form the basis are then real, a consequence of the
fact that the dif\/ference operator $\widehat{\Omega}^2$ has real coef\/f\/icients. In short, the
spectral resolution of the identity in the kinematical Hilbert space $\mathcal
H^+_{\varepsilon}$ associated with $\widehat{\Omega}^2$ can be expressed as
\begin{gather*}
\mathbbm{1}=\int_{\mathbb{R^+}} d\lambda
|e_{\lambda}^{\varepsilon}\rangle \langle
e_{\lambda}^{\varepsilon}|.
\end{gather*}

The behavior of the eigenfunctions $e^{\varepsilon}_{\lambda}(\varepsilon)$ in the limit $v\rightarrow\infty$ allows us to understand the relation between the quantization of the model within LQC and that of the standard WDW theory, where a Schr\"odinger-like representation is employed in the geometry sector, instead of polymeric. Let us study this limit.

In the WDW theory the analog to the operator $\widehat\Omega^2$ is simply given by \cite{mmo}
\begin{gather*}
 \widehat{\underline\Omega}^2=-\frac{\alpha^2}{4}\left[1+4v\partial_{v}+
4(v\partial_v)^2\right],
\end{gather*}
where $\alpha=4\pi\gamma l_{\text{Pl}}^2$. $\widehat{\underline\Omega}^2$ is well def\/ined on the Hilbert space $L^2(\mathbb{R}^+,dv)$. Moreover, it is essentially self-adjoint, and its spectrum is absolutely continuous with double degeneracy.
The generalized eigenfunctions corresponding to the eigenvalue $\lambda\in[0,\infty)$ will be labeled with
$\omega=\pm\sqrt{\lambda}\in\mathbb{R}$ and are given by
\begin{gather}\label{eq:wdw-eig-mmo}
\underline{e}_{\omega}(v)= \frac{1}{\sqrt{2\pi\alpha
|v|}}\exp\left({-i\omega\frac{\ln{|v|}}{\alpha}}\right).
\end{gather}
These eigenfunctions provide an orthogonal basis (in a generalized sense) for $L^2(\mathbb{R}^+,dv)$, with normalization
$\langle \underline{e}_{\omega} |\underline{e}_{\omega^{\prime}}\rangle=\delta(\omega-\omega^{\prime})$.

Using the results of \cite{kp-posL}, one can show that the loop basis eigenfunctions $e^{\varepsilon}_{\lambda}(v)$ converge for large $v$ to an eigenfunction of the WDW analog $\widehat{\underline\Omega}^2$. The WDW limit is explicitly given by~\cite{mmo}
\begin{gather*}
e^{{\varepsilon}}_{\lambda}(v) \xrightarrow{v\gg 1} r\big\{
\exp\left[{i\phi_{\varepsilon}(\omega)}\right]
\,\underline{e}_{\omega}(v) +
\exp\left[{-i\phi_{\varepsilon}(\omega)}\right]
\,\underline{e}_{-\omega}(v) \big\},
\end{gather*}
where $r$ is a normalization factor. In turn, the phase $\phi_{\varepsilon}(\omega)$ behaves as \cite{kp-posL, mop}
\begin{gather*}
\phi_{\varepsilon}(\omega) = T(|\omega|) +
c_{\varepsilon} +
R_{\varepsilon}(|\omega|),
\end{gather*}
where $T$ is a certain function of $|\omega|$, $c_{\varepsilon}$ is a constant, and
$\lim\limits_{\omega\rightarrow0}R_{\varepsilon}(|\omega|)=0$.

\subsection{Physical structure}
\label{1sec:phys}

\subsubsection{Physical Hilbert space}

We are now in a position to complete the quantization of the model. In order to do that, we can
follow two alternative strategies:
\begin{itemize}\itemsep=0pt
\item We can apply the group averaging procedure \cite{gave2,gave1d,gave1c,gave1b, gave1a}.
The physical states are the states invariant under the action of the group generated by the
self-adjoint extension of the constraint operator, and we can obtain them by averaging over that
group. In addition, this averaging determines a natural inner product
that endows the physical states with a~Hilbert structure.
 \item We can solve the constraint in the space
$\big(\widetilde{\text{Cyl}}_\text{S}\otimes\mathcal{S}(\mathbb{R})\big)^*$,
dual to the domain of def\/inition of the Hamiltonian constraint operator\footnote{We do not expect the solutions of the constraint to live in the kinematical Hilbert space $\mathcal H^\pm_{\varepsilon}\otimes L^2(\mathbb{R},d\phi)$, which is quite restricted, but rather in the larger space
\[\big(\widetilde{\text{Cyl}}_\text{S}\otimes\mathcal{S}(\mathbb{R})\big)^*\supset\mathcal H^\pm_{\varepsilon}\otimes L^2(\mathbb{R},d\phi)\supset\widetilde{\text{Cyl}}_\text{S}\otimes\mathcal{S}(\mathbb{R}) .
\]}. Namely, we can look for the elements $(\psi|\in\big(\widetilde{\text{Cyl}}_\text{S}\otimes\mathcal{S}(\mathbb{R})\big)^*$ that verify
$(\psi|\widehat{\mathcal C}^\dagger =0$.
Then,  in order to endow them with a Hilbert space structure, we
can impose self-adjointness in a complete set of observables. This determines the physical inner product~\cite{red2, red1}.
\end{itemize}

Both methods give the same result (up to unitary equivalence): the physical solutions are given by\footnote{See, e.g., \cite{aps3} for the application of the group averaging method, or~\cite{mmo} as an example of the second method.}
\begin{gather}\label{ga3}
 \Psi(v,\phi) =\int_0^\infty d\lambda\,
e^{\varepsilon}_\lambda(v)\big[\tilde\psi_+(\lambda)e^{i\nu(\lambda)\phi}
+\tilde\psi_-(\lambda)e^ {
-i\nu(\lambda)\phi }
\big],
\end{gather}
where
\begin{gather*}
 \nu(\lambda) :=\sqrt{\frac{3\lambda}{4\pi l_\text{Pl}^2\hbar\gamma^2}}.
\end{gather*}
In addition, the physical inner product is
\begin{gather*}
 \langle \Psi_1|\Psi_2 \rangle_{\text{phys}}= \int_0^\infty d\lambda\,\big[
\tilde\psi_{1+}^*(\lambda)\tilde\psi_{2+}(\lambda)+\tilde\psi_{1-}^*
(\lambda)\tilde\psi_{2-}(\lambda)\big ].
\end{gather*}
Therefore, the physical Hilbert space, where the spectral prof\/iles $\tilde\psi_\pm(\lambda)$ live,
is
\begin{gather*}
 \mathcal
H^{\varepsilon}_{\text{phys}}=L^2\left(\mathbb{R}^+,d\lambda\right).
\end{gather*}

\subsubsection{Evolution picture and physical observables}
\label{1subsec:evol}

In any gravitational system, as the one considered here, the Hamiltonian is a linear combination of
constraints, and thus it vanishes. In other words, the time coordinate of the metric is not a~physical time, and provides a~notion of ``frozen'' evolution, unlike what happens in theories, such as usual QFT, in
which the metric is a static background structure. With the aim of
interpreting the results in a time evolution picture, we need to def\/ine what this concept of evolution is.
To do that, we choose a suitable variable or a function of the phase space, and regard it as
internal time \cite{kuchar2}.

In the model that we are describing, it is natural to choose $\phi$ as the physical time. In this way, we
can regard the Hamiltonian constraint as an evolution equation $\phi$. In turn, $\nu$ plays
the role of frequency associated to that time.
As we see in equation~\eqref{ga3}, the solutions to the constraint can be decomposed in
positive and negative frequency components
\begin{gather*}
 \Psi_\pm(v,\phi) =\int_0^\infty d\lambda\,
e^{\varepsilon}_\lambda(v)\tilde\psi_\pm(\lambda)e^{\pm i\nu(\lambda)\phi},
\end{gather*}
that, moreover, are determined by the initial data $\Psi_\pm(v,\phi_0)$ via the unitary evolution
\begin{subequations}
\label{eq:evol}
\begin{gather}
 \Psi_\pm(v,\phi) = U_\pm(\phi-\phi_0)\Psi_\pm(v,\phi_0) ,\\
U_\pm(\phi-\phi_0) = \exp{\left[\pm
i\sqrt{\frac{3}{4\pi l_\text{Pl}^2\hbar\gamma^2} \widehat{\Omega}^2}(\phi-\phi_0)
\right]}.
\end{gather}
\end{subequations}

This allows us to def\/ine Dirac observables ``in evolution'', namely relational observables \cite{bianca-obs1,bianca-obs2, rovelli-obs}, and in turn, to interpret the physical results. Let us note f\/irst that,
in the classical theory, although~$v$ is not a constant of motion, $v(\phi)$ turns out to be a single-valued function of~$\phi$ in
each dynamical trajectory~\cite{aps3}, and then $v|_{\phi=\phi_0}$ is a well-def\/ined observable
for each f\/ixed value~$\phi_0$. It measures the volume at time $\phi_0$. The quantum analogue of that observable is the operator
\begin{gather*}
 \widehat{v}|_{\phi_0} \Psi(v,\phi)= U_+(\phi-\phi_0)v \Psi_+(v,\phi_0)+
U_-(\phi-\phi_0)v\Psi_-(v,\phi_0).
\end{gather*}
We see that, given a physical solution $\Psi(v,\phi)$, the action of this operator consists in:
\begin{enumerate}\itemsep=0pt
\item [i)] decomposing the solution in its positive and negative frequency components,
\item [ii)] freezing them at the initial time $\phi=\phi_0$,
\item [iii)] multiplying its initial datum by $v$, and
\item [iv)] evolving through equation~\eqref{eq:evol}.
\end{enumerate}
The result is again a physical solution, and then the operator $\widehat{v}|_{\phi_0}$ constructed in this way is indeed a Dirac observable.

Then, the constant of motion $\hat P_\phi=-i\hbar\partial_\phi$
and the operator $\widehat{v}|_{\phi}$ form a complete set of Dirac (and then physical) observables.
Note that both the physical observables and the physical inner product preserve not only the superselection sectors,
but also the subspaces of positive and negative frequency. Therefore, any of these subspaces provide an irreducible
representation of the observables algebra, and the analysis can be restricted, for instance, to the positive frequency sector.

The operator $\widehat{v}|_{\phi}$ allows to analyze the physical results in evolution.
Namely, one can compute the expectation value of that observable on physical states at dif\/ferent
times. We will carry out that analysis in  the next section for semiclassical
states, and see graphically the occurrence of the quantum bounce.

\subsection{Dynamical singularity resolution: quantum bounce}
\label{sec:bounce}

In the classical theory, when the volume of the universe vanishes, the energy density diverges, leading to a big bang singularity.
Now, in the quantum theory,  the structure of the superselection sectors, and more specif\/ically the form of the eigenfunctions
$e^{\varepsilon}_\lambda(v)$, ensures that the classical big bang singularity is replaced by a quantum bounce.
Actually, this result is a consequence of the following properties:
\begin{itemize}\itemsep=0pt
  \item {Exact standing-wave behavior}:
As we have seen, the eigenfunctions $e^{\varepsilon}_\lambda(v)$ converge in the large $v$ limit to a combination of two eigenfunctions of
the WDW theory. These eigenfunctions, given in equation~\eqref{eq:wdw-eig-mmo}, contract and expand in $v$, respectively, and can be interpreted as
incoming and outgoing waves. These components contribute with the same amplitude to the limit, and in this sense the limit is an exact standing-wave.
 \item {No-boundary description}: On the other hand, the eigenfunctions $e^{\varepsilon}_\lambda(v)$ have support in a single semiaxis that, moreover, does not contain the
putative singularity $v=0$. This feature is due just to the functional properties of the gravitational operator $\widehat{\Omega}^2$, and
not derived from imposing any particular boundary condition. In that sense, the eigenfunctions verify a no-boundary description\footnote{In quantum cosmology, the concept of no-boundary has been employed in a dif\/ferent sense of the one discussed here  \cite{haw2,haw1,haw3}.}.
\end{itemize}
These features imply that the incoming component must evolve into the outgoing one, and vice versa, since the f\/lux cannot escape through $v=0$.
Therefore, in the physical solution~\eqref{ga3}, restricted for instance to the positive frequency sector, the expanding and contracting components
must lead to two branches of a universe, one in expansion and one in contraction, that meet at some positive expectation value of
$\hat{v}|_{\phi_0}$ forming a quantum bounce. This result is then independent of the considered physical prof\/ile $\tilde\psi_+(\lambda)$.

\begin{figure}[t]
\centering
\includegraphics[width=10cm]{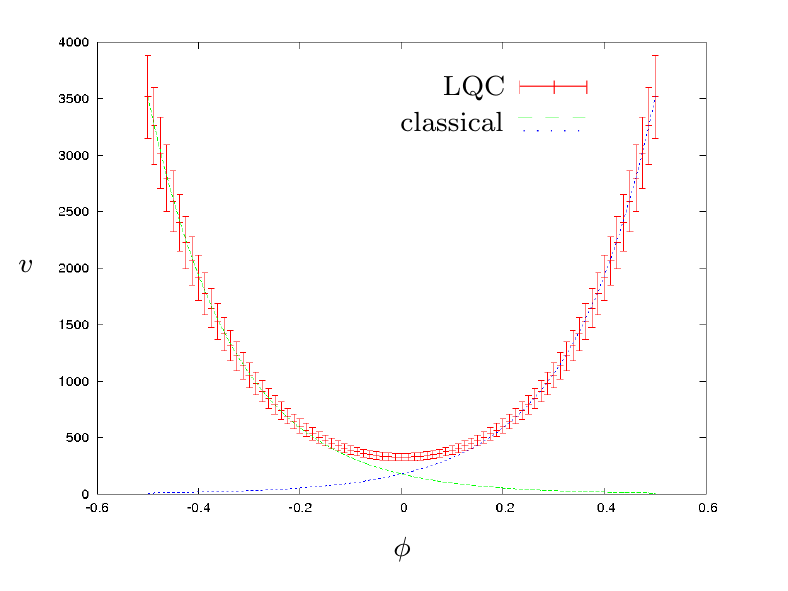}
\caption{Expectation values and dispersions of $\hat{v}|_{\phi}$ (in red, vertical bars) in the superselection
sector with $\varepsilon=1$, corresponding to the
physical prof\/ile given in equation~\eqref{profile} with $\langle \hat P_\phi \rangle=1000$ and
${\Delta\hat P_\phi}/{\langle \hat P_\phi \rangle}=0.1$. The quantum evolution is also compared with
the classical trajectories, one
in expansion (blue curve, increasing from left to right) and the other in contraction (green curve, decreasing). Graph by courtesy of J.~Olmedo.}\label{fig:bounce}
\end{figure}

For semiclassical states in the region of large $v$, the expectation value of $\hat{v}|_{\phi_0}$ is
peaked on trajectories that show the replacement of the classical big bang by a \emph{big bounce},
as depicted in Fig.~\ref{fig:bounce}. This example corresponds to a physical prof\/ile given by a
logarithmic normal distribution of the type
\begin{gather}\label{profile}
\tilde\psi_+(\lambda)=\frac1{(2\pi)^{1/4}\sqrt{\sigma\lambda}} e^{-[\ln(\lambda/\lambda_o)]^2/(4\sigma^2)}
\end{gather}
(as the ones considered in \cite{mop}), where the parameters $\lambda_o$ and $\sigma$ are related to the expectation value of
$\hat P_\phi$ and to its dispersion $\Delta\hat P_\phi$ by the relations~\cite{mop}
\begin{gather*}
\langle \hat P_\phi \rangle=\sqrt{12\pi G}\lambda_o e^{\sigma^2/2},\qquad \frac{\Delta\hat
P_\phi}{\langle \hat P_\phi \rangle}=\sqrt{e^{\sigma^2}-1} .
\end{gather*}
Around the bounce point, the expectation values approach the classical value very fast, so that the
semiclassical limit of the quantum theory agrees with general relativity, as desired. Furthermore,
it has been proven for quite a general class of states that semiclassicality is preserved
through the bounce~\cite{cs2, kp-posL}.

Another analytic result holds independently of the choice of
state, and can be illustrated in the $b$ representation~\cite{acs}. To this
purpose, we choose already classically the densitized Hamiltonian constraint, i.e., the total Hamiltonian such that the lapse function is equal to the volume, $N=\sqrt{|\det(E)|}=V$. Thus, one avoids the need to rewrite inverse powers of the volume in terms of Poisson brackets:
\begin{gather}
C  =  -\frac{1}{\g^2}{E_i^{\a}E_j^\g}\e^{ij}_{\ \ k}F^k_{\a\g}+ V C_{\rm
mat} =  -24(\pi G)^2\lim_{A_{\square}\to 0}\frac{\bu^2}{A_{\square}}v^2\sin^22 b +V C_{\rm
mat} .\label{SH3cls}
\end{gather}
At the quantum level, one should choose an operator ordering for the Hamiltonian costraint. Dif\/ferent orderings correspond to inequivalent def\/initions of the theory but they may lead anyway to very similar physics. In the absence of a guiding principle selecting one particular ordering over the others, one can make a choice convenient for calculational purposes. As an example illustrating this point, after regularizing equation~\eqref{SH3cls} ($A_\square\to \bu^2$), choosing the superselection sector $v=4n$, and quantizing in the $b$ representation ($v\to \hat V=-i\hbar\p_b$), the operator ordering can be arranged so that
\begin{gather*}
\left[3\pi l_\text{Pl}^2(\sin2b\p_b)^2-\p_\phi^2\right]\Psi[b,\phi]=0 .
\end{gather*}
Because one has a discrete one-dimensional lattice in $v$ space and the Fourier transform in $b$ space has support on the interval
$b\in(0,\pi/2)$ \cite{acs}, one can def\/ine
\begin{gather*}
z:= \frac{1}{\sqrt{12\pi l_\text{Pl}^2}} \ln\tan b ,
\end{gather*}
so that we get
\begin{gather*}
\left(\p_z^2-\p^2_\phi\right)\Psi[z,\phi]=0 .
\end{gather*}
This expression is formally identical to the Wheeler--DeWitt equation and the ensuing quantization follows step by step \cite{acs}.
A key dif\/ference, however, is that invariance of the wavefunction under parity (frame re-orientation) is not gauge-f\/ixed \emph{ab initio}
and physical states are required to satisfy $\Psi_+[-z,\phi]=-\Psi_+[z,\phi]$. It follows that the left- and right-moving sectors are not
superselected and must be considered together. In particular, we can write
\begin{gather*}
\Psi_+[z,\phi]=\Psi_{+,{\rm L}}[z_+]+\Psi_{+,{\rm R}}[z_-]=\xi(z_+)-\xi(z_-) ,
\end{gather*}
where $z_\pm=z\pm\phi$ and $\xi$ is some function.

This fact is crucial for the resolution of the big bang singularity. The volume operator in the $z$ variable is
\begin{gather*}
\hat V=-\rmi v_*\cosh(\k_0 z) \p_z ,
\end{gather*}
where $v_*$ is a positive constant and $\k_0=\sqrt{12\pi l_\text{Pl}^2}$. At any time $\phi$ and on any physical state, one can show that
the expectation value of the volume is
\begin{gather}
\langle |\hat V|\rangle = \langle\Psi_{+}|\,|\hat V|\,|\Psi_{+}\rangle= V_*\cosh(\k_0\phi) ,\label{finalbounce}
\end{gather}
where $V_*>0$ is the minimal volume at the bounce. Equation~\eqref{finalbounce} completes the proof that the big bang singularity is avoided in
minisuperspace LQC. Further evidence comes from noticing that matter energy density has an absolute upper bound (approximately equal
to~$0.41$ times the Planck density) on the whole physical Hilbert space~\cite{acs}. We can reach the same quantitative conclusion, albeit not as
robustly, when looking at the ef\/fective dynamics on semiclassical states (Section~\ref{hodi}).

\subsection{FRW models with curvature or cosmological constant}\label{klam}

In the previous sections, we ignored the contribution both of the intrinsic curvature $\Gamma_a^i=(\textsc{k}/2)\delta_a^i$ and of a
cosmological constant $\Lambda$. Here, we sketch scenarios where the universe is not f\/lat ($\textsc{k}=\pm 1$) and/or $\Lambda\neq 0$.
For more details, consult \cite{AsS}.

\subsubsection{Closed universe}

The case of a universe with positive-def\/inite spin connection, $\textsc{k}=1$, was studied in \cite{APSV,BoT,BoVa,LMNT,MHS,SiT,ViS,SKL}.
Due to the extra term in the connection, the form of the classical Hamiltonian constraint~\eqref{eq:ligFRW} as a function of $c$
(related to metric variables as $c=\g \dot a+\textsc{k}$, a~dot denotes derivative with respect to synchronous time) is modif\/ied by the
replacement $c^2\to c(c-V_o^{1/3})+(1+\g^2)V_o^{2/3}/4$. In the classical Friedmann equation, this replacement corresponds to
$H^2\to H^2+\textsc{k}/a^2$ with $\textsc{k}=1$, where $H:=\dot a/a$ is the Hubble parameter. The quantum constraint and the resulting dif\/ference
equation are modif\/ied accordingly. There is no arbitrariness in the f\/iducial volume~$V_o$, since it can be identif\/ied with the total volume of the
universe, which is f\/inite and well def\/ined. Then, the choice of elementary holonomy is more natural than in the f\/lat case and, locally, one can
distinguish between the group structure of~$SU(2)$ and $SO(3)$~\cite{SKL}. As in the f\/lat case, the constraint operator is essentially self-adjoint~\cite{SKL} and the singularity at~$v=0$ is removed from the quantum evolution~\cite{APSV, BoVa,SKL}. However, instead of a single-bounce event one now
has a cyclic model~\cite{APSV}. This can be traced back to the fact that the classical and quantum scalar constraint have both contracting and
expanding branches coexisting in closed-universe solutions, while these branches correspond to distinct solutions in the f\/lat case.

\subsubsection{Open universe}

Loop quantum cosmology of an open universe \cite{ViS,Szu07,van06} is slightly more delicate to deal with. In contrast with the f\/lat and closed
cases $\textsc{k}=0,1$, the spin connection is non-diagonal, so that also the connection is non-diagonal and it has two (rather than one) dynamical
components~$c(t)$ and~$c_2(t)$. The Gauss constraint f\/ixes $c_2=1$ and one ends up with the same number of degrees of freedom as usual. The volume
of the universe is inf\/inite as in the f\/lat case, and a~f\/iducial volume must be def\/ined. The classical Hamiltonian constraint is equation~\eqref{eq:ligFRW}
with $c^2\to c^2-V_o^{2/3}\g^2$. The quantum constraint is constructed after def\/ining a suitable holonomy loop; the bounce still takes place and the
$v=0$ big-bang state factors out of the dynamics.

\subsubsection{$\Lambda\neq 0$}

Another generalization is to add a cosmological constant term, positive \cite{BoVa,KaP,MHS} or nega\-tive~\cite{BeP,BoT,kale,Szu07}. At the level of the dif\/ference equation, these models have been studied in relation to the self-adjoint property.

For $\Lambda>0$, below a critical value $\Lambda_*$ (of order of the Planck energy), the Hamiltonian constraint operator admits many
self-adjoint extensions, each with a discrete spectrum. Above $\Lambda_*$, the operator is essentially self-adjoint but there are no physically
interesting states in the Hilbert space of the model~\cite{KaP}.

For $\Lambda<0$, the scalar constraint is essentially self-adjoint and its spectrum is discrete \cite{kale} (while, we recall, for
$\Lambda=0$ it is continuous and with support on the positive real line), also when $\textsc{k}=-1$ \cite{Szu07}.
As in the $\Lambda=0$, $\textsc{k}=1$ case, the universe undergoes cycles of bounces \cite{BeP}.

\section{Bianchi I model}
\label{chap:bianchi}

The next step in extending loop quantum cosmology to more general situations consists in the consideration of (still homogeneous but) anisotropic cosmologies. The simplest anisotropic spacetime is the Bianchi I model, since it has f\/lat spatial sections.
This model has been extensively studied, owing to its simplicity and applications in cosmology. In fact, prior to the development of loop quantum cosmology, its quantization employing Ashtekar variables was already analyzed \cite{aspu,neg1,neg2}. The f\/irst attempts of constructing a kinematical Hilbert space and the Hamiltonian constraint operator within a polymeric formalism were done in \cite{boj}. Then, soon after the quantization of the f\/lat FRW model was completed within the improved dynamics scheme \cite{aps3}, the same programme was applied to Bianchi~I, which we shall review now.

\subsection{Classical formulation in Ashtekar--Barbero variables}
\label{4sec:clas}

For simplicity, we will consider the model in vacuo. Unlike the FRW universe, which is static in vacuo,
the vacuum Bianchi I model has non-trivial dynamics. Its solutions are of Kasner type~\cite{kasner},
with two expanding scale factors and the third in contraction, or vice versa.

Moreover, for later convenience, we will consider a spatial three-torus topology. Therefore, it will
not be necessary to introduce any f\/iducial cell, since the model already provides a natural f\/inite
cell, that of the three-torus, described with angular coordinates $\{\theta,\sigma,\delta\}$
running from~0 to~$2\pi$.

Like in the isotropic case, we f\/ix the gauge and choose a diagonal f\/lat co-triad ${}^o
e_a^i=\delta_a^i$. The presence of three dif\/ferent directions requires three variables to describe
the Ashtekar--Barbero connection and three more for the densitized triad, that is\footnote{In the following, we will not use the Einstein summation convention, unless specif\/ied otherwise.}
\begin{equation*}
A_a^i=\frac{c^{i}}{2\pi}\delta_a^i, \qquad
E_i^a=\frac{p_{i}}{4\pi^2}\delta_i^a\sqrt{{}^oq},\qquad i=\theta,\sigma,\delta.
\end{equation*}
The Poisson brackets def\/ining the phase space are then $\{c^i,p_j\}=8\pi G\gamma\delta^i_j$. The spacetime metric in these variables reads
\begin{equation*} 
ds^2= -N^2
dt^2+\frac{|p_\theta p_\sigma p_\delta|}{4\pi^2}\left(
\frac{d\theta^2}{p_\theta^2}+\frac{d\sigma^2}{p_\sigma^2}+\frac{d\delta^2}{p_\delta^2}\right).
\end{equation*}
In turn, the phase space is constrained by the Hamiltonian constraint
\begin{gather}\label{eq:ligBianchi}
 C_\text{BI}=-\frac2{\gamma^2}\frac{c^\theta p_\theta c^\sigma p_\sigma+c^\theta p_\theta c^\delta
p_\delta+c^\sigma p_\sigma c^\delta p_\delta}{V}=0.
\end{gather}
In this expression, $V=\sqrt{|p_\theta p_\sigma p_\delta|}$ is the physical volume of the universe.

\subsection{Quantum representation}

In order to polymerically represent this system, we follow the approach described in
Section~\ref{1sec:kin}~\cite{chio}. Holonomies
$
h_i^{\mu_i}(c^i)=e^{\mu_{i}c^{i}\tau_{i}}
$
are def\/ined along straight edges of f\/iducial length
$2\pi\mu_i\in\mathbb{R}$ and oriented in the f\/iducial directions, here labeled by
$i=\theta,\sigma,\delta$. The f\/luxes of the densitized triad through rectangular surfaces of f\/iducial area
$A_{\square}^i$ and orthogonal to the $i$-th direction, given by
$
E(A_{\square}^i,f=1)=[p_{i}/(4\pi^2)]A_{\square}^i
$, complete the description of the phase space before quantization.
The conf\/iguration algebra is the tensor product of the algebras of quasi-periodic functions of the connection for each f\/iducial direction:
$
\text{Cyl}_\text{S}=\otimes_i\text{Cyl}_\text{S}^i=\text{lin}\{|\mu_\theta,\mu_\sigma,
\mu_\delta\rangle\}
$,
where the kets $|\mu_i\rangle$ denote the quantum states corresponding to the matrix elements of the holonomies
$\mathcal N_{\mu_i}(c^i)=e^{i\mu_{i}c^{i}/2}$ in momentum representation. Hence, the kinematical Hilbert space is the tensor product
$
\mathcal H_{\text{grav}}=\otimes_i\mathcal H_{\text{grav}}^i
$,
where $\mathcal H_{\text{grav}}^i$ is the Cauchy completion of $\text{Cyl}_\text{S}^i$ with respect
to the discrete inner product
$\langle\mu_i|\mu_i^\prime\rangle=\delta_{\mu_i\mu_i^\prime}$.

The basic operators are $\hat p_i$ and
$\hat{\mathcal N}_{\mu_i^\prime}$. Their action on the basis states
$|\mu_i\rangle$ is
\begin{gather*}
\hat p_i|\mu_i\rangle =p_i(\mu_i)|\mu_i\rangle,\qquad p_i(\mu_i)=4\pi\gamma l_\text{Pl}^2\mu_i,\qquad
\hat{\mathcal N}_{\mu_i^\prime}|\mu_i\rangle =|\mu_i+\mu_i^\prime\rangle,
\end{gather*}
such that $[\hat{\mathcal N}_{\mu_i},\hat p_j]=i\hbar \widehat{\{\mathcal N_{\mu_i}(c^i),p_j\}}$.

\subsection{Improved dynamics}
\label{4sec:improved}

The most involved aspect that one encounters when trying to adapt the quantization of the isotropic
case to the anisotropic case lies in the implementation of the improved dynamics, explained in Section~\ref{1sec:lig-ham-aps}. In the presence of anisotropies, we need to introduce three minimum f\/iducial
lengths $\bar\mu_i$, when def\/ining the curvature tensor in terms of a loop of holonomies.

Originally, a naive \emph{Ansatz} was chosen, given by
\begin{gather}\label{eq:mubarraA}
 {\frac1{\bar\mu_i'}}=\frac{\sqrt{|p_i|}}{\sqrt{\Delta}}.
\end{gather}
This is the simplest generalization of the ansatz of the isotropic case, equation~\eqref{mu}. As a~consequence, the operators entering the Hamiltonian constraint have the same form as those of the
f\/lat FRW model. Furthermore, operators corresponding to dif\/ferent directions commute among one
another. This allows to complete the quantization obtaining the physical Hilbert space~\cite{mmp}, in
the same way as for the FRW model. However, when the topology is non-compact and then a f\/inite
f\/iducial cell is introduced, the physical results depend on this f\/iducial choice~\cite{luc2}.
This drawback led to the revision of the def\/inition of~$\bar\mu_i$, and another \emph{Ansatz} free of these problems was proposed, this time given by%
\footnote{Whenever the three indices $i$, $j$, $k$ appear in the same expression, we will consider $\epsilon_{ijk}\neq0$, so that they are dif\/ferent.}
\begin{gather}\label{eq:mubarraB}
{\frac1{\bar\mu_i}}=\frac1{\sqrt{\Delta}}\sqrt{\left|\frac{p_j p_k}{p_i}\right|}.
\end{gather}
This choice is geometrically better justif\/ied (for a discussion about its derivation see~\cite{AsW}). Furthermore, this prescription is the only one verifying a remarkable property: For all
the f\/iducial directions, the exponents $\bar\mu_i c^i$ of the matrix elements $\mathcal
N_{\bar\mu_i}(c^i)$ have a constant and f\/ixed (up to a sign) Poisson bracket with the variable
\begin{gather}\label{v}
 v=\text{sgn}(p_\theta p_\sigma p_\delta)\frac{\sqrt{|p_\theta p_\sigma p_\delta|}}{2\pi\gamma
l_\text{Pl}^2\sqrt{\Delta}} ,
\end{gather}
which is proportional to the volume. Note that it coincides with the parameter $v(p)$ of the isotropic case if we identify the three
f\/iducial directions. As a consequence, as we will see, the volume will suf\/fer constant shifts in the quantum theory, as in the isotropic
case. Thanks to this property, the improved dynamics prescription \eqref{eq:mubarraB} nicely implements the interplay between the anisotropies
and the volume. Instead, within the naive prescription given by equation~\eqref{eq:mubarraA}, there is no interplay between the degrees of freedom
associated with dif\/ferent directions, because of the commutation between the operators acting on dif\/ferent f\/iducial directions. Thus, apart
from giving dependencies on f\/iducial choices, it also seems less physically motivated.

Because of these reasons, today it is generally accepted that the more correct improved dynamics
prescription is equation~\eqref{eq:mubarraB}, which we shall consider in this paper.

\subsection{Hamiltonian constraint operator}

As in the isotropic case, in order to obtain the Hamiltonian constraint operator we cannot
represent directly its classical form \eqref{eq:ligBianchi}, but its expression in terms of
the curvature tensor. For homogeneous models with vanishing spin connection, as Bianchi I, this expression was given in
equation~\eqref{eq:lig-escalar-lqg-hom}.

For simplicity, we will densitize the Hamiltonian constraint classically, by simply multiplying it by the volume $V$.
In this way we avoid the appearance of inverse powers of the volume that make the quantum theory complicated.
In any case, as seen in the f\/lat FRW model, the densitization could be carried out with no problem in the quantum theory.

In analogy with the isotropic case, but now taking into consideration that the three f\/iducial directions are dif\/ferent,
the curvature operator is the quantum counterpart of the classical expression
\begin{gather}\label{eq:curvatura-bianchi}
{F}^i_{ab}=-2\sum_{j,k}
\text{tr}\left(
\frac{h^{\bar\mu}_{\square_{jk}}-\delta_{jk}}{4\pi^2\bar\mu_j\bar\mu_k}\tau^i\right)\delta^j_a
\delta^k_b,\qquad h^{\bar\mu}_{\square_{jk}}=h_j^{\bar\mu_j} h_k^{\bar\mu_k} (h_j^{\bar\mu_j})^{-1}
(h_k^{\bar\mu_k})^{-1}.
\end{gather}
Taking into account equation~\eqref{eq:curvatura-bianchi}, the expression of the densitized triad and the def\/inition of~$\bar\mu_i$, the densitized Hamiltonian constraint for the Bianchi~I model reads
\begin{gather}\label{eq:ligadura-bianchi-B}
C_\text{BI}=\frac{2}{\gamma^2\Delta}V^2\sum_{i,j,k}\epsilon^{ijk}\text{sgn}
(p_j)\, \text{sgn}(p_k)\,\text{tr}\big(\tau_i h^{\bar\mu}_{\square_{jk}}\big).
\end{gather}

In order to represent this constraint as an operator, we f\/irst need to def\/ine the operators~$\hat{\mathcal N}_{\bar\mu_i}$, which represent the matrix elements of the holonomies~$h_i^{\bar\mu_i}$. To def\/ine them we follow a~similar strategy to that adopted in the isotropic
case. We start by reparametrizing $p_i(\mu_i)$ with a~pa\-ra\-me\-ter~$\lambda_i(p_i)$ such that the
vectorial f\/ield $\bar\mu_i\partial_{\mu_i}$ produces constant translations in~$\lambda_i$.
The solution is
$\lambda_i(p_i)={\text{sgn}(p_i)\sqrt{|p_i|}}/{(4\pi\gamma
l_\text{Pl}^2\sqrt{\Delta})^{1/3}}$ \cite{AsW}.
The dif\/ference with respect to the isotropic case is that the translations produced by
$\bar\mu_i\partial_{\mu_i}$, being constant with respect to the dependence on~$\lambda_i$, do depend
on the parameters~$\lambda_j$ and $\lambda_k$ associated with the other two directions. In fact, we
have $\bar\mu_i\partial_{\mu_i}=(2|\lambda_j\lambda_k|)^{-1}\partial_{\lambda_i}$.

As in the isotropic case, we def\/ine the operator $\hat{\mathcal N}_{\bar\mu_i}$ such that its action on the basis states~$|\lambda_i\rangle$ is the same as the transformation generated by $\bar\mu_i\partial_{\mu_i}$ on the parameter~$\lambda_i$, that is
\begin{gather*}
\hat{\mathcal
N}_{\pm\bar\mu_\theta}|\lambda_\theta,\lambda_\sigma,
\lambda_\delta\rangle=\bigg|\lambda_\theta\pm\frac1 {
2|\lambda_\sigma\lambda_\delta|},\lambda_\sigma,\lambda_\delta\bigg\rangle,
\end{gather*}
and similarly for $\hat{\mathcal N}_{\pm\bar\mu_\sigma}$ and $\hat{\mathcal N}_{\pm\bar\mu_\delta}$.
Moreover, inverting the change of variable, we obtain
\begin{gather*}
 \hat{p}_i|\lambda_\theta,\lambda_\sigma,\lambda_\delta\rangle=(4\pi\gamma
l_\text{Pl}^2\sqrt{\Delta})^{2/3}\text{sgn}(\lambda_i)\lambda_i^2|\lambda_\theta,\lambda_\sigma,
\lambda_\delta\rangle.
\end{gather*}
As explained in the f\/lat FRW case, one can always choose a suitable factor ordering for
the Hamiltonian constraint that allows to remove the kernel of the volume operator, generated by the
states with
$\lambda_\theta\lambda_\sigma\lambda_\delta=0$. This is what we will consider. Therefore, the
operators $\hat{\mathcal N}_{\bar\mu_i}$ are well def\/ined.

The action of $\hat{\mathcal N}_{\pm\bar\mu_i}$ can be slightly simplif\/ied by introducing the variable
$v$ def\/ined in equation~\eqref{v}, that in terms of $\lambda$'s is given by
$v=2\lambda_\theta\lambda_\sigma\lambda_\delta$.
Indeed, making the change from, e.g., the states
$|\lambda_\theta,\lambda_\sigma,\lambda_\delta\rangle$ to the states
$|v,\lambda_\sigma,\lambda_\delta\rangle$, one can check that, under the action of $\hat{\mathcal
N}_{\bar\mu_i}$, $v$~suf\/fers a constant shift equal to~$1$ or~$-1$ depending on the orientation of
the densitized triad coef\/f\/icients. On the other hand, the variables~$\lambda_\sigma$ and $\lambda_\delta$ suf\/fer a dilatation or contraction that only depends on their
own sign and on $v$ (see~\cite{AsW} for the details).

The variable $v$ is proportional to the volume
\begin{gather*}
 \hat{V}=\widehat{\sqrt{|p_\theta p_\sigma p_\delta|}}, \qquad
\hat{V}|v,\lambda_\sigma,\lambda_\delta\rangle=2\pi\gamma
l_{\text{Pl}}^2\sqrt{\Delta}|v||v,\lambda_\sigma,\lambda_\delta\rangle.
\end{gather*}
Therefore, as happened in the isotropic case, in this scheme the volume undergoes constant
translations. The other two variables measure the degree of anisotropy of the system.

Once we know how to represent the matrix elements of the holonomies, we can promote the Hamiltonian
constraint to an operator. When symmetrizing it, we will adopt the prescription of~\cite{gow-B},
whose factor ordering is analog to that considered in Section~\ref{chap:1-flatFRW} (see
equation~\eqref{eq:operador-grav-mmo}).  Explicitly, it is given by~\cite{gow-B,mmw}
\begin{gather}\label{densCB}
\widehat{C}_{\text{BI}} =-\frac1{\gamma^2}(\widehat{\Omega}_\theta\widehat{\Omega}_\sigma+
\widehat{\Omega}_\sigma\widehat{\Omega}_\theta+\widehat{\Omega}_\theta\widehat{\Omega}_\delta+
\widehat{\Omega}_\delta\widehat{\Omega}_\theta+\widehat{\Omega}_\sigma\widehat{\Omega}_\delta+
\widehat{\Omega}_\delta\widehat{\Omega}_\sigma),
\end{gather}
where
\begin{gather}\label{cp-quantum}
\widehat\Omega_i =\frac1{4i\sqrt{\Delta}}\widehat{\sqrt{V}}
\left[\big(\hat{\mathcal
N}_{2\bar\mu_i}-\hat{\mathcal
N}_{-2\bar\mu_i}\big)\widehat{\text{sgn}(p_i)}
+\widehat{\text{sgn}(p_i)}\big(\hat{\mathcal
N}_{2\bar\mu_i}-\hat{\mathcal
N}_{-2\bar\mu_i}\big)\right]\widehat{\sqrt{V}}.
\end{gather}
This operator dif\/fers from that of \cite{AsW} in the treatment applied to the signs of $p_i$ when
symmetrizing.
Then, $\widehat{{C}}_{\text{BI}}$ not only decouples the zero-volume states, but also it does not
relate states with opposite orientation of any of the triad coef\/f\/icients,
namely, states $|v,\lambda_\sigma,\lambda_\delta\rangle$ with opposite sign in any of their quantum
numbers. Therefore, $\widehat{{C}}_{\text{BI}}$ leaves invariant all the octants in the
tridimensional space def\/ined by $v$, $\lambda_\sigma$ and $\lambda_\delta$. Hence, we can restrict
the study to any of them. We will restrict ourselves to the subspace of positive densitized triad
coef\/f\/icients, given~by
\begin{gather*}
{\text{Cyl}}_\text{S}^+ =\text{lin}\{|v,\lambda_\sigma,\lambda_\delta\rangle;\,
v,\lambda_\sigma,\lambda_\delta>0\}.
\end{gather*}

The action of $\widehat{{C}}_{\text{BI}}$ on the states of
${\text{Cyl}}_\text{S}^+$ turns out to be
\begin{gather*}
\widehat{{C}}_{\text{BI}}|v,\lambda_\sigma,\lambda_\delta\rangle =\frac{(\pi
l_{\text{Pl}}^2)^2}{4}\big[x_-(v)|v-4,\lambda_\sigma,\lambda_\delta\rangle_-
-x^-_0(v)|v,\lambda_\sigma,\lambda_\delta\rangle_{-}\nonumber\\
\phantom{\widehat{{C}}_{\text{BI}}|v,\lambda_\sigma,\lambda_\delta\rangle =}{} -
x^+_0(v)|v,\lambda_\sigma,\lambda_\delta\rangle_{+}+x_+(v)|v+4,\lambda_\sigma,
\lambda_\delta\rangle_+\big],
\end{gather*}
where we have introduced the coef\/f\/icients
\begin{subequations}\label{coeficientes}
\begin{alignat}{3}
& x_-(v)=2\sqrt{v}(v-2)\sqrt{v-4}[1+\text{sgn}(v-4)],\qquad
&&  x_+(v)=x_-(v+4), & \label{coefficient1}\\
& x^-_0(v)=2(v-2)v[1+\text{sgn}(v-2)],\qquad &&x^+_0(v)=x^-_0(v+2), &\label{coefficient3}
\end{alignat}
\end{subequations}
and the following linear combination of states
\begin{gather}
|v\pm n,\lambda_\sigma,\lambda_\delta\rangle_\pm =\bigg|v\pm n,\lambda_\sigma,\frac{v\pm n}
{v\pm2}\lambda_\delta\bigg\rangle+\bigg|v\pm n,\frac{v\pm n}
{v\pm2}\lambda_\sigma,\lambda_\delta\bigg\rangle\nonumber\\
\hphantom{|v\pm n,\lambda_\sigma,\lambda_\delta\rangle_\pm =}{}
+\bigg|v\pm n,\frac{v\pm2}
{v}\lambda_\sigma,\lambda_\delta\bigg\rangle+\bigg|v\pm n,\lambda_\sigma,
\frac{v\pm2} {v}\lambda_\delta\bigg\rangle\nonumber\\
\hphantom{|v\pm n,\lambda_\sigma,\lambda_\delta\rangle_\pm =}{}
+\bigg|v\pm n,\frac{v\pm2}
{v}\lambda_\sigma,\frac{v\pm n} {v\pm2}\lambda_\delta\bigg\rangle+\bigg|v\pm n,\frac{v\pm n}
{v\pm2}\lambda_\sigma,\frac{v\pm2}{v}\lambda_\delta\bigg\rangle.\label{eq:combi}
\end{gather}
Note that, in fact, the operator $\widehat{{C}}_{\text{BI}}$ is well def\/ined
in ${\text{Cyl}}_\text{S}^+$, since $x_-(v)=0$ if $v\leq4$, and $x^-_0(v)=0$ if $v\leq2$. Since there is no $v=0$ state, the singularity has no longer
analog in the kinematical Hilbert space, and then is resolved already kinematically.

\subsection{Superselection sectors}\label{superselection}

The analysis of the action of $\widehat{C}_{\text{BI}}$ on a generic basis state
$|v,\lambda_\sigma^\star,\lambda_\delta^\star\rangle$ shows that:
\begin{itemize}\itemsep=0pt
 \item [i)] Concerning the variable $v$, it suf\/fers a constant shift equal to $4$ or $-4$, the
latest only if  $v>4$. Therefore,
$\widehat{{C}}_{\text{BI}}$ preserves the subspace of states whose quantum number $v$ belongs to
any of the semilattices of step four
\begin{equation*}
\mathcal
L_{\varepsilon}^+=\{\varepsilon+4k,k=0,1,2,\dots\},\qquad\varepsilon\in(0,4].
\end{equation*}
\item [ii)] Concerning the anisotropy variables, $\lambda_\sigma$ and $\lambda_\delta$,
the ef\/fect upon them does not depend on the initial quantum numbers $\lambda_\sigma^\star$ and
$\lambda_\delta^\star$, but only on $v=\varepsilon+4k$. Moreover, this dependence occurs
via fractions whose denominator is two units bigger or smaller than the numerator.
As a consequence, the iterative action of the constraint operator on
$|v,\lambda_\sigma^\star,\lambda_\delta^\star\rangle$, only relates this state with states whose
quantum numbers
$\lambda_\sigma$ and $\lambda_\delta$ are of the form
$\lambda_a=\omega_{\varepsilon}\lambda_a^\star$, with~$\omega_{\varepsilon}$
belonging to the set
\begin{gather}
\mathcal
W_{\varepsilon}=\bigg\{\left(\frac{\varepsilon-2}{\varepsilon}\right)^z\prod_{m,
n\in\mathbb{N}}\left(\frac{\varepsilon+2m}{\varepsilon+2n}\right)^{k_n^m};\nonumber\\
\hphantom{W_{\varepsilon}=\bigg\{}{}
k_n^m\in\mathbb{N},\;  z\in\mathbb{Z}\text{ if } \varepsilon>2,\;\;z=0\text{ if }
\varepsilon<2\bigg\}.\label{W-set}
\end{gather}
\end{itemize}

The set $\mathcal W_{\varepsilon}$ is inf\/inite and, moreover, one can prove that it is dense in
the positive real line \cite{gow-B}. Nonetheless, it is countable.
Therefore, while the variable $v$
has support in simple semilattices of constant step, the variables $\lambda_a$ take
values belonging to complicated sets, but they also provide separable subspaces. As a concrete
example,
we see that, if both $\varepsilon$ and
$\lambda_a^\star$ are integers, then $\lambda_a$ take values in the positive rational numbers.

In conclusion, the operator $\widehat{{C}}_{\text{BI}}$ leaves invariant the Hilbert
subspaces $\mathcal H^+_{\varepsilon,\lambda_\sigma^\star,\lambda_\delta^\star}$,
def\/ined as the Cauchy completion of
\begin{gather*}
\text{Cyl}^+_{\varepsilon,\lambda_\sigma^\star,\lambda_\delta^\star}=\text{lin}\big\{
 |v,\lambda_\sigma,\lambda_\delta\rangle;\;
v\in\mathcal L_{\varepsilon}^+,\;\lambda_a=\omega_{\varepsilon}\lambda_a^\star,\;
\omega_{\varepsilon}\in\mathcal W_{\varepsilon},\;\lambda_a^\star\in\mathbb{R}^+\big\},
\end{gather*}
with respect to the discrete inner product
$
\langle
v,\lambda_\sigma,\lambda_\delta|v',\lambda'_\sigma,\lambda'_\delta\rangle=\delta_{vv'}\delta_{
\lambda_\sigma \lambda'_{\sigma}}\delta_{\lambda_\delta \lambda'_{\delta}}
$.
As we will see, physical observables also preserve these separable subspaces
$\mathcal H^+_{\varepsilon,\lambda_\sigma^\star,\lambda_\delta^\star}$, and therefore they
provide sectors of superselection, and we can restrict the study to any of them.

\subsection{Physical Hilbert space}

The Hamiltonian constraint operator $\widehat{{C}}_{\text{BI}}$ is quite complicated and,
unlike the isotropic case (and unlike previous quantizations of the model~\cite{mmp}),
its spectral properties have not been determined. Consequently, it has not been diagonalized
either. Therefore, the group averaging approach is not useful in this situation, and to analyze the
physical solutions one has to impose the constraint directly on the dual space
$(\text{Cyl}^{+}_{\varepsilon,\lambda_\sigma^\star,\lambda_\delta^\star})^*$. The elements
$(\psi|$ of that space have the formal expansion
\begin{gather*}
(\psi| =\sum_{v\in\mathcal
L_{\varepsilon}^+}\sum_{\omega_{\varepsilon}\in\mathcal
W_{\varepsilon}}\sum_{\bar\omega_{\varepsilon}\in\mathcal W_{\varepsilon}}
\psi(v,\omega_{\varepsilon}\lambda_\sigma^\star,\bar\omega_{\varepsilon}
\lambda_\delta^\star)
\big\langle
v,\omega_{\varepsilon}\lambda_\sigma^\star,\bar\omega_{\varepsilon}
\lambda_\delta^\star\big|.
\end{gather*}
From the action of $\widehat{C}_\text{BI}$, one obtain that the constraint
$\big(\psi\big|\widehat{\mathcal
C}_\text{BI}^\text{B}{}^\dagger=0$ leads to the following recurrence relation,
\begin{gather*}
\psi_+(v+4,\lambda_\sigma,\lambda_\delta)=\frac1{x_+(v)}
\big[{x^-_0(v)}\psi_{-}(v,\lambda_\sigma,\lambda_\delta)+{x^+_0(v)}\psi_{+}(v,
\lambda_\sigma,\lambda_\delta)\nonumber\\
\hphantom{\psi_+(v+4,\lambda_\sigma,\lambda_\delta)=\frac1{x_+(v)}
\big[}{} -{x_-(v)}\psi_-(v-4,\lambda_\sigma,\lambda_\delta)\big].
\end{gather*}
In this expression, in order to simplify the notation, we have introduced the projections of
$(\psi|$ on the linear combinations of six states def\/ined in equation~\eqref{eq:combi}, namely,
\begin{gather*}
 \psi_\pm(v\pm n,\lambda_\sigma,\lambda_\delta)=
(\psi|v\pm n,\lambda_\sigma,\lambda_\delta\rangle_\pm .
\end{gather*}

Owing to the property $x_-(\varepsilon)=0$, the above recurrence relation, that is of order 2 in
the variable $v$, becomes a f\/irst-order equation if
$v=\varepsilon$:
\begin{gather*}
\psi_+(\varepsilon+4,\lambda_\sigma,\lambda_\delta)=\frac1{x_+(\varepsilon)}
\big[{x^-_0(\varepsilon)}\psi_{-}(\varepsilon,\lambda_\sigma,\lambda_\delta)+{
x^+_0(\varepsilon)}\psi_{+}(\varepsilon,\lambda_\sigma,\lambda_\delta)\big].
\end{gather*}
Therefore, if we know all the data in the initial section $v=\varepsilon$, we obtain all the combinations of six terms given by
\begin{gather*}
\psi_+(\varepsilon+4,\lambda_\sigma,\lambda_\delta)
 =\psi\left(\varepsilon+4,\lambda_\sigma,\frac{\varepsilon+4}{\varepsilon+2}
\lambda_\delta\right)
+\psi\left(\varepsilon+4,\frac{\varepsilon+4}{\varepsilon+2}\lambda_\sigma,
\lambda_\delta\right)
\nonumber\\
\hphantom{\psi_+(\varepsilon+4,\lambda_\sigma,\lambda_\delta)=}{}
+\psi\left(\varepsilon+4,\lambda_\sigma,
\frac{\varepsilon+2}{\varepsilon}\lambda_\delta\right)
+\psi\left(\varepsilon+4,\frac{\varepsilon+2}{\varepsilon}\lambda_\sigma,
\lambda_\delta\right)
\nonumber\\
\hphantom{\psi_+(\varepsilon+4,\lambda_\sigma,\lambda_\delta)=}{}
+\psi\left(\varepsilon+4,\frac{\varepsilon+2}{\varepsilon}\lambda_\sigma,
\frac{\varepsilon+4}{\varepsilon+2} \lambda_\delta\right)
+\psi\left(\varepsilon+4,\frac{\varepsilon+4}{\varepsilon+2}\lambda_\sigma,
\frac{\varepsilon+2}{\varepsilon} \lambda_\delta\right).
\end{gather*}

From the combinations $\psi_+(\varepsilon+4,\lambda_\sigma,\lambda_\delta)$, it is possible to
determine any of the individual terms
$\psi(\varepsilon+4,\lambda_\sigma,\lambda_\delta)$ that compose them, since it has been shown that
the system of equations that relate the formers with the latters is formally invertible~\cite{mmw}. This has been proven not only for $v=\varepsilon+4$ but also for all
$v\in\mathcal L_{\varepsilon}^+$. In conclusion, the physical solutions of the
Hamiltonian constraint are completely determined by the set of initial data
\begin{gather*}
 \{\psi(\varepsilon,\lambda_\sigma,
\lambda_\delta)=\psi(\varepsilon,\omega_\varepsilon\lambda_\sigma^\star,
\bar\omega_\varepsilon\lambda_\delta^\star),\
\omega_\varepsilon,\bar\omega_\varepsilon\in\mathcal W_{\varepsilon},\
\lambda_\sigma^\star,\lambda_\delta^\star\in\mathbb{R}^+\} ,
\end{gather*}
and we can identify solutions with this set. We can also characterize the physical Hilbert space as the Hilbert space of the initial data.
In order to endow the set of initial data with a Hilbert structure, one can take a complete set of
observables forming a closed algebra, and impose that the quantum counterpart of their
complex conjugation relations become adjointness relations between operators. This determines a
unique (up to unitary equivalence) inner product.

\looseness=-1
Before doing that, it is suitable to change the notation. Following \cite{mmw}, let us
introduce the variables $x_a=\ln(\lambda_a)=\ln(\lambda_a^\star)+\rho_{\varepsilon}$. Note that
$\rho_{\varepsilon}$ takes values in a dense set of the real line, given by the logarithm of the
points in the set $\mathcal W_{\varepsilon}$. We will denote that set by $\mathcal Z_{\varepsilon}$.
For each direction $a=\sigma$ or $\delta$, we consider the linear span
$\text{Cyl}_{\lambda_a^\star}$ of the states whose support is just
one point $x_a$ of the superselection sector def\/ined by taking the product of
$\lambda_a^\star$ with all the points in the set $\mathcal Z_{\varepsilon}$.
We call $\mathcal H_{\lambda_a^\star}$
the Hilbert completion of this vector space with the discrete inner product.

Then, a set of observables acting on the initial data
$\tilde\psi(x_\sigma,x_\delta):=\psi(\varepsilon,x_\sigma,
x_\delta)$ is that formed by the operators
$\widehat{e^{ix_a}}$ and $\widehat{U}_a^{\rho_a}$, with $\rho_a\in\mathcal
Z_{\varepsilon}$ and $a=\sigma,\delta$, def\/ined as
\begin{gather}\label{conjunto-completo}
 \widehat{e^{ix_\sigma}}\tilde\psi(x_\sigma,x_\delta)={e^{ix_\sigma}}\tilde\psi(x_\sigma,x_\delta),
\qquad
\widehat{U}_\sigma^{\rho_\sigma}\tilde\psi(x_\sigma,x_\delta)=
\tilde\psi(x_\sigma+\rho_\sigma,x_\delta),
\end{gather}
and similarly for $\widehat{e^{ix_\delta}}$ and $\widehat{U}_\delta^{\rho_\delta}$.
These operators provide an overcomplete set of observables and are unitary in
$\mathcal H_{\lambda_\sigma^\star}\otimes\mathcal H_{\lambda_\delta^\star}$, according with their
reality conditions.
Therefore, we conclude that this Hilbert space is precisely the
physical Hilbert space of the vacuum Bianchi I model.

Owing to the complicated form of the solutions, together with the fact that a basis of
eigenfunctions of the Hamiltonian constraint operator is not known, the evolution picture of this
model has not been studied yet. It is expected that, as in the isotropic models, a quantum bounce
solves dynamically the big bang singularity. Indeed, this has been already checked using an
ef\/fective dynamics~\cite{chi2}, in the case of the model coupled to a massless scalar f\/ield, as in
the FRW case. In this respect, it is worth commenting that in analyses where a massless
scalar f\/ield is introduced, the latter serves as internal time to describe the notion of evolution, as
seen in Section~\ref{1subsec:evol}. This f\/ield (unlike the geometric degrees of freedom)
is quantized adopting a standard Schr\"odinger-like representation, which makes straightforward the construction of a~family of unitarily related
observables parametrized by the internal time. However, in vacuum cases such as the Bianchi~I model we have just described, where
a suitable matter f\/ield is not at hand, it would be interesting to describe the evolution regarding as
internal time one of the geometry degrees of freedom. There is a complication because of the fact that such internal time is
polymeric, since the geometry degrees of freedom are polymerically quantized.
Although such description has not been carried out for the Bianchi~I model within the current improved dynamics, it has been
nonetheless constructed for the vacuum Bianchi~I model quantized within the naive
improved dynamics given in equation~\eqref{eq:mubarraA}. That was done in~\cite{mmp2}, and an
analog construction could just as well serve to describe the evolution of the current vacuum Bianchi~I
model (quantized within the scheme given in equation~\eqref{eq:mubarraB}), using
either the volume variable~$v$ or its momentum as internal time.

\subsection{Loop quantization of other Bianchi models}

Bianchi models are characterized by possessing three spatial Killing f\/ields. In the case of the
Bianchi I model, the Killing f\/ields commute and, then, it is the simplest of the Bianchi models.
This is the reason why it has been extensively studied in the literature, and in particular in the
framework of LQC. Nevertheless, loop quantization has been also extended to other Bianchi
models (with non-commuting Killing f\/ields), in particular to Bianchi~II and Bianchi~IX. A~preliminary loop quantization of these models was already considered just after the
birth of LQC~\cite{boj,boj-date}. More recently, their quantization has been achieved implementing
the improved dynamics developed for Bianchi~I, in~\cite{AsW2} for Bianchi~II and in~\cite{we} for Bianchi~IX. In the following we summarize the main
characteristics of these works. We refer the reader to those references for further details.

Both the Bianchi II and the Bianchi IX models possess non-commuting Killing f\/ields. As a~consequence, the f\/iducial triad and co-triad cannot be chosen to be diagonal, a f\/irst feature that
complicates the analysis in comparison with Bianchi~I. Moreover, the spin connection of
those models is non-trivial. After choosing a suitable gauge and appropriate parameterizations for
the Ashtekar--Barbero variables, the Hamiltonian constraint of
both Bianchi~II and Bianchi~IX consists in that of the (analog) Bianchi~I model plus an extra term,
of course dif\/ferent for each model, but that in both cases involves components of the connection and
inverses of the densitized triad coef\/f\/icients.

The kinematical Hilbert space of the Bianchi II and IX models is identical to that of
Bianchi~I. The main dif\/f\/iculty lies in the representation of the components of the connection
that appear in the extra term of the Hamiltonian constraint\footnote{The inverse of the components
of the densitized triad can be regularized using commutators with holonomies, in an analog way as
that employed in the FRW model.}. To tackle this issue, the strategy adopted is to def\/ine an
operator representing the connection~\cite{AsW2,ydm2}, choosing a suitable loop of holonomies and
implementing the ideas employed when constructing the curvature operator for Bianchi~I. As
a result, the components $c^i$ of the connection turn out to be represented by the polymeric
operator
\begin{gather*}
 \hat c_i=\widehat{\frac{\sin \bar \mu_i c^i}{\bar \mu_i}},
\end{gather*}
where $\bar\mu_i$ is the minimum length def\/ined in equation~\eqref{eq:mubarraB}.

Once the above operator is def\/ined, the representation of the Hamiltonian constraint as a~symmetric
operator follows straightforwardly, in the same way as for the Bianchi~I model. Once again,
classical singularities are avoided in both models, since the kernel of the volume operator
can be removed from their quantum theories.

\part{Midisuperspace models in loop quantum gravity}\label{part2}

In the previous part we have discussed the loop quantization of several homogeneous cosmological
models. In all of them the classical cosmological singularity is avoided. Nonetheless, it is
natural to ask whether the resolution of the singularity is an intrinsic feature of the loop
quantization or, on the contrary, if it is a result due to the high symmetry of the homogeneous
models. In order to answer this question, it seems inevitable to extend the loop quantization to inhomogeneous
systems. Furthermore, it is essential to make this step in order to develop a realistic theory of
quantum cosmology. Indeed, as the results obtained in modern cosmology indicate, in the early
universe the inhomogeneities played a fundamental role in the formation of the cosmic structures
that we observe today.

The quantization of inhomogeneous models is technically more complicated than that of minisuperspaces, since they possess f\/ield degrees of freedom, as
the full theory. Hopefully, facing the loop quantization of midisuperspaces we will get insights about the open problems present in loop quantum gravity.

In this section we will review the status of loop quantization of midisuperspace models. Classically, these models have Killing vectors which reduce
the degrees of freedom of the metric but the number of Killing vectors are low enough to ensure that the remaining
degrees of freedom are local. The metrics are therefore parametrized by functions of time and spatial coordinate(s).
Thus, unlike minisuperspace scenarios, midisuperspace models are f\/ield theories (for a comprehensive review, see~\cite{midisuperspacereview}). Questions beyond the reach of minisuperspace models can be addressed in the study of midisuperspaces both classically
as well as in the context of a~particular quantization scheme. In the context of LQG, these issues may include the construction of Dirac
observables or of quasilocal observables, the closure of the quantum constraint algebra and, most importantly, whether the singularity resolution
mechanism of LQC continues to be valid in the f\/ield theory context.

There exist a number of midisuperspace models but the loop quantization procedure has been attempted only in a few of them so far.
One important class of models is obtained by symmetry reduction of GR.
\begin{itemize}\itemsep=0pt
\item {\em Spherical symmetry.} In $3+1$ dimensions, a spacetime is called spherically symmetric if its isometry group contains a subgroup isomorphic to $SO(3)$,
and the orbits of this subgroup are 2-spheres such that the induced metric thereon is Riemannian and proportional to the unit round
metric on $S^2$. These are, in a sense, midway between the minisuperspace models and models with an inf\/inite number of physical degrees of freedom.
Spherically symmetric spacetime metrics depend on the radial coordinate, and therefore these models have to be treated as f\/ield theories. However, in
vacuum, the physical solutions are characterized by only a single parameter according to Birkhof\/f's theorem. In that respect they are
{\em dynamically trivial}, although the gauge-f\/ixing procedure is extremely non-trivial. To make them into physical f\/ield theories, we need to add
matter in the form of dust shells.

Here we are mainly interested in cosmological models while the spherical symmetric models are mostly black hole solutions, in particular
the Schwarzschild solution. Hence, we will brief\/ly mention the progress made in the context of LQG for completeness.

In \cite{martinsph1} the kinematical framework for studying spherically symmetric models in LQG was introduced. The volume operator was constructed in
\cite{martinsph2} and it was shown that the volume eigenstates are not eigenstates of the f\/lux operator. Consequently, the standard prescription of
constructing the Hamiltonian operator cannot be used. This problem was circumvented in
\cite{martinsph3}, where the Hamiltonian constraint operator was constructed in terms of non-standard variables which mix the connection and the
extrinsic curvature. This formulation was extended to explore the question of singularity resolution of Schwarzschild black holes in
\cite{martinsph4} and the Lema\^itre--Toleman--Bondi collapse of a spherical inhomogeneous dust cloud in \cite{martinsph5}. The choice of
variables made in the above programme is similar to the polymer quantization of the Gowdy $T^3$ model, which will be described in detail later.
Since the basic quantum variables used in these constructions are dif\/ferent from the basic quantum variables of the full
theory, another approach in loop quantization of the spherically symmetric models was explored in~\cite{pullinsph1}. In this approach, the dif\/feomorphism
constraint is f\/ixed leaving the Gauss and the Hamiltonian constraints. The latter is then applied to the Schwarzschild solution. The exterior solutions
agree with the ones obtained in geometrodynamics. The interior solution was studied in~\cite{pullinsph2} where it was shown that, after a partial gauge
f\/ixing, it can be mapped to the minisuperspace Kantowski--Sachs model. After loop quantization, the singularity is replaced by a bounce.
In \cite{pullinsph3} the issue of the residual dif\/feomorphism invariance of loop-quantized spherical symmetric models has been investigated.
Although the quantization programme is still incomplete, the studies done so far indicate that the singularity resolution mechanism described in the
previous section may be a robust feature of loop quantization.
\end{itemize}

Another class of midisuperspace models can be roughly classif\/ied on the isometry group of the metric. In most cases, this is equivalent to a classif\/ication
based on the number of Killing vectors of the metric.
\begin{itemize}\itemsep=0pt
\item {\em Spacetimes with one Killing vector.}
These models are obtained by symmetry reduction corresponding to one-dimensional spatial isometry groups taken to be ${\mathbb R}$ or $U(1)$.
Some of the important features of GR are retained, such as dif\/feomorphism invariance and the f\/ield-theoretic non-linear nature of the physical
degrees of freedom. It was shown in \cite{geroch} that, locally, it is possible to interpret these $3+1$ dimensional models as \mbox{$2+1$} dimensional
general relativity coupled to two matter f\/ields, a scalar f\/ield and a one-form f\/ield corresponding, respectively, to the norm and the twist of the
four-dimensional Killing vector f\/ield. However, not much work has been done so far in loop quantizing this reduced system apart from the preliminary
analysis carried out in \cite{husainpullin} in terms of complex Ashtekar variables.

\item {\em Spacetimes with two Killing vectors.}
The next level of simplif\/ication is to consider two-dimensional spacelike isometry groups. Two types of group action and spatial topologies have been
studied in LQG:
\begin{itemize}\itemsep=0pt
\item Isometry group of ${\mathbb R} \times U(1)$ with the spatial topology being ${\mathbb R}^3$. These correspond to Einstein--Rosen cylindrical
waves. An preliminary attempt has been made to construct the kinematic Hilbert space and def\/ine the volume operator (see Section~5.11 of~\cite{lqc1}),
although there has not been much progress so far.
\item Isometry group of $U(1) \times U(1)$ with the spatial topology being $T^3$. These correspond to Gowdy models.
\end{itemize}
If we impose the additional condition that the Killing vectors are mutually orthogonal, we obtain {\em polarized} models.
These are the simplest midisuperspace models with only one f\/ield theory degree of freedom. An example is the
polarized Gowdy $T^3$ model. This has been studied in some detail in LQG from two dif\/ferent perspectives, both of which will
be described in detail in the next two sections.
\end{itemize}

\section{Hybrid quantization of the polarized Gowdy $T^3$ model}\label{hybrid-gowdy}

Gowdy models are among the best known inhomogeneous cosmologies \cite{gowdy1,gowdy2}. They represent
globally hyperbolic vacuum spacetimes, with compact spatial sections and with two spatial Killing
f\/ields. The simplest example is the linearly polarized Gowdy $T^3$
model. Indeed, its classical solutions are exactly known \cite{ise, mon1, mon2}. They represent
gravitational waves propagating in a closed expanding universe. Its standard quantization was already
considered since the 70's \cite{berger1,berger2,berger3,ccq,men0,misner,pierri,torre}. Moreover, a Fock quantization of the model, in which the dynamics is implemented unitarily, was
achieved~\cite{men1b, men1a}, and it has been shown that this quantization is essentially
unique~\cite{men2,men3} (in a sense that we will explain later).

In order to apply the quantization employed in LQC to this model, the simplest possibility is to
carry out a hybrid quantization, that combines the polymeric quantization of the degrees of freedom
that parametrize the homogeneous solutions, with the Fock quantization for the inhomogeneities.
This hybrid quantization was developed in~\cite{gow-B,gow-let,gow-ijmp}. Here we
summarize its construction and main results.

\subsection{Classical description of the Gowdy $T^3$ model}

The Gowdy $T^3$ model represents vacuum solutions to the Einstein equations, with two spatial
Killing f\/ields that commute, and with spatial hypersurfaces homeomorphic to a three-torus. As said
before, we will consider the linearly polarized model, that possesses an additional symmetry: the
Killing vectors are orthogonal to hypersurfaces and, therefore, are mutually orthogonal everywhere.
Letting $\partial_\sigma$ and $\partial_\delta$ be the Killing vectors, the model admits global
coordinates $\{t,\theta,\sigma,\delta\}$ adapted to the symmetries, with $\theta,\sigma,\delta\in
S^1$.

After a 3+1 decomposition, we can describe the spacetime metric in terms of the three-metric~$q_{ab}$ induced in the spatial sections foliating the four-dimensional manifold, the densitized
lapse ${N_{_{_{\!\!\!\!\!\!\sim}}\;}}=N/\sqrt{q}$, and the shift vector $N^{a}$, with
$a,b\in\{\theta, \sigma,\delta\}$. Owing to the isometries~$\partial_\sigma$ and~$\partial_\delta$,
the Gowdy model verif\/ies $q_{\theta\sigma}=0=q_{\theta\delta}$. This condition f\/ixes the gauge
freedom associated with the momentum constraint in those directions, and implies
$N^\sigma=0=N^\delta$~\cite{man}. As a~consequence, the metric components only depend on $t$ and
$\theta$ and are periodic in the latter. This periodicity allows us to decompose the metric
components in Fourier modes\footnote{We adopt the following convention to
def\/ine the Fourier modes $\phi_{m}$ of a generic f\/ield $\phi(\theta)$:
\begin{gather*}
\phi(\theta)=\sum_{m\in\mathbb{Z}}\frac1{\sqrt{2\pi}}\phi_{m}e^{im\theta},
\qquad \phi_{m}=\frac1{\sqrt{2\pi}}\oint d\theta
\phi(\theta)e^{-im\theta}.
\end{gather*}}.
On the other hand, the condition of linear polarization imposes
$q_{\sigma\delta}=0$. Therefore, the three-metric is diagonal and can be described by three f\/ields
$(\tau,\xi,\bar\gamma)$, that essentially characterize the area of the isometry group orbits, the
norm of one of the Killing vectors, and the scale factor of the metric induced on the set of group
orbits. The phase space is then parametrized by those f\/ields and by their momenta
$(P_{\tau},P_{\xi},P_{\bar\gamma})$, and constrained by the $\theta$-momentum constraint and by the
Hamiltonian constraint~\cite{men1b}.

In order to prepare the model for its quantization, the gauge is further reduced. One
imposes that the generator of the conformal transformations,~$P_{\bar\gamma}$, and the area of
the isometry group orbits,~$\tau$, are homogeneous functions. These conditions f\/ix the gauge freedom
associated with the non-zero Fourier modes of the $\theta$-momentum constraint and of the
densitized Hamiltonian constraint, and imply that the functions $N_\theta$ and
${N_{_{_{\!\!\!\!\!\!\sim}}\;}}$ are homogeneous \cite{men1b,gow-let,gow-ijmp}.
Then, two global constraints remain in the model, the spatial average of the
$\theta$-momentum constraint, generating rigid rotations in the circle, and the spatial average
of the densitized Hamiltonian constraint. We will denote them by~$C_\theta$ and~$C_{\text{G}}$,
respectively.

The classically reduced phase space can be split into homogeneous and inhomogeneous sectors.
The homogeneous sector coincides with the phase space of the Bianchi~I spacetime with three-torus
topology. This sector will be quantized \emph{\`a la} LQC, and
therefore is parametrized by the variables $\{(c^i,p_i),i=\theta,\sigma,\delta\}$, with Poisson
bracket $\{c^i,p_j\}=8\pi G\gamma\delta^i_j$, as described in Section~\ref{chap:bianchi}.
The inhomogeneous sector is given by the non-zero (inhomogeneous) modes of the f\/ields unaf\/fected by
the gauge f\/ixing, namely
$\{(\xi_{_m},P_{\xi_{_m}}),m\in\mathbb{Z}\setminus \{0\}\}$. This sector will be quantized employing the
Fock quantization of \cite{men1b}.
To employ this Fock representation, the above inhomogeneous modes are in turn described by
annihilation and creation variables~$(a_m,a_m^*)$, def\/ined as those related to a free
massless scalar f\/ield.
This quantization is preferred as long as is the only Fock quantization of the
deparametrized system in which the dynamics is unitary and with a vacuum invariant under the~$S^1$
translations (the remaining symmetry after gauge f\/ixing)~\cite{men2,men3}.

The three-metric in terms of the chosen variables reads~\cite{gow-B}
\begin{gather}
ds^2 =-q_{\theta\theta}\left(\frac{|p_\theta|}{4\pi^2}\right)^2
{N_{_{_{\!\!\!\!\!\!\sim}}\;}}^2dt^2+q_{\theta\theta}d\theta^2+q_{\sigma\sigma}d\sigma^2
+q_{\delta\delta}d\delta^2,\nonumber\\
q_{\theta\theta} =\frac1{4\pi^2}\left|\frac{p_\sigma
p_\delta}{p_\theta}\right|\exp\left\{\frac{2\pi}{\sqrt{|p_\theta|}}\frac{c^\delta p_\delta-c^\sigma
p_\sigma}{c^\sigma p_\sigma+c^\delta p_\delta}\tilde\xi(\theta)-\frac{\pi^2}{|p_\theta|}[
\tilde\xi(\theta)]^2 -\frac{8\pi G\gamma}{c^\sigma
p_\sigma+c^\delta p_\delta} \zeta(\theta)\right\},\nonumber\\
q_{\sigma\sigma} =\frac1{4\pi^2}\left|\frac{p_\theta
p_\delta}{p_\sigma}\right|\exp\left\{-\frac{2\pi}{\sqrt{|p_\theta|}}\tilde\xi(\theta)\right\},
\nonumber\\
q_{\delta\delta} =\frac1{4\pi^2}\left|\frac{p_\theta
p_\sigma}{p_\delta}\right|\exp\left\{\frac{2\pi}{\sqrt{|p_\theta|}}\tilde\xi(\theta)\right\},\label{newmetric}
\end{gather}
where the inhomogeneities are encoded in the terms
\begin{gather*}
\tilde\xi(\theta) =\frac1{\pi}\sum_{m\neq0}\sqrt{\frac{G}{|m|}}(a_m+a_{-m}^*)e^{im\theta},
 \\
\zeta(\theta) =i\sum_{m\neq0}\sum_{\tilde m\neq0}\text{sgn}(m+\tilde m)\frac{\sqrt{|m+\tilde
m||\tilde m|}}{|m|}(a_{-\tilde {m}}-a^*_{\tilde m}\big)\big(a_{m+\tilde m}+a^*_{-(m+\tilde
m)}\big)e^{im\theta}.
\end{gather*}

On the other hand, the remaining constraints have the following form:
\begin{gather}
C_\theta =\sum_{m=1}^\infty
m(a_m^*a_m-a_{-m}^*a_{-m})=0,\nonumber \\ 
C_{\text{G}} = C_\text{BI}+C_\xi=0,\quad C_\xi={G}\left[\frac{(c^\sigma
p_\sigma+
 c^\delta p_\delta)^2}{\gamma^2|p_\theta|}
H_\text{int}^\xi+32\pi^2|p_\theta|
H_0^\xi\right].\label{Cclas}
\end{gather}
In the above expression,
\begin{gather} \label{HIclas}
H_\text{int}^\xi
  =\sum_{m\neq
0}\frac{1}{2|m|}\left[2a^*_ma_m+
a_ma_{-m}+a^*_ma^*_{-m}\right], \qquad H_0^\xi=\sum_{m\neq
0}|m|a^*_ma_m,
\end{gather}
and $C_\text{BI}$ is the (densitized) Hamiltonian constraint of the Bianchi~I model given in equation~\eqref{eq:ligadura-bianchi-B}.
In the Hamiltonian constraint, the inhomogeneities appear in the term $H_0^\xi$, that corresponds
to the Hamiltonian of a free massless scalar f\/ield, and in the term~$H_\text{int}^\xi$, that
represents an interaction term. The inhomogeneities are coupled to the homogeneous sector in a
non-trivial way, so that the feasibility of the hybrid quantization is not straightforward \emph{a priori}.

\subsection{Fock quantization of the inhomogeneous sector}\label{Fphysi}

Once the inhomogeneous sector is described with the appropriate annihilation and creation like
variables,~$a_m$ and~$a^*_m$, it is straightforward to get its Fock quantization.
With that aim, we promote the variables~$a_m$ and~$a^*_m$ to annihilation and creation operators,~$\hat a_m$ and~$\hat a^\dagger_m$ respectively, such that
$[\hat a_m,\hat a^\dagger_{\tilde m}]=\delta_{m\tilde m}$. From the vacuum state
$|0\rangle$, characterized by the equations
\begin{equation*}
\hat a_m|0\rangle=0,\qquad \forall \, m\in\mathbb{Z},
\end{equation*}
we construct the one-particle Hilbert space, and the associated symmetric Fock space
$\mathcal{F}$ \cite{wald2}.
The annihilation and creation operators are densely def\/ined in the
subspace of $\mathcal{F}$ given by f\/inite linear
combinations of $n$-particle states
\begin{gather*}
 |\mathfrak n \rangle:=|\dots,n_{-2},n_{-1},n_1,n_2,\dots\rangle,
\end{gather*}
such that $\sum_mn_m<\infty$, being $n_m\in\mathbb{N}$ the occupation number (or number of
particles) of the $m$-th mode. We will denote that space by $\mathcal
S$. Note that the $n$-particle states provide a basis for the Fock space, orthonormal with respect
to the inner product $\langle\mathfrak n^\prime|\mathfrak n\rangle=\delta_{\mathfrak
n^\prime\mathfrak n}$.
The action of $\hat a_m$ and $\hat a^\dagger_m$ on these states is
\begin{gather*}
  \hat a_m |\dots,n_m,\dots\rangle=\sqrt{n_m}|\dots,n_m-1,\dots\rangle ,\\
  \hat a_m^\dagger|\dots,n_m,\dots\rangle=\sqrt{n_m+1}|\dots,n_m+1,\dots\rangle .
\end{gather*}

\subsubsection{Generator of translations in the circle}

The constraint that generates translations in the circle, $C_\theta$, does not af\/fect the
homogeneous sector, and then it is represented on the above Fock space. Taking normal ordering, the
corresponding operator is
\begin{equation*}
\widehat C_\theta=\hbar\sum_{m>0}^\infty m(\hat
a^\dagger_m \hat a_m-\hat a^\dagger_{-m} \hat a_{-m}).
\end{equation*}
This operator is self-adjoint in the Fock space $\mathcal F$.

The $n$-particle states annihilated by $\widehat C_\theta$ are those that satisfy the condition
\begin{equation*}
\sum_{m>0}^\infty m X_m=0,\qquad X_m=n_m-n_{-m}.
\end{equation*}
They provide a basis for a proper subspace of the Fock space, that we will denote by  $\mathcal
F_f$.

\subsection{Hamiltonian constraint operator}

Physical states must be annihilated as well by the quantum counterpart of the Hamiltonian
constraint $C_\text{G}$, given in equation~\eqref{Cclas}, which involves both homogeneous and
inhomogeneous sectors.

In the previous section, we have already described the representation of the inhomogeneous sector,
with basic operators $\hat a_m$ and $\hat a^\dagger_m$ acting on the Fock space $\mathcal F$, which
thus constitutes the inhomogeneous sector of the kinematical Hilbert space.
On the other hand, the homogeneous sector is quantized following LQC, namely, it is given by the loop
quantization of the Bianchi~I model.
 As we discussed in Section \ref{chap:bianchi}, in the
literature two dif\/ferent implementations of the improved dynamics has been applied to the Bianchi~I
model. Therefore, there exist also two dif\/ferent descriptions for the hybrid Gowdy model, one
adopting the naive {\em Ansatz}~\eqref{eq:mubarraA} \cite{gow-B,gow-let,gow-ijmp}, and another
adopting the improved {\em Ansatz}~\eqref{eq:mubarraB} \cite{gow-B,mmw}. Here we will just explain the
second description, which adopts the quantization of the Bianchi~I model described in Section~\ref{chap:bianchi} when representing the homogeneous sector. This sector of the
kinematical Hilbert space will be the kinematical Hilbert space
$\mathcal H^+_{\varepsilon,\lambda_\sigma^\star,\lambda_\delta^\star}$, def\/ined in Section~\ref{superselection}.

The f\/irst term of the Hamiltonian constraint operator, $\widehat C_\text{G}=\widehat
C_\text{BI}+\widehat C_\xi$, is thus the Bianchi~I operator~\eqref{densCB}. We just need to construct the operator $\widehat C_\xi$ that couples homogeneous and inhomogeneous sector.

Let us f\/irst focus on the inhomogeneous terms.
In order to represent the free Hamiltonian~$H_0^\xi$ and the
interaction term $H_\text{int}^\xi$, def\/ined in equation~\eqref{HIclas}, we choose normal ordering. Then,
their quantum analogs are given by
\begin{gather*}
\widehat{H}_0^\xi =\sum_{m>0}^\infty m\hat N_m,\qquad\hat N_m=
\hat{a}^{\dagger}_m \hat{a}_m+\hat{a}^{\dagger}_{-m} \hat{a}_{-m},
\\
\widehat{H}_\text{int}^\xi
  =\sum_{m>0}^\infty\frac{\hat N_m+\hat Y_m}{m},\qquad\hat Y_m=\hat{a}_m
\hat{a}_{-m}+\hat{a}^{\dagger}_m\hat{a}^{\dagger}_{-m},
\end{gather*}
both densely def\/ined in the space $\mathcal S$ of $n$-particle states.
The operator $\widehat{H}_0^\xi$ acts diagonally on the $n$-particle states, and then it is
well-def\/ined in the Fock space $\mathcal F$. On the contrary, $\widehat{H}_\text{int}^\xi$ does
not leave invariant the domain $\mathcal S$. Indeed, the operator $\hat Y_m$
annihilates and creates pairs of particles in modes with the same wavenumber~$|m|$, and then
$\widehat{H}_\text{int}^\xi$ creates an inf\/inite number of particles. However, one can prove~\cite{gow-B} that the norm of $\widehat{H}_\text{int}^\xi|\mathfrak n\rangle$ is f\/inite for all
$\mathfrak n\in\mathcal S$, and therefore this operator, with domain $\mathcal S$, is also well
def\/ined in the Fock space~$\mathcal F$.

For the homogeneous terms, we recall that the operator $\widehat\Omega_i$, def\/ined in equation~\eqref{cp-quantum}, is the loop quantum analogue of the classical term~$c^ip_i$, and that the inverse
powers of~$|p_i|$ can be regularized taking commutators of $p_i$ with holonomies.
In view of these prescriptions, $C_\xi$ can be represented by the symmetric operator~\cite{gow-B,mmw}
\begin{gather*}
 \widehat{\mathcal{C}}_{\xi}
=l_{\text{Pl}}^2\left\{
\widehat{\left[\frac{1}{|p_\theta|^{\frac1{4}}}\right]}^2
\frac{(\widehat\Omega_\sigma+\widehat\Omega_\delta)^2}{\gamma^2}
\widehat{\left[\frac{1}{|p_\theta|^{\frac1{4}}}\right]}^2
\widehat{H}_\text{int}^\xi+32\pi^2\widehat{|p_\theta|}
\widehat{H}_0^\xi\right\},
\end{gather*}
where
\begin{gather}
 \widehat{\left[\frac1{|p_\theta|^{\frac1{4}}}\right]}|v,\lambda_\sigma,\lambda_\delta\rangle
 =\frac{b_\theta^\star(v,\lambda_\sigma,\lambda_\delta)}{(4\pi\gamma
l_{\text{Pl}}^2\sqrt{\Delta})^{\frac1{6}}}|v,\lambda_\sigma, \lambda_\delta\rangle,\nonumber\\
b_\theta^\star(v,\lambda_\sigma,\lambda_\delta) =
\sqrt{2|\lambda_\sigma\lambda_\delta|}\left|\sqrt{|v+1|}-\sqrt{|v-1|}\right|.\label{btheta}
\end{gather}
The operator $\widehat{\mathcal{C}}_{\xi}$, so constructed, leaves the sectors of
superselection of the Bianchi~I model~in\-variant, and then it is in fact well def\/ined on the separable
kinematical Hilbert space $\mathcal
H^+_{\varepsilon,\lambda_\sigma^\star,\lambda_\delta^\star}\otimes\mathcal F$.

\subsection{Physical Hilbert space}

In order to impose the Hamiltonian constraint
$\big(\psi\big|\widehat{\mathcal{C}}_{\text{G}}^\text{B}{}^\dagger=0$,
we expand a general state $(\psi|$ in the basis of states
$|v,\omega_{\varepsilon}\lambda_\sigma^\star,
\bar\omega_{\varepsilon}
\lambda_\delta^\star\rangle$ of the homogeneous sector.
That is,
\begin{gather*}
(\psi| =\sum_{v\in\mathcal
L_{{\varepsilon}}}\sum_{\omega_{\varepsilon}\in\mathcal
W_{\varepsilon}}\sum_{\bar\omega_{\varepsilon}\in
\mathcal W_{\varepsilon}}\langle
v,\omega_{\varepsilon}\lambda_\sigma^\star,\bar\omega_{\varepsilon}
\lambda_\delta^\star|\otimes(
\psi(v,\omega_{\varepsilon}\lambda_\sigma^\star,
\bar\omega_{\varepsilon}\lambda_\delta^\star)| ,
\end{gather*}
where, let us recall, $\mathcal W_{\varepsilon}$ is the set~\eqref{W-set}.
In the above expression, \[
(\psi(v,\lambda_\sigma,\lambda_\delta)|=
(\psi(v,\omega_{{\varepsilon}}\lambda_\sigma^{\star},
\bar\omega_{{\varepsilon}}\lambda_\delta^{\star})|
\] is the projection of $(\psi|$ on the
state
$|v,\lambda_\sigma,\lambda_\delta\rangle=
|v,\omega_{\varepsilon}\lambda_\sigma^\star,\bar\omega_{\varepsilon}
\lambda_\delta^\star\rangle$ of the homogeneous sector and, in principle, it
must belong to the dual space of some appropriate dense domain of the Fock space~$\mathcal F$.

If we substitute the above expansion in the constraint, and take into account the action of the
operators af\/fecting the homogeneous sector, we obtain that the projections
$(\psi(v,\lambda_\sigma,\lambda_\delta)|$ satisfy dif\/ference equations in~$v$ that, generically,
relate data on the section $v+4$ with data on the sections $v$ and $v-4$, as it happened in the
Bianchi~I model. Following~\cite{gow-B}, to simplify the notation of the resulting equation,
we introduce the projections of $(\psi|$ on the linear combinations given in equation~\eqref{eq:combi}.
Namely, we def\/ine $(\psi_\pm(v\pm n,\lambda_\sigma,\lambda_\delta)|
=(\psi|v\pm n,\lambda_\sigma,\lambda_\delta\rangle_\pm$.
Similarly, it is convenient to introduce the combinations of states
\begin{gather*}
|v\pm n,\lambda_\sigma,\lambda_\delta\rangle_\pm' =\bigg|v\pm n,\lambda_\sigma,\frac{v\pm n}
{v}\lambda_\delta\bigg\rangle+\bigg|v\pm n,\frac{v\pm n}
{v\pm2}\lambda_\sigma,\frac{v\pm2}
{v}\lambda_\delta\bigg\rangle\nonumber\\
\phantom{|v\pm n,\lambda_\sigma,\lambda_\delta\rangle_\pm' =}{}
+\bigg|v\pm n,\frac{v\pm n}{v}\lambda_\sigma,\lambda_\delta
\bigg\rangle+\bigg|v\pm n,\frac{v\pm2}
{v}\lambda_\sigma,\frac{v\pm n}
{v\pm2}\lambda_\delta\bigg\rangle,
\end{gather*}
and to def\/ine the projections of $(\psi|$ on them:
$ (\psi'_\pm(v\pm n,\lambda_\sigma,\lambda_\delta)|=(\psi|v\pm n,\lambda_\sigma,
\lambda_\delta\rangle_\pm'$.
With this notation, the solutions of the Hamiltonian constraint satisfy the explicit relation
\begin{gather}
(\psi_+ (v+4,\lambda_\sigma,\lambda_\delta)|-\eta
[b_\theta^\star(v,\lambda_\sigma,\lambda_\delta)b
_\theta^\star(v+4,\lambda_\sigma,\lambda_\delta)]^2\frac{
v+4}{v}(\psi{}_+'(v+4,\lambda_\sigma, \lambda_\delta)
|\widehat{H}_\text{int}^\xi\nonumber\\
\qquad{} =-\frac{1}{\eta}\frac{32v^2}{\lambda_\sigma^2
\lambda_\delta^2x_+(v)}(\psi(v,
\lambda_\sigma,\lambda_\delta)|\widehat{H}_0^\xi
+\frac{x^-_0(v)}{x_+(v)}(\psi_{-}
(v,\lambda_\sigma,\lambda_\delta)|+
\frac{x^+_0(v)}{x_+(v)}(\psi_{+}
(v,\lambda_\sigma,\lambda_\delta)|
\nonumber\\
\qquad{} -\frac{x_-(v)}{x_+(v)}
(\psi_-(v-4,\lambda_\sigma,\lambda_\delta)|
+\eta[b_\theta^\star(v,\lambda_\sigma,
\lambda_\delta)]^4\bigg\{
\left[\frac{b_\theta^\star(v-4,\lambda_\sigma,\lambda_\delta)}{b_\theta^\star(v,\lambda_\sigma,
\lambda_\delta)}\right]^2
\frac{v-4}{v}\frac{x_-(v)}{x_+(v)}\nonumber\\
\qquad{} \times(\psi_-'(v-4,\lambda_\sigma,
\lambda_\delta)|-
\bigg[\frac{x_{0}^{-}(v)}{x_+(v)}(\psi'_{-
}(v,\lambda_\sigma,\lambda_\delta)|+
\frac{x_{0}^{+}(v)}{x_+(v)}(\psi'_{+
}(v,\lambda_\sigma,\lambda_\delta)|\bigg]\bigg\}
\widehat{H}_\text{int}^\xi,\label{solutionB}
\end{gather}
where $\eta=\left(\frac{l_{\text{Pl}}}
{4\pi\gamma\sqrt{\Delta}}\right)^{2/3}$ is a dimensionless parameter,
$b_\theta^\star(v,\lambda_\sigma,\lambda_\delta)$ is the function~\eqref{btheta}, and the
coef\/f\/icients $x_\pm(v)$ and $x^\pm_0(v)$ were def\/ined in equation~\eqref{coeficientes}.

Similarly to the analysis done in the Bianchi~I model, it has been investigated whether the
solution is totally determined (at least formally) by the data in the initial section
$v={\varepsilon}$. The presence of the interaction term in the left-hand side of equation~\eqref{solutionB} complicates a direct demonstration of the above statement. However, it is
possible to obtain such result in terms of an asymptotic analysis of the solutions. Note that the
model provides a dimensionless parameter $\eta$ that can be used to develop an asymptotic
procedure, without the need to introduce any external parameter by hand. This analysis was carried
out in~\cite{gow-B}, and we refer to it for the details. The main result of this analysis is
that, in fact, the initial data $(\psi({\varepsilon},\lambda_\sigma,\lambda_\delta)|$
(where $\lambda_\sigma$ and $\lambda_\delta$ run over all possible values in their corresponding
superselection sectors) completely determine the solution. The solutions turn out to be formal,
in the sense that the states $(\psi(v+4,\lambda_\sigma,\lambda_\delta)|$ do not belong in general
to the dual space of~$\mathcal S$, owing to the presence of $\widehat{H}_\text{int}^\xi$ in their
expression.

The physical Hilbert space can be characterized, even though the solutions are formal. Indeed, once
we justify that the set of initial data
$
\{\left(\psi({\varepsilon},\omega_{{\varepsilon}}\lambda_\sigma^{\star},
\bar\omega_{\varepsilon}\lambda_\delta^{\star})\right|;\;
\omega_{{\varepsilon}},\bar\omega_{\varepsilon}\in\mathcal W_{{\varepsilon}}\}
$
specif\/ies the solution, we can identify solutions with their corresponding initial data, and the
physical Hilbert space with the Hilbert space of such initial data, exactly as we proceeded with
the Bianchi I model.

Once again, the reality conditions over a complete set of observables, acting on the initial data,
univocally determines the inner product that provides the Hilbert structure. Such observables are
given, for instance, by the overcomplete set of observables of the Bianchi I model, given in equation~\eqref{conjunto-completo}, together with a suitable complete set of observables for the
inhomogeneous sector, given by~\cite{gow-B}
\begin{gather*}
\big\{(\hat a_m+\hat a_m^\dagger)\pm(\hat a_{-m}+\hat a_{-m}^\dagger), \
i[(\hat a_m-\hat a_m^\dagger)\pm(\hat a_{-m}-\hat a_{-m}^\dagger)]; \  m\in\mathbb{N}^+\big\}.
\end{gather*}
These operators represent the real Fourier coef\/f\/icients of the non-zero modes of the f\/ield
$\xi(\theta)$ and of its momentum $P_\xi(\theta)$, and in fact they are self-adjoint in the Fock
space $\mathcal F$.

Finally, imposing the remaining symmetry of translations on $S^1$, the result is that the physical
Hilbert space of the Gowdy model is \cite{gow-B}
\begin{gather*}
\mathcal H_{\text{phys}}=\mathcal H_{\lambda_\sigma^\star,\lambda_\delta^\star}
\otimes\mathcal F_f .
\end{gather*}
Namely, it is the tensor product of the physical Hilbert space of the Bianchi I model times the
physical Fock space for the inhomogeneities (def\/ined in Section~\ref{Fphysi}). We note that~$\mathcal F_f$ is unitarily equivalent
to the physical space of the Fock quantization of the deparametrized system~\mbox{\cite{men1b, men1a}}.
Therefore, the standard quantum f\/ield theory for the inhomogeneities is recovered, and they can be
seen as propagating over a polymerically quantized Bianchi~I background. This result supports the
validity of the hybrid quantization, since this should lead to the standard quantization of the
system in the limit in which the ef\/fects coming from the discreteness of the geometry are
negligible. This result is not trivial, since the hybrid quantization is introduced in the
kinematical setting, and the relation between kinematical and physical structures cannot be
anticipated before the quantization is completed.

\subsubsection{Singularity resolution}

The classical solutions of the linearly polarized Gowdy $T^3$ model generically display a
cosmological singularity~\cite{mon2}. In the parametrization employed for the hybrid quantization
of the model, this classical singularity corresponds to vanishing values of the coef\/f\/icients $p_i$.
In the quantum theory, the kernel of the operators~$\hat p_i$ is removed and, as a consequence,
there is no analog of the classical singularity. This resolution of the singularity is kinematical
and, therefore, independent of the dynamics. It persists in the Hilbert space of the physical
states since they do not have projection on the kernel of the operators~$\hat p_i$. Moreover, they
only have support in a sector with positive orientation of the coef\/f\/icients~$p_i$ and, then, they
do not cross the singularity towards other branches of the universe corresponding to dif\/ferent
orientations.

A description of the evolution picture of the model is missing, owing to its high complication. It
is worthy to note that, at least for the choice of the original naive improved dynamics, the
ef\/fective dynamics of the model has been thoroughly analyzed~\cite{eff,eff1}. In particular, it has
been studied how the inhomogeneities af\/fect the dynamics of the Bianchi~I background. Numerical
simulations show that the ef\/fect of the inhomogeneities does not destroy the bounce. For the improved dynamics discussed here, a similar analysis has not been done yet,
but we can expect similar results, since the bounce mechanism appears for both improved schemes.

\subsection{The Gowdy $T^3$ model coupled to a massless scalar f\/ield}

So far we have discussed the hybrid quantization of the linearly polarized Gowdy model in vacuo.
This model allows, almost straightforwardly, for the introduction of a minimally coupled free
massless scalar f\/ield with the same symmetries as the metric \cite{gow-matter}. Indeed, after a suitable rescaling of matter modes, these contribute to
the constraints $C_\theta$ and $C_\text{G}$ exactly in the same manner as the gravitational f\/ield
$\xi$. Also, the Fock quantization of that system (after a complete deparametrization) enjoys the
same uniqueness results as that of the model in vacuo and, hence, there is a preferred Fock
description also for the inhomogeneities of the matter f\/ield. Therefore, the hybrid approach
follows exactly in the same way as for the vacuum case.

The interest in considering the model f\/illed with matter lies in the fact
that FRW-type solutions are then allowed. Indeed, as we saw before, the vacuum Gowdy model can be
seen as a~vacuum Bianchi~I background f\/illed with inhomogeneities propagating in one direction, and
the subclass of isotropic solutions of the vacuum Bianchi I model represent trivial Minkowski
spacetimes rather than f\/lat FRW universes. Nonetheless, in the presence of matter, the f\/lat-FRW becomes
the isotropic sector of the Bianchi~I model. In that sense, there is a subclass of solutions of the
Gowdy model coupled to matter that can be regarded as a f\/lat FRW background f\/illed with
inhomogeneities propagating in one direction. Therefore, the linearly polarized Gowdy~$T^3$~model
coupled to a massless scalar f\/ield provides a simple
laboratory where to study, at the quantum level (by means of the hybrid quantization), interesting physical
phenomena such as the backreaction of the (quantum) inhomogeneities on
(polymerically quantized) f\/lat FRW cosmologies or, vice versa, the ef\/fect of the
quantum background geometry on the propagation of the inhomogeneities~\cite{gow-matter2}. This
analysis is intended to be a f\/irst step towards a quantum theory of FRW plus inhomogeneities. Its character is quite preliminary, since the inhomogeneities of Gowdy are just a subclass of the
inhomogeneities that one would introduce in the FRW model to account for the inhomogeneities that we observe in our universe. Nonetheless, a complete quantization of such a
system has not been yet achieved, and the hybrid Gowdy model of\/fers a~suitable setting to start
with. Actually, by employing the same hybrid procedure, the f\/lat FRW model plus perturbations is being
analyzed~\cite{frw-matter}. The hybrid quantization also applies in this more realistic system as
long as a unique Fock quantization for the inhomogeneities is at hand as well~\cite{frw-uni2,frw-uni3, frw-uni1}.

\section{Polymer quantization of the polarized Gowdy $T^3$ model}

In the previous section, we saw a successful quantization scheme of the linearly polarized Gowdy
$T^3$ model where the degrees of freedom were split into homogeneous
and inhomogeneous sectors. The homogeneous sector was quantized using the LQC techniques, while the inhomogeneous part was Fock quantized. One of the
signif\/icant advantages of the hybrid quantization is that the calculations are tractable and the tools developed and studied in LQC can be used to
address questions even in the midisuperspace context. While it is an extremely useful f\/irst step in quantization of midisuperspace models, it crucially
depends on the fact that the inhomogeneous degrees of freedom can be treated perturbatively, and it
is assumed that there exits a regime in which the most important ef\/fects emerging from the discretization of the geometry are those that af\/fect the
homogeneous subsystem. Ideally, we would like to loop quantize the full polarized Gowdy $T^3$ model without separating the
degrees of freedom. In this section we review the work that has been done so far in that direction, which has been carried out in~\cite{kinjal1,kinjal2}.

\subsection{Classical theory}
The variables chosen in this section are signif\/icantly dif\/ferent from the ones used in the rest of the review so far. We shall therefore indicate the
steps followed in obtaining these variables.

\subsubsection{Gowdy $T^3$ model in Ashtekar variables}

In order to loop quantize, we f\/irst need to rewrite the Gowdy $T^3$ model in terms of real Ashtekar
variables. Canonical quantization of the {\em unpolarized} Gowdy $T^3$ model in terms of the complex
Ashtekar variables has been given in \cite{HusainSmolin,Menamarugan} which we will
brief\/ly sketch below in terms of the real Ashtekar variables.

Recall that, owing to global hyperbolicity, spacetime can be decomposed as \mbox{${\mathcal M} = \Sigma_t \otimes {\mathbb R}$}, where $\Sigma_t$
is homeomorphic to a three-tours. As in the previous section, let the angular coordinates of $\Sigma_t$ be ($\theta,\sigma,\delta$), and
the two commuting Killing vectors be $\xi_1^a = \p_\sigma$ and $\xi_2^a = \p_\delta$.
These isometries imply that the Lie derivatives along these two Killing vectors vanish, i.e.
\begin{gather*}
{\mathcal L}_{\xi_1}A_a^i   =  0   =    {\mathcal L}_{\xi_1}E^a_i,\\
{\mathcal L}_{\xi_2}A_a^i   =  0  =    {\mathcal L}_{\xi_2}E^a_i .
\end{gather*}
The phase-space variables are therefore only functions of $\theta$. The Gauss and the dif\/feomorphism constraint reduce to
\begin{gather*}
G_i = \partial_\theta E^\theta_i + \epsilon_{ij}^k A_a^j E^a_k, \\
V_a = (\partial_a A_b^i)E^b_i - (\partial_\theta A_a^i) E^\theta_i + \epsilon_{jk}^i  A_a^j  A_b^k E^b_i .
\end{gather*}
The vector constraint given by $C_a = A_a^i G_i - V_a $ generates spatial dif\/feomorphisms.

We now impose the following gauge-f\/ixing conditions:
\begin{gather*}
E^{\theta}_I  =  0  =  E^{\rho}_3 ,\qquad \rho  =  \sigma,  \delta ,\qquad I  =  1,  2 .
\end{gather*}
The constraints $G_I$ and $C_\rho$ are then solved by $A_{\theta}^I = 0 = A_{\rho}^3$.

Thus, only one component of the Gauss constraint ($G_3 $) and one of the dif\/feomorphism constraint along the $\theta$ direction
($C_{\theta} =: C$) survive together with the Hamiltonian constraint. Since none of the quantities depend on $\sigma$ or $\delta$, we can
integrate over the torus $T^2$ and write the symplectic structure as\footnote{In this section, sum over repeated indices is understood.}
\begin{gather}
\Omega =  \frac{4\pi^2}{\kappa\gamma}\int {d}\theta\big( {d}A_{\theta}^3 \wedge {d}E^{\theta}_3 +
{d}A_{\rho}^I \wedge {d}E^{\rho}_I \big). \label{symplectic1}
\end{gather}
This is the classical phase phase in terms of real Ashtekar variables. One important observation is that this is basically a
one-dimensional theory. This is useful because in one dimension, under orientation-preserving coordinate transformations, a tensor density of
contravariant rank~$p$, covariant rank~$q$ and weight~$w$, can be thought of as a scalar density of weight $= w + q - p$. Hence, under a~$\theta$ coordinate transformation, $E^{\theta}_3$  transforms as a scalar, $E^{\rho}_I$'s transform as scalar densities of weight 1,
$A_{\theta}^3$ transforms as a scalar density of weight~1, and $A_{\rho}^I$'s transform as scalars.

\subsubsection{Choice of new variables}

It turns out that these variables are not suitable for loop quantization and we need to make
canonical transformations similar to those performed
for the spherical symmetric case in~\cite{martinsph1}.  Note that, for each $\rho$, the $A_{\rho}^I$ and $E_I^{\rho}$ rotate among themselves
under the $U(1)$ gauge transformations generated by the Gauss constraint. These suggest that we can perform canonical transformations to def\/ine the
following variables:
\begin{alignat*}{3}
& E^{{\sigma}}_1=  E^\sigma \cos \beta , \qquad &&  E^{{\sigma}}_2 = E^\sigma \sin \beta, & \\ 
& E^{{\delta}}_1 = - E^\delta \sin \bar\beta,\qquad &&  E^{{\delta}}_2 =
E^\delta \cos \bar\beta, & \\ 
& A_{{\sigma}}^1  =  A_{{\sigma}}  \cos ( \alpha + \beta) , \qquad &&
A_{{\sigma}}^2 = A_{{\sigma}}  \sin ( \alpha + \beta) , & \\ 
& A_{{\delta}}^1 = - A_{{\delta}}  \sin ( \bar\alpha + \bar\beta) , \qquad &&
A_{{\delta}}^2 = A_{{\delta}} \cos (\bar\alpha +\bar\beta). & 
\end{alignat*}
The angles for the connection components are introduced in a particular fashion for later convenience.

The radial coordinates, $E^\sigma$, $E^\delta$, $A_\sigma$, $A_\delta$, are gauge invariant and always strictly positive (va\-nishing radial
coordinates correspond to a trivial symmetry orbit which is ignored).

In terms of these variables, the symplectic structure (\ref{symplectic1}) gets expressed as
\begin{gather*}
\Omega=\frac{4\pi^2}{\kappa\gamma}\int {d}\theta\left[ {d}A^3_{\theta} \wedge {d}E_3^{\theta} +
{d} X \wedge {d} E^{\sigma} + {d} Y \wedge {d} E^{\delta} + {d} \beta
\wedge {d} P^{\beta} +
{d} \bar\beta \wedge {d} \bar P^{\beta}\right] ,
\end{gather*}

where
\begin{alignat*}{3}
& X:=  A_\sigma \cos (\alpha), \qquad &&  Y := A_\delta \cos (\bar\alpha ), & \\
& P^{\beta} := - E^\sigma A_\sigma \sin(\alpha), \qquad &&
\bar P^{\beta} :=  - E^\delta A_\delta \sin ( \bar\alpha ).
\end{alignat*}
It is convenient to make a further canonical transformation:
\begin{alignat*}{3}
& \xi = \beta - \bar\beta, \qquad &&  \eta = \beta + \bar\beta, & \\ 
& P^{\xi} = \frac{P^{\beta} -\bar P^{\beta}}{2}, \qquad && P^{\eta} = \frac{P^{\beta} + \bar P^{\beta}}{2}. & 
\end{alignat*}
The constraints are greatly simplif\/ied and their detailed expressions can be found in~\cite{kinjal1}. This completes the description of the
unpolarized Gowdy $T^3$ model in the variables we have def\/ined. The number of canonical f\/ield variables is 10 while there is a threefold
inf\/inity of f\/irst-class constraints. There are therefore 2 f\/ield degrees of freedom.  We now need to impose two second-class constraints such that
the number of f\/ield degrees of freedom are reduced from two to one (as it should be in the polarized case).

\subsubsection{Reduction to polarized model}

In terms of the variables def\/ined above, the spatial three-metric is given by
\begin{gather*}
{d}s^2= \cos\xi \frac{E^\sigma E^\delta}{E_3^\theta} {d}\theta^2 +
\frac{E_3^\theta}{\cos\xi} \frac{E^\delta}{E^\sigma}{d}\sigma^2 +
\frac{E_3^\theta}{\cos\xi} \frac{E^\sigma}{E^\delta}{d}\delta^2 - 2
\frac{E_3^\theta}{\cos\xi}\sin\xi \ {d}\sigma {d}\delta.
\end{gather*}

For the Killing vectors  $\p_\sigma$ and $\p_\delta$ to be
orthogonal to each other, the ${d}\sigma {d}\delta$
term in the metric should be zero.  This implies that the polarization condition is implemented by restricting to
the $\xi = 0$ sub-manifold of the phase space of the unpolarized model\footnote{Actually there are two possible choices, $\xi = 0$ and $\xi = \pi$.
We shall take the constraint to be $\xi = 0$, which implies $E^{\theta}_3 > 0$.}.
In order to get a non-degenerate symplectic structure, we need one more condition. We expect the two conditions to reduce a f\/ield degree of freedom.
This turns out to be
\begin{gather*}
\chi(\theta) := 2 P^{\xi} + E_3^\theta \partial_\theta \ln\frac{E^\delta}{E^\sigma} ~ \approx 0 . 
\end{gather*}
Thus, the reduction to the polarized model is obtained by imposing the two {\em polarization constraints}
\begin{equation*}
\xi \approx 0 ,\qquad \chi \approx 0 ,\qquad
 \{\xi(\theta), \chi(\theta')\} = 2 \kappa\gamma\delta(\theta - \theta')  .
\end{equation*}
We can solve the polarization constraints strongly and use Dirac brackets. Since the polarization constraints weakly commute with all the other
constraints, the constraint algebra in terms of Dirac brackets is same as that in terms of the Poisson brackets and remains unaf\/fected.
Furthermore, equations of motions for all the variables other than~$\xi$,~$P_{\xi}$ also remain unaf\/fected. We can thus set the polarization
constraints strongly equal to zero in all the expressions and continue to use the original Poisson brackets.
It also turns out that the basic variables~$X$,~$Y$, unlike in the full theory, are actually the extrinsic curvature components in the~$\sigma$ and $\delta$ direction, respectively.

The above construction can be carried out equivalently using $SU(2)$ variables by using $\eta$-dependent $\tau$ matrices:
\begin{gather}
\tau_\sigma(\theta)   :=   \cos\eta(\theta) \ \tau_1 + \sin\eta(\theta)   \tau_2 , \nonumber\\
\tau_\delta(\theta)   :=   -\sin\eta(\theta) \ \tau_1 + \cos\eta(\theta)   \tau_2 , \nonumber\\
\tau_3(\theta)   :=  \tau_3 .\label{taumatrix}
\end{gather}
The $SU(2)$ formulation is useful in the quantum theory especially while constructing the Hamiltonian constraint operator.

Let us review the classical phase space we have constructed. For convenience of notation we rename $E_3^{\theta} :=  {\mathcal E}$ and
$A^3_{\theta}:= {\mathcal A}$. The basic conf\/iguration variables are $X$, $Y$, ${\mathcal A}$, $\eta$ and the momentum variables are
$E^\sigma$, $E^\delta$, ${\mathcal E}$, $P^{\eta}$, with Poisson brackets of the form $\{X, E^\sigma\}
= (2 G /\pi) \gamma \delta(\theta - \theta')$.\footnote{Only in this section we will absorb the $4\pi^2$ and use $\kappa':= 2G/\pi$.}
The spatial metric is given by
\begin{gather}
{d}s^2= \frac{E^\sigma E^\delta}{\mathcal E} {d}\theta^2 +
{\mathcal E} \frac{E^\delta}{E^\sigma} {d}\sigma^2 +
{\mathcal E}\frac{E^\sigma}{E^\delta} {d}\delta^2  .
\label{polymergowdymetric}
\end{gather}
The expressions of the constraints are greatly simplif\/ied:
\begin{gather}
G_3   =   \frac{1}{\kappa'\gamma}\left[\partial_{\theta}{\mathcal E} + P^{\eta}\right], \label{ch4gauss} \\
C_{\theta}  =  \frac{1}{\kappa'\gamma}\big[E^\sigma \partial_{\theta}X + E^\delta
\partial_{\theta}Y -
{\mathcal A} \partial_{\theta}{\mathcal E} + P^{\eta} \partial_{\theta}{\eta} \big], \label{ch4diffeo} \\
H   =  - \frac{1}{\kappa'}\frac{1}{\sqrt{E}} \left[ \frac{1}{\gamma^2} \big(  X E^\sigma Y
E^\delta + {\mathcal A} {\mathcal E}( X E^\sigma  + Y E^\delta)
+  {\mathcal E} \partial_\theta \eta (X E^\sigma  + Y E^\delta) \big) - E^\sigma\Gamma_\sigma E^\delta\Gamma_\delta\right] \nonumber \\
\phantom{H   =}{}
+ \frac{1}{2\kappa'}\partial_\theta\left\{\frac{2 {\mathcal E} \left(\partial_\theta
{\mathcal E} \right) }{\sqrt{E}}\right\}  - \frac{\kappa'}{4} \frac{G^2}{\sqrt{E}} -
\frac{\gamma}{2} \partial_{\theta}\left(\frac{G}{\sqrt{E}}\right) , \label{ch4hamiltonian}
\end{gather}
where $E=|{\mathcal E}| E^\sigma E^\delta$.

It is obvious from these def\/initions that $X$, $Y$, ${\mathcal E}$, $\eta$ are scalars while $E^\sigma$,
$E^\delta$, ${\mathcal A}$, $P^{\eta}$ are
scalar densities of weight~1.  The Gauss constraint shows that~${\mathcal A}$ transforms as a~$U(1)$ connection, while~$\eta$ is {\em translated} by the gauge parameter. All other variables are gauge invariant.

This completes the process of symmetry reduction from the unpolarized to the polarized case. This is a consistent symmetry reduction as can be checked
by verifying the constraint algebra. It is also possible to show that the solutions of the equations of motion are equivalent to the standard Gowdy
solutions. In this construction, the goal has been to express the polarized Gowdy $T^3$ in terms of
variables which are suitable for loop quantization. In particular, they allow a simpler choice of edge and point holonomies, a~simpler form for the volume operator, and also a more tractable expression for the Hamiltonian constraint. Although it may not be possible to make the same
choice of variables for other midisuperspace models, similar variables have been used in the preliminary steps of loop quantization of another
midisuperspace model which we describe in brief below.

{\bf Plane gravitational waves.}
The dif\/ference between the polarized Gowdy $T^3$ model and the plane polarized~(pp) gravitational waves is in the global topology. While the Gowdy
model has a compact topology, pp waves have the global topology of Minkowski space. The coordinates are no longer angular but due to the homogeneity
it is possible to choose an arbitrary f\/inite area from the plane wavefronts and consider only f\/inite wave packets. The classical phase space in
Ashtekar variables is constructed in \cite{ppwavesloops,ppwavesloops1} in a similar way as done for the polarized Gowdy $T^3$ model described above.

Waves travelling only in one direction are considered to avoid the problem of wave collision. Finite pulses of pp waves travelling in the positive or
in the negative $z$ direction are characterized by a null Killing vector $k_\mu$ satisfying $\nabla_{(\mu}k_{\nu)} = 0$.
This gives rise to two new constraints:
\begin{gather*}
U_+  =  E^\sigma K_\sigma + E^\delta K_\delta - \partial_z{\mathcal E} , \qquad
U_-  =   E^\sigma K_\sigma - E^\delta K_\delta - {\mathcal
E} \ln\frac{E^\delta}{E^\sigma} .
\end{gather*}
It can be also shown that the constraint $U_-$ is identically zero on the constraint surface and has weakly vanishing Poisson brackets with all the
other constraints, i.e.\ it is gauge invariant and conserved under spatial dif\/feomorphisms and time evolutions. The constraint~$U_+$ Poisson commutes with
the Gauss, dif\/feomorphism and Hamiltonian constraints. This constraint can be added as a new f\/irst-class constraint and the standard
constraint algebra can be enlarged. This is an additional ingredient in the analysis of pp waves and the system can now be loop quantized.

\subsection{Quantum theory}

In this section, we review the loop quantization of the Gowdy model~\cite{kinjal2}. The methods and steps used here closely follow those used in LQG and
are to be viewed as f\/irst steps towards constructing a quantum theory of the Gowdy model where all the gravitational degrees of freedom are loop
quantized.

\subsubsection{Basic states}

Since this is a one-dimensional theory, the graphs are just $n$ arcs with $n$ vertices. The conf\/iguration variable ${\mathcal A}$ is a $U(1)$
connection 1-form, so we integrate it along an edge (an arc along~$S^1$) and by taking its exponential we def\/ine the (edge) holonomy variable
valued in $U(1)$:
\begin{equation*}
h^{(k)}_e({\mathcal A}) := \exp\left(i \frac{k}{2} \int_e {\mathcal A} \right) ,\qquad k \in \mathbb{Z} .
\end{equation*}
The conf\/iguration variables $X, Y \in \mathbb{R}$ and $\eta \in \mathbb{R}/\mathbb{Z}$ are scalars and hence no smearing is
needed. For these we def\/ine the point holonomies (at points $v$)
\begin{gather*}
h_v^{(\mu)}(X)  :=  \exp\left[i\frac{\mu}{2} X(v)\right] , \qquad
h_v^{(\nu)}(Y)  :=  \exp\left[i\frac{\nu}{2} Y(v)\right] , \qquad
h_v^{\lambda}(\eta)  :=  \exp\left[i\lambda \eta(v)\right] ,
\end{gather*}
where $\mu,\nu \in \mathbb{R}$ and $\lambda \in \mathbb{Z}$. The $X$, $Y$ point holonomies are interpreted as unitary representations of the
compact Abelian group $\mathbb{R}_{\mathrm{Bohr}}$, which is the Bohr compactif\/ication of the additive group of real numbers $\mathbb{R}$.

The kinematical Hilbert space is thus a tensor product of the Hilbert spaces constructed for~${\mathcal A}$,~$X$, $Y$, $\eta$ variables.
For ${\mathcal A}$, the Hilbert space can be constructed using $U(1)$ holonomies in a procedure similar to full LQG.
For $X$, $Y$ and~$\eta$, we can use the point holonomies as in minisuperspace LQC, where the quantum conf\/iguration space is taken to be the
Bohr compacti\-f\/i\-cation $\mathbb{R}_{\mathrm{Bohr}}$. By contrast, $\eta$ is an angle variable, so the
corresponding point holonomy is valued in~$U(1)$.

An orthonormal basis on the tensor-product Hilbert space is provided by the ``charge network functions''.
They are labelled by a close, oriented graph $\Gamma$ with $n$ edges $e$ and $n$ vertices~$v$, a~$U(1)$~representation~$k_e$
for each edge, a~$U(1)$~representation $\lambda_v \in \mathbb{Z}$ for each vertex, and $\mathbb{R}_{\mathrm{Bohr}}$
representations $\mu_v$, $\nu_v$ for each vertex:
\begin{gather}
T_{\Gamma, \vec k, \vec \mu, \vec \nu, \vec \lambda}({\mathcal A}, X, Y, \eta) := \prod_{e\in \Gamma} k_e[h^{(e)}]~
\prod_{v \in {V}(\Gamma)} \mu_v [h_v(X)] \nu_v [h_v(Y)] \lambda_v [h_v(\eta)] \label{BasisStates} \\
\hphantom{T_{\Gamma, \vec k, \vec \mu, \vec \nu, \vec \lambda}({\mathcal A}, X, Y, \eta)}{}
=  \prod_{e\in \Gamma} \exp \left(i \frac{k_e}{2} \int_e {\mathcal A} \right) \prod_{v \in {V}(\Gamma)}
 \left[\exp \left(i \frac{\mu_v}{2} X \right) \exp \left(i \frac{\nu_v}{2} Y \right) \exp \left( i \lambda_v
\eta \right) \right] ,\nonumber 
\end{gather}
where $V(\Gamma)$ represents the set of vertices belonging to the graph~$\Gamma$. Functions where any of the labels are dif\/ferent are
orthogonal~-- in particular, two graphs must coincide for non-zero inner product. These basis states provide an orthogonal decomposition
for the kinematical Hilbert space when all the representation labels are non-zero.

Note that, unlike in the full theory, in this model we have both point and edge holonomies. This construction is also signif\/icantly dif\/ferent
from the hybrid quantization of the previous section. There, the loop Hilbert space has only the homogeneous part represented by point holonomies
similar to LQC, while the inhomogeneous part is Fock quantized. In that case, the full Hilbert space is a tensor product of the two.

\subsubsection{Flux operators}

The conjugate variables are represented as
\begin{equation*}
E^\sigma (\theta) \sim -i \gamma l_{\rm Pl}^2 \frac{\delta h_{\theta}(X)}{\delta
X(\theta)}\frac{\partial}{\partial h_{\theta}(X)},
\end{equation*}
where $l_{\rm Pl}^2 := \kappa'\hbar$.

The f\/lux variables corresponding to $E^\sigma$, $E^\delta$, $P^{\eta}$ are def\/ined by integrating these
densities on an interval
${\mathcal I}$ of the circle, eg ${\mathcal F}_{\sigma,{\mathcal I}} := \int_{\mathcal I} E^\sigma,
{\mathcal F}_{\delta,{\mathcal I}} := \int_{\mathcal I} E^\delta$.
${\mathcal E}$, being a scalar, is already a suitable variable.  Their actions on the basis functions~(\ref{BasisStates}) are
\begin{gather*}
\hat{\mathcal E}(\theta) T_{\Gamma, \vec k, \vec \mu, \vec \nu, \vec \lambda}  =   \frac{\gamma l_{\rm Pl}^2}{2}
 \frac{k_{e^+(\theta)} + k_{e^-(\theta)}}{2} T_{\Gamma, \vec k, \vec \mu, \vec \nu, \vec \lambda}, \\
\int_I \hat{E}^\sigma T_{\Gamma, \vec k, \vec \mu, \vec \nu, \vec \lambda}   =   \frac{\gamma l_{\rm Pl}^2}{2} \sum_{v \in V(\Gamma) \cap
{\mathcal I}} \mu_v T_{\Gamma, \vec k, \vec \mu, \vec \nu, \vec \lambda}, \\
\int_I \hat{E}^\delta T_{\Gamma, \vec k, \vec \mu, \vec \nu, \vec \lambda}  =  \frac{\gamma l_{\rm Pl}^2}{2} \sum_{v \in V(\Gamma) \cap
{\mathcal I}} \nu_v T_{\Gamma, \vec k, \vec \mu, \vec \nu, \vec \lambda}, \\
\int_I \hat{P}^{\eta} T_{\Gamma, \vec k, \vec \mu, \vec \nu, \vec \lambda}  =
\gamma l_{\rm Pl}^2 \sum_{v \in V(\Gamma) \cap {\mathcal I}} \lambda_v T_{\Gamma, \vec k, \vec \mu, \vec \nu, \vec \lambda} ,
\end{gather*}
where ${\mathcal I}$ is an interval on $S^1$. The symbols $e^{\pm}(\theta)$ either refer to the two oriented edges of the graph $\Gamma$, meeting
at $\theta$ if there is a vertex at $\theta$, or they denote two parts of the same edge if there is no vertex at $\theta$. In such a case,
the $k$ labels are the same. In case a vertex is an end-point of the interval, there is an additional factor of $1/2$ for its contribution to the sum\footnote{This follows from
\begin{gather*}
\int_a^b dx\, \delta(x - x_0)   =   \begin{cases}
 1 &    {\rm if} \  x_0   \in   (a, b); \\
 \frac{1}{2} & {\rm if} \ x_0 =  a  \ \mathrm{or}  \  x_0  =  b;
 \\
 0 &  {\rm if} \  x_0  \notin  [a, b].
 \end{cases}
\end{gather*}}.
Note that classically the triad components $E^\sigma$ and $E^\delta$ are positive. Fluxes, however, can take both signs since they
involve integrals which depend on the orientation. We have thus constructed the kinematical Hilbert space together with the representation of the
basic variables. Next, we turn to the construction of composite operators.

\subsubsection{Construction of more general operators}

The dif\/feomorphism covariance requires that all operators of interest are integrals of expressions in terms of the basic operators.
They also involve products of elementary operators at the same point (same $\theta$) and thus need a ``regularization''. As in LQG, the
general strategy to def\/ine such operators is:
\begin{enumerate}\itemsep=0pt
\item replace the integral by a Riemann sum using a ``cell-decomposition'' (or partition) of $S^1$;
\item for each cell, def\/ine a regulated expression choosing suitable ordering of the basic operators, and evaluate the
action on basis states;
\item check ``cylindrical consistency'' of this action so that the (regulated) operator can be densely def\/ined on
the kinematical Hilbert space via projective limit;
\item f\/inally, one would like to remove the regulator.
\end{enumerate}

Since our model is one-dimensional, both the cell-decomposition and the graphs underlying the basis states are characterized by f\/initely many
points and the arcs connecting the consecutive points.  Adapting the techniques used in LQG~\cite{lqg3, lqg1}, the products of elementary variables are
regulated by using a point splitting and then expressing the f\/ields in terms of the appropriate holonomies and f\/luxes.

A regulator, for each given graph $\Gamma$, then consists of a family of partitions
$\Pi^{\Gamma}_{\epsilon}$ such that, for each $\epsilon$, each vertex of $\Gamma$ is contained in exactly one cell. There is also a
choice of representation labels $k_0$, $\mu_0$, $\nu_0$, $\lambda_0 $ which can be taken to be the same for all $\epsilon$. Since each $\Pi^\Gamma$ can
also be thought of as being def\/ined by a set of points such that each vertex is f\/lanked by two points, any orientation-preserving dif\/feomorphism
will automatically preserve the order of the vertices and cell boundaries. Every suf\/f\/iciently ref\/ined partition then automatically becomes a~dif\/feomorphism-covariant regulator. We shall assume that the parameter $\epsilon$ denoting a family of partitions are suf\/f\/iciently ref\/ined and also
plays the role of a dif\/feo-covariant regulator. The regulated expressions depend on $\epsilon$ and we recover the classical expressions as
$\epsilon \to 0$.

As in LQG, the issue of cylindrical consistency is automatically sorted out by referring to the orthogonal decomposition of
${\mathcal H}_{\mathrm{kin}}$, i.e., by specifying the action of the operators on basis states with all representation labels
being non-zero. A few comments about the subsequent construction:
\begin{enumerate}\itemsep=0pt
\item We have assumed the ``length of the intervals'' to be same and equal to $\epsilon$. This corresponds to a ``cubic'' partition and
is chosen for convenience only.
\item The charges $\mu_v$, $\nu_v$ can take both signs depending on the orientation of the interval. However, the eigenvalues of the volume operator
must have explicit absolute values.
\item ${\mathcal I}_i$ denotes the $i$-th cell of the partition. For a given graph, the partition is so chosen that each vertex is included in one
and only one interval ${\mathcal I}_i$. The intervals which do not contain any vertex of the graph, do not contribute to the summation
owing to the property of f\/lux operators. Hence, the sum collapses to contributions only from the vertices, independent of
the partition. The action is manifestly independent of $\epsilon$ and even though the number of intervals go to inf\/inity as
$\epsilon \to 0$, the action remains f\/inite and well def\/ined.
\item Because of this property of the f\/luxes, we can choose the $\bar \theta_i$ point in a cell to coincide with a vertex of a graph if
${\mathcal I}_i$ contains a vertex, or with an arbitrary point if ${\mathcal I}_i$ does not contain a vertex.
\item The measure of the integrals in this section is sometimes suppressed for brevity of notation but can be clearly understood from the context.
\end{enumerate}

 {\bf Volume operator.}
In the classical expression for the Hamiltonian constraint, powers of $E := |{\mathcal E}|E^\sigma E^\delta$ occur in the same manner
as in the full theory. It is therefore natural to consider the expression for the volume of a region ${\mathcal I}\times T^2$
and construct the corresponding operator. The classical volume operator written in terms of basic variables is
\begin{gather}
\mathcal{V} \big(\mathcal{I} \times T^2\big) = \int_{\mathcal{I} \times T^2} d^3 x \sqrt {g}
= 4 \pi ^2 \int_\mathcal{I} d\theta \sqrt { | {\mathcal E} |  E^\sigma
E^\delta}. \label{volumeexpression}
\end{gather}
To obtain the quantum operator, we f\/irst rewrite equation~\eqref{volumeexpression} as a Riemann sum of volume of the cells, which we denote as
\begin{gather*}
\mathcal{V}_{\epsilon} (\mathcal{I})   \approx    \sum_{i = 1}^n \int_{\theta_i}^{\theta_{i} + \epsilon} d\theta
\sqrt{|\mathcal{E}|E^\sigma E^\delta} .
\end{gather*}
This has to be written in terms of the f\/lux variables:
\begin{gather*}
\mathcal{V}_{\epsilon} (\mathcal{I})  \approx   \sum_{i = 1}^n \epsilon \sqrt{|\mathcal{E}(\bar \theta_i)|}
\sqrt{E^\sigma(\bar \theta_i)} \sqrt{E^\delta(\bar \theta_i)} \\
\hphantom{\mathcal{V}_{\epsilon} (\mathcal{I})}{}
 \approx    \sum_{i = 1}^n \sqrt{|\mathcal{E}|} \sqrt{\epsilon E^\sigma} \sqrt{\epsilon E^\delta}
 \approx   \sum_{i = 1}^n \sqrt{|\mathcal{E}|(\bar \theta_i)}
\sqrt{\left|\int_{\theta_i}^{\theta_i + \epsilon} d\theta E^\sigma\right|}
\sqrt{\left|\int_{\theta_i}^{\theta_i + \epsilon} d\theta E^\delta\right|} .
\end{gather*}
The right-hand side is now expressed in terms of f\/lux variables. The regulated volume operator can be def\/ined as:
\begin{gather*}
\hat{\mathcal{V}_{\epsilon}} (\mathcal{I}) :=  \sum_{i = 1}^n \sqrt{\hat {|\mathcal{E}|}(\bar \theta_i)} \sqrt{\widehat
{\left|\int_{ {\mathcal I}_i} E^\sigma\right|} } \sqrt{\widehat {\left|\int_{ {\mathcal I}_i}
E^\delta\right|}}. 
\end{gather*}
Clearly, this is diagonal in the basis states and its action on a basis state
$T_{\Gamma, \vec{k}, \vec{\mu}, \vec{\nu}, \vec{\lambda} }$ yields the eigenvalue
\begin{equation}
V_{ \vec k, \vec \mu, \vec \nu, \vec \lambda} =  \frac{1}{\sqrt{2}} \left(\frac{\gamma l_\Pl^2}{2}\right)^{3/2}
\sum_{v \in {\mathcal I} \cap V(\Gamma)}\bigg[|\mu_v|\ |\nu_v|\ | k_{e^{+}(v)} + k_{e^{-}(v)}|\bigg]^{\frac{1}{2}}.
\label{volumeeigenvalues}
\end{equation}
Thanks to our choice of basic variables, the eigenstates of the f\/lux operators are also volume eigenstates.

 {\bf Gauss constraint.}
Consider the Gauss constraint (\ref{ch4gauss}):
\begin{gather*}
G_3 = \int_{S^1} d\theta (\partial_{\theta} {\mathcal E} +  P^{\eta})
\approx \sum_{i = 1}^n \left[\int_{{\mathcal I}_i} P^{\eta} + {\mathcal E}(\theta_i + \epsilon) - \mathcal{E}(\theta_i)\right], \\
\hat {G_3^{\epsilon}}  :=   \sum_{i = 1}^n \left[\widehat{\int_{ {\mathcal I}_i } P^{\eta}}  + \hat {\mathcal E}(\theta_i +
\epsilon) - \hat{\mathcal E}(\theta_i)\right].
\end{gather*}
Again, this is easily quantized with its action on a basis state $T_{\Gamma, \vec{k}, \vec{\mu}, \vec{\nu}, \vec \lambda}$, giving the eigenvalue
\begin{equation*}
{\gamma l_{\rm Pl}^2}\sum_{v \in V(\Gamma)} \left[ \lambda_v + \frac{k_{e^{+}}(v) - k_{e^{-}}(v)}{2}\right].
\end{equation*}
Notice that in the limit of inf\/initely f\/ine partitions, for a given graph, if there is a vertex $v \in {\mathcal I}_i$, then
there is {\em no vertex} in the adjacent cells. As a result, ${\mathcal E}(\theta_{i + 1})$ gives $k_{e^+}(v)/2$ and
$- {\mathcal E}(\theta_i)$ gives $- k_{e^-}(v)/2$, since $\theta_i$ divides the same edge and so does $\theta_{i + 1}$.

Once again, the eigenvalues are manifestly independent of $\epsilon$ and the action is dif\/feo-invariant. Imposition of the Gauss constraint can be done
simply by restricting to basis states with labels satisfying $\lambda_v = - (k_{e^+(v)} - k_{e^-(v)})/2$, $\forall\, v \in V(\Gamma)$.
Since $\lambda_v \in \mathbb{Z}$, the dif\/ference in the $k$ labels at each vertex must be an {\em even} integer. We will assume these restrictions on the
representation labels and from now on deal with gauge-invariant basis states. Explicitly,
\begin{gather*}
T_{\Gamma,\vec k,\vec \mu,\vec \nu}= \prod_{e\in \Gamma} \exp \left\{ i \frac{k_e}{2} \int_e \left[ {\mathcal A} (\theta) -
\partial_{\theta} \eta \right] \right\}   \prod_{v \in {V}(\Gamma)}\left\{ \exp \left[i \frac{\mu_v}{2} X(v) \right]
\exp \left[ i \frac{\nu_v}{2} Y(v) \right]  \right\}.
\end{gather*}
We have also used $\eta(v^+(e)) - \eta(v^-(e)) = \int_e \partial_{\theta} \eta$,  where $v^{\pm}(e)$ denote the tip and tail of the edge~$e$.

\subsubsection{Hamiltonian constraint}

 {\bf Preliminaries.}
The Hamiltonian constraint is much more complicated. After some manipulation, we can write~(\ref{ch4hamiltonian})
as a sum of a kinetic and a potential term,
\begin{gather*}
H    :=   - \frac{1}{\kappa'}[ H_K + H_P ], \\
H_K   :=   \frac{1}{\gamma^2} \int_{S^1} {d} \theta N (\theta) \frac{1}{\sqrt{E}} \big[  X
E^\sigma Y E^\delta + \left({\mathcal A} +
\partial_{\theta}\eta\right) {\mathcal E}\big( X E^\sigma  + Y E^\delta\big)  \big], \\ 
H_P  :=   -\int_{S^1} {d} \theta N (\theta) \frac{1}{\sqrt{E}} \left[ -\frac{1}{4} \left( \partial_\theta {\mathcal E}
\right)^2 + \frac{({\mathcal E})^2}{4} \left(\frac{\partial_\theta E^\sigma}{ E^\sigma} -
\frac{\partial_\theta E^\delta}{ E^\delta}\right)^2 \right]
 \\
\hphantom{H_P  :=}{}  -\int_{S^1} {d} \theta N (\theta) \frac{1}{2}\partial_\theta\left[\frac{2 {\mathcal E} \left(\partial_\theta {\mathcal E}
\right) }{\sqrt{E}}\right] .
\end{gather*}
Here it is more convenient to use $SU(2)$-valued holonomies using the $\eta$-dependent $\tau$ matrices def\/ined equation~(\ref{taumatrix}):
\begin{gather}
h_\theta (\mathcal{I})   :=   \exp \left[\tau_3 \int_{\mathcal{I}} d \theta' {\mathcal A} (\theta') \right]  =
\cos\left(\frac{1}{2}\int_{\mathcal{I}} \mathcal{A}\right) + 2 \tau_3   \sin\left(\frac{1}{2}\int_{\mathcal{I}}
\mathcal{A}\right), \nonumber \\
h_\sigma(\theta)  :=  \exp \left[ \mu_0 X(\theta) \tau_\sigma (\theta) \right]  =
\cos\left[\frac{\mu_0}{2} X(\theta)\right] + 2\tau_\sigma(\theta) \sin\left[\frac{\mu_0}{2}
X(\theta)\right], \label{holonomies} \\
h_\delta(\theta)  :=  \exp \left[ \nu_0 Y(\theta)  \tau_\delta (\theta)  \right]  =
\cos\left[\frac{\nu_0}{2} Y(\theta)\right] + 2\tau_\delta (\theta)  \sin\left[\frac{\nu_0}{2}
Y(\theta)\right]. \nonumber
\end{gather}
Each of the $SU(2)$-valued holonomies, as well as the sine and cosine, are well def\/ined on the kinematical Hilbert space. The
interval~$\mathcal{I}$ will typically be a cell of a partition, $(\theta_i, \theta_i + \epsilon)$. The parameters~$\mu_0$,~$\nu_0$ are the chosen and f\/ixed representations of $\mathbb{R}_{\mathrm{Bohr}}$, $k_0 = 1$ is the f\/ixed
representation of the~$U(1)$. As before the parameter $\epsilon$ which denotes a family of partitions, also plays the role of a dif\/feo-covariant
regulator. We brief\/ly describe the steps in obtaining a~well-def\/ined quantum operator; the details can be found in \cite{kinjal2}.

Consider an expression of the form Tr($h_ih_jh_i^{-1}h_j^{-1}h_k\{h_k^{-1},\sqrt{E}\}$), for distinct $i$, $j$, $k$ taking
values $\theta$, $\sigma$, $\delta$. For small values of $X$, $Y$, $\int_{\mathcal{I}}\mathcal{A}$, we can make the following approximations:
\begin{gather*}
h_\sigma(\theta)\big\{h_\sigma(\theta)^{-1}, V(\mathcal{I})\big\}  =
-\frac{\kappa'\gamma}{2}\mu_0\tau_\sigma \frac{ \mathcal{E}(\theta)
\int_{\mathcal{I}}E^\delta }{V(\mathcal{I})}  \approx -\frac{\kappa'\gamma}{2}\mu_0\tau_\sigma
\frac{E^\delta(\theta)
\mathcal{E}(\theta)}{\sqrt{E(\theta)}}, \nonumber \\
h_\delta(\theta)\big\{h_\delta(\theta)^{-1}, V(\mathcal{I})\big\}  =
-\frac{\kappa'\gamma}{2}\nu_0\tau_\delta \frac{ \mathcal{E}(\theta)
\int_{\mathcal{I}}E^\sigma }{V(\mathcal{I})}  \approx  -\frac{\kappa'\gamma}{2}\nu_0\tau_\delta
\frac{E^\sigma(\theta)
\mathcal{E}(\theta)}{\sqrt{E(\theta)}}, \nonumber \\
h_{\theta}\big\{h_{\theta}^{-1}, V(\mathcal{I})\big\}  = -\frac{\kappa'\gamma}{2}\tau_3
\frac{\int_{\mathcal{I}}E^\sigma
\int_{\mathcal{I}}E^\delta}{V(\mathcal{I})}  \approx -\frac{\kappa'\gamma}{2}\epsilon\tau_3
\frac{E^\sigma(\theta) E^\delta(\theta)}
{\sqrt{E(\theta)}}, \\ 
\int_{\mathcal{I}}\mathcal{A}  \approx  \epsilon \mathcal{A}(\theta) , \qquad
\int_{\mathcal{I}}E^\sigma ~ \approx ~ \epsilon E^\sigma(\theta)   ,\qquad
\int_{\mathcal{I}}E^\delta ~ \approx ~ \epsilon E^\delta(\theta) ,
\end{gather*}
where  $V(\mathcal{I})$ is the volume of the interval $\mathcal{I}$.

Then, the holonomies can be expanded in a power series. Because of the trace, it is enough to expand each holonomy up to f\/irst order. The surviving
terms are quadratic terms arising from products of the linear ones and a linear term coming from~$h_k$. If one interchanges the
$i \leftrightarrow j $ holonomies, the linear term retains the sign while the quadratic one changes the
sign. Thus, taking the dif\/ference of the two traces leaves us only with the quadratic terms, which are exactly of the form
needed in~$H_K$.
There are derivatives of $\eta$ which arise from the position dependence of the $\tau_\sigma$,
$\tau_\delta$ matrices:
\begin{gather*}
\tau_\sigma(\theta + \epsilon) - \tau_\sigma(\theta)   \approx
\epsilon\partial_{\theta}\tau_\sigma  =
\epsilon\partial_{\theta}\eta   \tau_\delta(\theta) , \nonumber \\
\tau_\delta(\theta + \epsilon) - \tau_\delta(\theta)   \approx
\epsilon\partial_{\theta}\tau_\delta
  =   - \epsilon\partial_{\theta}\eta   \tau_\sigma(\theta) . 
\end{gather*}

In the quantization of the $H_P$, we also need to use the following identities repeatedly (in the form LHS/RHS = 1):
\begin{gather}
\mathcal{Z}(\mathcal{I})   :=   \epsilon^{abc}\mathrm{Tr}\left[  h_a\{h_a^{-1}, V(\mathcal{I})\}
h_b\{h_b^{-1}, V(\mathcal{I})\}  h_c\{h_c^{-1}, V(\mathcal{I})\}  \right] \nonumber \\
\hphantom{\mathcal{Z}(\mathcal{I})}{}
 =   \frac{3}{2} \left(\frac{\kappa'\gamma}{2}\right)^3 \mu_0\nu_0 V(\mathcal{I}) , \label{Identity1} \\
\mathcal{Z}_{\alpha}(\mathcal{I})   :=  \epsilon^{abc}\mathrm{Tr}\left[  h_a\{h_a^{-1},({V(\mathcal{I}))^{\alpha}}\}
h_b\{h_b^{-1},({V(\mathcal{I}))^{\alpha}}\}  h_c\{h_c^{-1},({V(\mathcal{I}))^{\alpha}}\}  \right] \nonumber \\
\hphantom{\mathcal{Z}_{\alpha}(\mathcal{I})}{}
 =   \frac{3}{2} \left(\frac{\kappa'\gamma}{2}\right)^3 \mu_0\nu_0   \alpha^3 [V(\mathcal{I})]^{3 \alpha - 2}
 =  \alpha^3 [V(\mathcal{I})]^{3(\alpha - 1)} \mathcal{Z}(\mathcal{I}). \label{Identity2}
\end{gather}
These are essentially versions of the identity $ 1 = (|\det(e_a^i)|/\sqrt{E})^n$ \cite{QSD5}.

It is also convenient to def\/ine the following families of operators:
\begin{gather*}
\hat {\mathcal O}_{\alpha}^\sigma({\mathcal I},\theta)   :=   \left\{ \cos \left[ \frac{1}{2} \mu_0 X(\theta)\right] \hat V^{\alpha}
({\mathcal I}) \sin \left[\frac{1}{2} \mu_0 X(\theta)\right] \right. \nonumber \\
\left. \hphantom{\hat {\mathcal O}_{\alpha}^\sigma({\mathcal I},\theta)   := }{}
-\sin\left [\frac{1}{2} \mu_0 X(\theta) \right] \hat V^{\alpha}({\mathcal I}) \cos \left[
\frac{1}{2} \mu_0 X(\theta) \right] \right\}, \nonumber \\
\hat {\mathcal O}_{\alpha}^\delta({\mathcal I},\theta)  :=  \left\{ \cos \left[ \frac{1}{2} \mu_0 Y(\theta)\right] \hat V^{\alpha}
({\mathcal I}) \sin \left [\frac{1}{2} \mu_0 Y(\theta)\right] \right. \nonumber \\
\left. \hphantom{\hat {\mathcal O}_{\alpha}^\delta({\mathcal I},\theta)  :=}{}
-\sin\left [\frac{1}{2} \mu_0 Y(\theta) \right] \hat V^{\alpha}({\mathcal I}) \cos \left[ \frac{1}{2} \mu_0
Y(\theta) \right] \right\} , \nonumber \\
\hat {\mathcal O}_{\alpha}^{\theta}({\mathcal I},\theta)  :=  \left[ \cos \left( \frac{1}{2} \int_{ {\mathcal I} }{\mathcal A} \right)
\hat V^{\alpha}({\mathcal I}) \sin \left (\frac{1}{2} \int_{ {\mathcal I} }{\mathcal A} \right)
 -\sin\left (\frac{1}{2} \int_{ {\mathcal I} }{\mathcal A} \right) \hat V^{\alpha}({\mathcal I}) \cos
\left( \frac{1}{2} \int_{ {\mathcal I}}{\mathcal A} \right) \right]  .
\end{gather*}
Above, $\theta$ is a point in the interval ${\mathcal I}$ and $\alpha > 0$ is the power of the volume operator. Again,
for simplicity of notation we will suppress the $\theta$ labels in the above operators.

The operators ${\mathcal O}_{\alpha}^{a} := [\cos(\cdots) \hat V^{\alpha} \sin(\cdots) - \sin(\cdots) \hat V^{\alpha}
\cos(\cdots)]$, $a =\theta,\sigma,\delta $ appear in all the terms and are functions of both
holonomies and f\/luxes. To see that
this is actually diagonal in the charge network basis, write the cos and sin operators as sums and dif\/ferences of the
exponentials (i.e., holonomies).  It then follows that
\begin{gather}
\cos(\cdots) \hat V^{\alpha} \sin(\cdots) - \sin(\cdots) \hat V^{\alpha}
\cos(\cdots)   =   \frac{1}{2i}\big[ e^{-i(\cdots)} \hat V e^{+i(\cdots)} - e^{+i(\cdots)} \hat V e^{-i(\cdots)} \big].
  \label{diagonalO}
\end{gather}
It is now obvious that the operators are diagonal and thus commute with all the f\/lux operators.
Finally, the operator form of ${\mathcal Z}_{\alpha}({\mathcal I})$ can be obtained as
\begin{gather*}
\hat{\mathcal Z}_{\alpha}({\mathcal I})   :=    \epsilon^{abc} \,\mbox{Tr}
\big\{ \hat{h}_a [ \hat{h}_a^{-1} , {\hat{V}(\mathcal I)}^{\alpha} ]
\hat{h}_b [\ \hat{h}_b^{-1} , {\hat{V}(\mathcal I)}^{\alpha}  ]
\hat{h}_c [\ \hat{h}_c^{-1} , {\hat{V}(\mathcal I)}^{\alpha}  ]  \big\}  \\
\hphantom{\hat{\mathcal Z}_{\alpha}({\mathcal I})}{}
 =   -12  \hat {\mathcal O}_{\alpha}^\sigma({\mathcal I}) \hat {\mathcal O}_{\alpha}^\delta({\mathcal I})
\hat {\mathcal O}_{\alpha}^{\theta}({\mathcal I}).
\end{gather*}

Having noted the ingredients common to the quantization of the dif\/ferent pieces of the Hamiltonian constraint, we turn to
each one in some detail.

{\bf Quantization of $\boldsymbol{H_K}$.}
Choosing a partition of $S^1$ with a suf\/f\/iciently large number of $n$ points at
$\theta_i$, $i = 1, \dots, n$, $\theta_n = 2\pi$, $\epsilon = \theta_{i + 1} - \theta_i $, we write the integral as a sum,
\begin{gather*}
H_K   \approx    \frac{1}{\gamma^2} \sum_{i = 1}^n  \epsilon N (\bar \theta_i) \frac{1}{\sqrt{E}(\bar \theta_i)}
\left[  X E^\sigma Y E^\delta +
\left({\mathcal A} + \partial_{\theta}\eta\right) {\mathcal E}( X E^\sigma  + Y E^\delta)
\right](\bar \theta_i)  \nonumber\\
\phantom{H_K }{}   =
\frac{1}{\gamma^2} \sum_{i = 1}^n N (\bar \theta_i) \frac{1}{V(\mathcal{I}_i)} \left\{  X(\bar \theta_i)
\left(\int_{\mathcal{I}_i}E^\sigma\right) Y(\bar \theta_i) \left(
\int_{\mathcal{I}_i}E^\delta\right)  \right. \\
\left. \phantom{H_K   \approx }{} +\left(\int_{\mathcal{I}_i}{\mathcal A} + \partial_{\theta}\eta\right) {\mathcal E}(\bar \theta_i)
\left[ X(\bar \theta_i) \int_{\mathcal{I}_i}E^\sigma  + Y(\bar \theta_i)
\int_{\mathcal{I}_i}E^\delta \right] \right\} . 
\end{gather*}

For small values of the extrinsic curvature components ($ \sim X, Y$, classical regime) and suf\/f\/iciently ref\/ined partition
($\epsilon \ll 1$, continuum limit), the $i$-th term in the sum can be written in terms of the traces
of the $SU(2)$-valued holonomies. The expression in terms of holonomies and f\/luxes goes over to the
classical expression in the classical regime. It can be promoted to an operator by putting hats on the holonomies and
f\/luxes and replacing Poisson brackets by $(i\hbar)^{-1}$ times the commutators.  Here, the standard choice of putting the holonomies on the left
is made. Then, we use the expressions for the holonomies in terms of the trigonometric operators given in equation~(\ref{holonomies}),
and evaluating the traces we get the quantum operator as
\begin{gather*}
\widehat H_K^{\mathrm{reg}}   =
- i\frac{4}{l_{\rm Pl}^2\gamma^3}\frac{1}{\mu_0\nu_0} \sum_{i = 1}^n N (\bar \theta_i) \Bigg( \left\{  \sin\left[\mu_0 X(\bar \theta_i)
\right]\sin \left[\nu_0 Y(\bar \theta_i) \right]  \right\} \times {\mathcal O}_1^{\theta}({\mathcal I}_i)\\
\phantom{\widehat H_K^{\mathrm{reg}}   = }{}
+\left\{2 \sin \left [\frac{1}{2} \nu_0 Y(\bar \theta_i + \epsilon) \right]\cos \left[\frac{1}{2} \nu_0
Y(\bar \theta_{i})\right]\sin \left( \int_{\mathcal{I}_i} {\mathcal A} -\Delta_i\right)\right\}\times
{\mathcal O}_1^{x}({\mathcal I}_i)\\ 
\phantom{\widehat H_K^{\mathrm{reg}}   = }{}
 +\left\{2 \sin \left [\frac{1}{2} \mu_0 X (\bar \theta_i + \epsilon) \right]\cos \left[\frac{1}{2} \mu_0
X(\bar \theta_{i})\right]\sin \left( \int_{\mathcal{I}_i} {\mathcal A} -\Delta_i\right)\right\} \times
{\mathcal O}_1^{y}({\mathcal I}_i) \Bigg) , 
\end{gather*}
where $\Delta_i := \eta(\bar \theta_i) - \eta(\bar \theta_i + \epsilon)$ is outside the integrals.

  {\bf Quantization of $\boldsymbol{H_P}$.}
All the three terms of~$H_P$ are functions only of the momenta, but there are a couple of obstacles in a straightforward transcription of~$H_P$.
These have to be expressed in terms of basic variables, i.e., in terms of f\/luxes and holonomies. Also, the power(s) of momenta in the denominators
will make the action on some states singular. The f\/irst part is easy to take care of thanks to the density weight~1. For the second part, we use the
identities~(\ref{Identity1}) and~(\ref{Identity2})\footnote{In this one-dimensional model, this procedure is equivalent to the
point-splitting procedure of \cite{QSD5}.}.

The common strategy followed for these terms is:
\begin{enumerate}\itemsep=0pt
\item introduce a suf\/f\/iciently large number $k > 0$ of positive powers of
\[
1 = 16[ 3 (\kappa'\gamma)^3 \mu_0\nu_0]^{-1}\mathcal{Z}(\mathcal{I})/V(\mathcal{I}),
\]
and express $\mathcal{Z}$ in terms of $\mathcal{Z}_{\alpha}$. This introduces further powers of the volume;
\item choose $\alpha(k)$ such that explicit multiplicative factors of the volume become 1 and further choose $k$.
\end{enumerate}

Now the expression can be promoted to an operator.

 {\underline{\it First term of $H_P$}.}
We f\/irst rewrite this in terms of the basic variables:
\begin{gather*}
-\int_{S^1} d \theta N (\theta) \frac{1}{\sqrt{E(\theta)}} \left[ -\frac{1}{4} \left( \partial_\theta {\mathcal E} \right)^2 \right]
  \approx    \frac{1}{4} \sum_{i = 1}^n N(\bar \theta_i) \frac{ \left[ \mathcal{E}(\bar \theta_i + \epsilon) - \mathcal{E}(\bar \theta_i)
\right]^2}{\sqrt{\mathcal{E}(\bar \theta_i)   \int_{ {\cal I}_i} E^\sigma   \int_{ {\cal I}_i}
E^\delta }} .
\end{gather*}
We now follow the strategy mentioned for the terms in the denominator. After some manipulation, the right-hand side becomes
\begin{gather*} 
 \left. \frac{1}{4} \left[\frac{16}{3 (\kappa'\gamma)^3\mu_0\nu_0 \alpha^3}\right]^k \sum_{i = 1}^n N(\bar \theta_i)
\left[\mathcal{E}(\bar \theta_i + \epsilon) - \mathcal{E}(\bar \theta_i)\right]^2 \left[\mathcal{Z}_{\alpha}
(\mathcal{I}_i)\right]^k \right|_{\alpha := \frac{2}{3} - \frac{1}{3 k}}  .
\end{gather*}
The choice $\alpha := 2/3 - 1/(3 k)$ removes explicit factors of the volume.
The choice of $k > 0$ is limited by $\alpha > 0$ (being it a power of the volume appearing in $\mathcal{Z}_{\alpha}$). Some
convenient choices would be $k = 1$ $(\alpha = 1/3)$, $k = 2$ $(\alpha = 1/2)$, and so on. For all of them, the
above expression can be promoted to a well-def\/ined operator.

 {\underline{\it Second term of $H_P$}.}
To begin with, one observes that $E^\delta/E^\sigma$ is a scalar and
$\partial_{\theta}\ln(E^\delta/E^\sigma)$ is a scalar density. This term is then
manipulated as:
\begin{gather*}
- \frac{1}{4} \int_{S^1} N (\theta) \frac{[{\mathcal E}(\theta)]^2}{\sqrt{E(\theta)}} \left(
\frac{\partial_\theta E^\sigma}{E^\sigma} -
\frac{\partial_{\theta} E^\delta}{E^\delta} \right)^2
 =  - \frac{1}{4} \int_{S^1} N (\theta) \frac{[{\mathcal E}(\theta)]^2}{\sqrt{E(\theta)}} \left(\partial_\theta \ln
\frac{E^\delta}{E^\sigma}\right)^2.
\end{gather*}
We write the right-hand side in terms of f\/lux variables as
\begin{gather*} 
\mathrm{RHS}  =  - \frac{1}{4} \sum_{i = 1}^n N(\bar \theta_i) \frac{[{\mathcal E}(\bar \theta_i)]^2}{V(\mathcal{I}_i)} \left[ \frac{\int_{
{\mathcal I}_i} E^\sigma}{\int_{ {\mathcal I}_i} E^\delta}\ \left( \frac{\int_{ {\mathcal I}_{i +
1} } E^\delta}{\int_{ {\mathcal
I}_{i + 1} } E^\sigma}  - \frac{\int_{ {\mathcal I}_{i} } E^\delta}{\int_{ {\mathcal I}_{i} }
E^\sigma} \right) \right]^2.
\end{gather*}
Now we have the f\/luxes in the denominator which can be def\/ined exactly as the inverse triad operators of LQC.
Denoting the f\/luxes as ${\mathcal F}_{\sigma,{\mathcal I}} := \int_{\mathcal I}E^\sigma, {\mathcal
F}_{\delta,{\mathcal I}} := \int_{\mathcal I} E^\delta$,
\begin{gather*}
{\mathcal F}^{-1}_{\sigma,{\mathcal I}}   =   \left(\frac{1}{\kappa'\gamma l} \right)^{\frac{1}{1 -
l}} \big\{ X(v),  {\mathcal F}^l_{\sigma,{\mathcal I}}
\big\}^{\frac{1}{1 - l}}   \\
\phantom{{\mathcal F}^{-1}_{\sigma,{\mathcal I}}}{}
=   \left(\frac{2 i}{\kappa'\gamma l \mu_0} \right)^{\frac{1}{1 - l}} \left(h_v^{(\mu_0/2)}(X)\big\{ h_v^{(- \mu_0/2)}(X),
{\mathcal F}^l_{\sigma,{\mathcal I}} \big\}\right)^{\frac{1}{1 - l}} , \qquad l \in (0, 1) ,
\end{gather*}
and similarly for ${\mathcal F}^{-1}_{\delta,{\mathcal I}}$. These can be promoted to well-def\/ined
operators. Then, following our strategy for the inverse
volume factors, we get
\begin{gather*} 
\mathrm{RHS}  = - \frac{1}{4} \sum_{i = 1}^n N(\bar \theta_i) \frac{[{\mathcal E}(\bar \theta_i)]^2}{V(\mathcal{I}_i)}
\left[ {\mathcal F}^{-1}_{\delta,{\mathcal I}_i} {\mathcal F}_{\sigma,{\mathcal I}_i} \big(
{\mathcal F}^{-1}_{\sigma,{ {\mathcal I}_{i + 1}}}
{\mathcal F}_{\delta,{\mathcal I}_{i + 1}}  - {\mathcal F}^{-1}_{\sigma,{\mathcal I}_{i}} {\mathcal
F}_{\delta,{\mathcal I}_{i}} \big) \right]^2
\\
\phantom{\mathrm{RHS}}{}
 =   - \frac{1}{4} \left[\frac{16}{3 (\kappa'\gamma)^3\mu_0\nu_0 \alpha^3}\right]^k   \sum_{i = 1}^n N(\bar \theta_i)
[{\mathcal E}(\bar \theta_i)]^2 \\
\phantom{\mathrm{RHS}=}{}
\times
\left[ {\mathcal F}^{-1}_{\delta,{\mathcal I}_i} {\mathcal F}_{\sigma,{\mathcal I}_i} \big(
{\mathcal F}^{-1}_{\sigma,{ {\mathcal I}_{i + 1}}}
{\mathcal F}_{\delta,{\mathcal I}_{i + 1}}  - {\mathcal F}^{-1}_{\sigma,{\mathcal I}_{i}} {\mathcal
F}_{\delta,{\mathcal I}_{i}} \big) \right]^2
\left.  \left[\mathcal{Z}_{\alpha}(\mathcal{I}_i) \right]^k \right|_{\alpha = \frac{2}{3} - \frac{1}{3 k}}.
\end{gather*}

The choice of $\alpha$ would be same as that in the f\/irst term.

{\underline {\em Third term of $H_P$}.} We can rewrite
\begin{gather*}
H_T   =   - \int_{S^1} N(\theta)\partial_\theta\left[\frac{\mathcal{E}\partial_{\theta}\mathcal{E}}{\sqrt{E(\theta)}}\right]
  \approx   - \sum_{i = 1}^n N(\bar \theta_i) \epsilon \partial_{\theta}\left[\frac{\mathcal{E}(\bar
\theta_i)\partial_{\theta}\mathcal{E}}{\sqrt{E(\bar \theta_i)}}\right]
\end{gather*}
as
\begin{gather*} 
\mathrm{RHS}   =   - \left[\frac{16}{3 (\kappa'\gamma)^3\mu_0\nu_0 \alpha^3}\right]^k
 \sum_{i = 1}^n N(\bar \theta_i) \left\{
{\mathcal{E}(\bar \theta_i + \epsilon) \left[ \mathcal{E}(\bar \theta_i + 2\epsilon) - \mathcal{E}(\bar \theta_i +
\epsilon) \right] }[\ \mathcal{Z}_{\alpha}(\mathcal{I}_{i + 1})\ ]^{k} \right.   \\
\left. \left.\hphantom{\mathrm{RHS}   =}{}   - \mathcal{E}(\bar \theta_{i}) \left[ \mathcal{E}(\bar \theta_i + \epsilon) -
\mathcal{E}(\bar \theta_{i})~ \right] [\ {\mathcal Z}_{\alpha}(\mathcal{I}_{i})\ ]^k \right\}
\right|_{\alpha = \frac{2}{3} - \frac{1}{3 k}},
\end{gather*}
where the choice of $\alpha$ is as before.

We have expressed $H_P$ in terms of the holonomy-f\/lux variables. Quantization can be carried out
simply via the replacement $({\mathcal Z}_{\alpha})^{k} \to (-i/\hbar)^{3k} (\hat {\mathcal Z}_{\alpha})^{k}$.

We have thus managed to write the Hamiltonian constraint as a well-def\/ined operator on the kinematic Hilbert space. Obviously, there are operator
ordering ambiguities in the quantization of~$H_K$. However, because of equation~(\ref{diagonalO}), there are no ordering ambiguities in~$H_P$.
It is also straightforward to verify that the action of the operator is well def\/ined on the states in the Hilbert space~\cite{kinjal2}.

\subsubsection{Ambiguities in the quantization scheme}

In the course of the above construction, there have been obvious issues of quantization ambiguities. These are in fact ambiguities in the ordering of
the operators, in the transcription in terms of basic quantum variables as well as in the choice of partitions.

Let us review them and the choices we made.
\begin{enumerate}\itemsep=0pt
\item In $H_K$ we chose to keep the holonomies to the left. Then, the term containing the volume operator acts f\/irst on the states
and the pieces which do not have any vertices give zero.
\item In the regularization of $H_K$ we used the inverse volume and plaquette holonomies. We could have
introduced inverse f\/lux operators and $\hat {\mathcal E}$ operators to replace $1/\sqrt{E}$ and also replaced the
$X$, $Y$, $\int_{ {\mathcal I}_i}{\mathcal A}$ by $\sin(\mu_0X)/\mu_0$, and similarly for the others. Such a replacement
would still give the classical expression back in the limit of small $X$, $Y$, $\epsilon$. The quantum operator, however, would
be dif\/ferent.
\item The second term in the $H_P$ could be manipulated in terms of inverse powers of $\sqrt{E}$ instead of introducing
inverse f\/lux operators (e.g., by replacing $1/E^\sigma = {\mathcal E} E^\delta/(\sqrt{E})^2$). This
would lead to
${\mathcal E}^2\left({\mathcal F}_{\sigma, {\mathcal I}_i}{\mathcal F}_{\delta, {\mathcal I}_{i +
1}}  - {\mathcal F}_{\delta, {\mathcal
I}_i}{\mathcal F}_{\sigma, {\mathcal I}_{i + 1}}\right)^2$ and $\alpha(k) = 2/3 - 5/(3k)$.
In the limit of inf\/inite ref\/inement, each cell would contain {\em at most} one vertex and the cells adjacent to such a~cell
would always be empty. Consequently, the second term of $H_P$, regulated in the above manner, would always give a zero action.
\item Over and above these dif\/ferent transcriptions, we also have the ambiguities introduced by the arbitrary positive power
$k$ (and $\alpha(k)$) and by the arbitrary power $l \in (0, 1)$ in the def\/inition of inverse f\/lux
operators, which is similar to the one in the minisuperspace models described before.
\end{enumerate}

There are also issues related to the choice of partitions, the subsequent $\epsilon \to 0$ limit, and the presence/absence of
local degrees of freedom.  This is most dramatically brought out by the second term of~$H_P$. Classically, this is the term
which reveals spatial correlations in a~solution spacetime through $\partial_{\theta} \ln
(E^\delta/E^\sigma)$~\cite{kinjal1} and
ref\/lects the presence of inf\/initely many physical solutions. In the (vacuum) spherically-symmetric case, such a  term is absent and
so are local physical degrees of freedom. We would like to see if there is a quantization of this term which ref\/lects these
correlations. This can only be ensured if we chose a partition such that every cell has  {exactly} one vertex.
Then, the contributions will explicitly depend upon $\mu$, $\nu$ labels of adjacent vertices and, in this sense, spatial correlations will
survive in the constraint operator. However, the price to pay for that is that we cannot take inf\/inite ref\/inement~($\epsilon \to 0$).

An even more restrictive choice would be to pick the partition def\/ined by the graph itself~-- cells def\/ined by the edges and the
boundary points of cells as vertices. In this case, the new vertices created by~$H_K$ would be the already present vertices
and the constraint equation would lead to a (partial) dif\/ference equation among the labels. The~$\epsilon \to 0$ limit may then be
thought to be relevant when states have support on graphs with a very large (but f\/inite) number of vertices; heuristically, for semiclassical states.
However, more work needs to be done to determine the validity of this proposal.

This completes the kinematic framework of the polymer quantization of this model. Due to the complicated nature of the expressions,
not much progress has been made beyond that so far. A few directions in which there is ongoing further research are:
\begin{itemize}\itemsep=0pt
\item verif\/ication of the quantum constraint algebra,
\item obtaining the spectrum and checking the self-adjointness of the Hamiltonian constraint operator,
\item construction of Dirac and quasi-local energy observables, at least on the kinematic Hilbert space,
\item exploring the possibility of $\bar{\mu}$-type quantizations in the homogeneous directions,
\item checking whether the Bianchi~I cosmological model can be viewed as a sector of this model.
\end{itemize}

\section{Comparison with the hybrid quantization}

As we have seen in the previous two sections, the polarized Gowdy $T^3$ model has been quantized in two dif\/ferent ways within the framework of LQG. While
it would be good to have a procedure to compare the physical results of the two frameworks, it is not possible currently.
Although both quantization schemes start from the classically-reduced phase space of the model, there are signif\/icant dif\/ferences in the treatment,
both in the classical and the quantum theory.

The motivation of the hybrid quantization programme is to utilize the tools of LQC on midi\-superspace models. It tries to determine whether the
singularity resolution in LQC is a feature of the quantization scheme or an artifact of the high degree of symmetry of minisuperspace.
Moreover, it provides a suitable arena to analyze the back-reaction between inhomogeneities and quantum background geometry.
On the other hand, the polymer approach tries to construct a~loop-quantized theory \textit{ab initio}, trying to mimic the procedures of the
full theory. In addition to the fate of the classical singularity in polymer quantization, its aim is to provide a toy model
where some problems of the full theory, such as the verif\/ication of the quantum constraint algebra and the construction of observables, can be explored.

Let us review the progress made in the two quantization schemes so far.

\begin{itemize}\itemsep=0pt
\item The hybrid quantization scheme employs the machinery developed in minisuperspace LQC described before, in order to study the midisuperspace model.
After a partial gauge f\/ixing, it can be easily seen that the polarized Gowdy $T^3$ model can be thought of as a Bianchi I model f\/illed with
inhomogeneities propagating in one direction. This fact is exploited to break up the degrees of freedom into a homogeneous Bianchi I part and an
inhomogeneous scalar f\/ield. In addition, as a consequence of the partial gauge f\/ixing, only two global constraints remain in the model.
The polymer quantization programme, on the other hand, aims to implement the loop quantization programme by def\/ining suitable Ashtekar variables for the
entire model. Unlike in the hybrid approach, the system is symmetry reduced, but no further gauge is f\/ixed at the classical level, so
that the constraints are not global but depend on the point.

The dif\/ference between the two approaches can be seen, on the one hand, in the dif\/ferent way the spatial metric is parametrized, by
comparing equations~(\ref{newmetric}) and (\ref{polymergowdymetric})\footnote{Note that there is
no relation between the scalar f\/ield $\tilde\xi$ in~(\ref{newmetric}) and the angle~$\xi$ in~(\ref{polymergowdymetric}).} and, on the other hand,
in the dif\/ferent constraints surviving in the classically-reduced model.

\item The quantization that is subsequently carried out is also dif\/ferent. In hybrid quantization, the homogeneous Bianchi I is loop quantized while the
inhomogeneous scalar f\/ield is Fock quantized using creation/annihilation operators. The full kinematic Hilbert space is a~tensor product of the two.
There is a non-trivial interaction term in the Hamiltonian which couples the homogeneous and inhomogeneous modes. However, all the constraints
can be expressed as densely def\/ined operators on the tensor product Hilbert space.
In the polymer quantization, all the degrees of freedom are loop quantized but, unlike the full theory, there are both point and edge holonomies. The
techniques of full LQG are used to def\/ine the kinematic polymer Hilbert space. Subsequently, more general operators are constructed including the
Hamiltonian constraint operator and it can be shown that they are well def\/ined on the kinematic Hilbert space and do not depend on the regulator.

Thus, we have two dif\/ferent quantum theories of the same classical system, both of which have a well-def\/ined action of the constraint
operators. However, these operators are def\/ined on very dif\/ferent Hilbert spaces, and the quantum
theories may not be unitarily equivalent. While the polymer quantization scheme is closer in spirit to LQG, progress has stalled
beyond this point because of the extremely complicated nature of the expressions. On the other hand, signif\/icant progress has been made in the
hybrid quantization programme.

\item The construction of the physical Hilbert space has not been carried out so far in the polymer scheme. On the other hand, in the hybrid approach
the physical Hilbert space has been constructed, which turns out to be tensor product of the physical Hilbert space of
the Bianchi I model and the physical Fock space for the inhomogeneities, which are not neglected in this model. Rather, it is possible to
view this system as some inhomogeneous scalar f\/ield on a polymer-quantized Bianchi I space. The classical singularity is absent in the physical
Hilbert space even in the presence of inhomogeneities.
\end{itemize}

\part{Ef\/fective dynamics}\label{part3}

In the last part, we discuss two aspects of loop quantum cosmology which play an important role in
the connection between theory and phenomenology. Section \ref{hodi} presents the ef\/fective FRW
dynamics obtained by evaluating the scalar constraint on semiclassical states. The choice of
parametrization in minisuperspace is determined by rather robust arguments, but the latters undergo
several modif\/ications in the context of inhomogeneous models; this is the subject of Section
\ref{latti}, where the lattice ref\/inement framework is introduced.


\section{Homogeneous ef\/fective dynamics}\label{hodi}

The exact and numerical discrete dynamics stemming from the quantum Hamiltonian constraint provides important information about the singularity resolution in LQC, but it does not yield itself to manipulations suitable for the extraction of inf\/lationary dynamics and observables in a~semiclassical limit. This can be achieved by evaluating the Hamiltonian constraint on semiclassical states, resulting in \emph{continuous} Friedmann equations corrected by quantum terms. At this point, standard analysis techniques developed in classical FRW cosmology can be applied to these equations. The study of linear perturbations (which shall not be reviewed here) requires, however, some extra ef\/fort.

Ef\/fective cosmological equations of motion are derived from the expression of the Hamiltonian constraint on a semiclassical state. The latter is typically decomposed into a gravitational and matter sector, $|\Psi_{\rm sc}\rangle=\sum_{A,B}|{\rm grav}\rangle_A\otimes |{\rm mat}\rangle_B$. In general, geometrical and matter operators do not act separately on physical states because solutions to the Hamiltonian constraint already incorporate correlations between the two sectors. So operators on such states are in general complicated, entangled observables. However, on a semiclassical state geometrical and matter operators commute and they can be treated separately.

Before discussing how semiclassical states determine an ef\/fective dynamics, it is convenient to generalize the Hamiltonian constraint and introduce some ambiguity parameters which were previously kept f\/ixed. This is done in order to accommodate results which will be later obtained in an inhomogeneous setting. In this part we set $\hbar=1$.

\subsection{Parametrization of the Hamiltonian constraint}

Let $\k^2=8\pi G$. As before, we def\/ine a pair of variables
\begin{gather*}
b:= \frac{\bu c}2 ,\qquad v:= \frac{6}{(1+n)\g\k^2}\frac{p}{\bu} ,
\end{gather*}
where $\bu$, however, is now an arbitrary dimensionless function of the densitized triad:
\begin{gather}\label{bu2}
\bu=\left(\frac{p_*}{p}\right)^n = \left(\frac{a_*}{a}\right)^{2n} ,
\end{gather}
where $n\in\mathbb{R}$ (until now it was f\/ixed to $n=1/2$) and $p_*$ and $a_*$ are, respectively, constants of dimension $[p_*]=-2$ and $[a_*]=0$. Then,
\begin{gather*}
\{b,v\} =1 .
\end{gather*}
In a purely homogeneous model, there is no reason in favour of (and, in fact, there are some against) taking $n\neq 1/2$, but for the time being we do not attempt to justify this generalization. Another ambiguity parameter~$q\in\mathbb{R}$ can arise when writing down Thiemann's identity
\begin{gather*}
\e^{ijk}\frac{E_{i}^{a}E_{j}^{b}}{\sqrt{|\det
E}|} = 2\e^{abc}\frac{V^{1-q}}{q}\frac{\delta V^q}{\delta
E^c_k} .
\end{gather*}
Denote with $\ell_0^2=A_\square$ the area of an elementary plaquette. For a f\/lat homogeneous background, the classical scalar constraint becomes \cite{cqc}
\begin{gather}
C  =  -\frac{1}{\g^2}\frac{E_i^{a}E_j^b}{\sqrt{|\det E|}}\e^{ij}_{\ \ k}F^k_{ab}+C_{\rm mat}\nonumber\\
\phantom{C}{} =  \frac{8(1+n)}{\g^2q} \left[\frac{(1+n)\g\k^2}{3}\right]^{\frac{1-2n}{2(1+n)}}
\lim_{\ell_0\to 0}\frac{1}{\ell_0^3}\left(\frac{p_*^{n}}{2}\right)^{\frac{3}{2(1+n)}} v^{\frac{3(1-q)}{2(1+n)}}\sin^22 b\nonumber\\
\phantom{C=}{} \times\Big[\sin b  \big\{\cos b,v^{\frac{3q}{2(1+n)}}\big\}-\cos b  \big\{\sin b,v^{\frac{3q}{2(1+n)}}\big\}\Big]+C_{\rm mat} .\label{SH3cl}
\end{gather}
The gravitational sector is only a function of $b$ and $v$. The scalar f\/ield part (with potential $U(\phi)$) only contains volume factors,
\begin{gather*}
C_{\rm mat}=\k^2\frac{\Pi_\phi^2}{p^{3/2}}+p^{3/2}U(\phi) .
\end{gather*}
As before, the quantum constraint is regularized by assuming that holonomy plaquettes cannot be shrunk indef\/initely, replacing the limit $\ell_0\to0$ in equation~\eqref{SH3cl} with $\ell_0\to V_o^{1/3}\bu$. With this substitution, the quantum Hamiltonian operator corresponding to equation~\eqref{SH3cl} is well def\/ined:
\begin{gather}\label{cig}
\hat C = -4 \widehat{\sin2b} \hat A  \widehat{\sin2b}+\hat C_{\rm mat} ,
\end{gather}
where
\begin{gather*}
\hat A  =
\frac{\rmi(1+n)}{4q V_o\g^2}\left[\frac{(1+n)\g\k^2}{3}\right]^{\frac{1+4n}{2(1+n)}}
\left(\frac{p_*^n}{2}\right)^{-\frac{3}{2(1+n)}} \nonumber\\
\phantom{\hat A  = }{}
\times\widehat{|v|^{\frac{3(1+2n-q)}{2(1+n)}}}\bigg[\widehat{\cos b} \widehat{|v|^{\frac{3q}{2(1+n)}}}\widehat{\sin b}-\widehat{\sin b} \widehat{|v|^{\frac{3q}{2(1+n)}}}\widehat{\cos b}\bigg]\nonumber\\
\phantom{\hat A }{}
=\frac{1+n}{8qV_o\g^2}\left[\frac{(1+n)\g\k^2}{3}\right]^{\frac{1+4n}{2(1+n)}}\left(\frac{p_*^n}{2}\right)^{-\frac{3}{2(1+n)}} \nonumber\\
\phantom{\hat A  = }{}
 \times\widehat{|v|^{\frac{3(1+2n-q)}{2(1+n)}}}\bigg[\widehat{\rme^{-\rmi b}} \widehat{|v|^{\frac{3q}{2(1+n)}}}\widehat{\rme^{\rmi b}}-\widehat{\rme^{\rmi b}} \widehat{|v|^{\frac{3q}{2(1+n)}}}\widehat{\rme^{-\rmi b}}\bigg].
\end{gather*}
We continue to use the notation $|v\rangle$ as the eigenstates of $\hat v$ upon which holonomies act as translations,
\begin{gather*}
\hat v |v\rangle =v |v\rangle ,\qquad \widehat{\rme^{\rmi v' b}}|v\rangle=|v+v'\rangle .
\end{gather*}
These states are also eigenstates of $\hat A$,
\[
\hat A|v\rangle=A_v|v\rangle ,
\]
with eigenvalues
\begin{gather*}
A_v =
\frac{1+n}{8qV_o\g^2}\left(\frac{p_*^n}{2}\right)^{-\frac{3}{2(1+n)}}
\left[\frac{(1+n)\g\k^2}{3}\right]^{\frac{1+4n}{2(1+n)}}
|v|^{\frac{3(1+2n-q)}{2(1+n)}}\left(|v+1|^{\frac{3q}{2(1+n)}}-|v-1|^{\frac{3q}{2(1+n)}}\right) .
\end{gather*}

\subsection{Minisuperspace parametrization}

While in LQG the area spectrum is bounded from below by the minimum area $\Delta$, due to the symmetry reduction the same property is not shared by loop quantum cosmology. Nonetheless, we have seen that one may draw inspiration from the full theory and \emph{assume} that the kinematical area of any loop inside the comoving volume $V_o$ is bounded by the area gap for the gauge invariant states which are likely to be realized in a homogeneous context. This value is (twice) the LQG area gap \cite{AsW},
\begin{gather}\label{minar2}
\Delta= 4\sqrt{3}\pi \g l_\Pl^2 ,
\end{gather}
so that
\begin{gather}\label{ineqal}
(a \ell_0)^2\geq \Delta .
\end{gather}
This step is rather speculative inasmuch as it borrows a result of the background-independent framework and forces it into the symmetry-reduced model. It is necessary, however, because the quantum scalar constraint in minisuperspace would be singular if one maintained the limit $\ell_0\to0$. Moreover, the semiclassical limit and the Wheeler--DeWitt equation are reproduced correctly.

In a general background, the edges of a spin-network state would intersect a given cell only once. By symmetry, the edges of a spin-network state in minisuperspace should traverse the f\/iducial cell, rather than intersecting it from one side; hence the factor of two in equa\-tion~\eqref{minar2}~\cite{AsW}. Notice, however, that there is no unique way of f\/ixing the value of $\Delta$, and calculations accounting for dif\/ferent details can produce dif\/ferent numerical prefactors. At any rate, these dif\/ferences are not so large as to give qualitatively inequivalent physical ef\/fects.

If the inequality \eqref{ineqal} is saturated (smallest possible holonomy path), the comoving cell area is also the comoving area gap, that is, the smallest non-vanishing eigenvalue of the area operator measuring comoving surfaces. In particular,
\begin{gather}\label{buimp}
\frac{\ell_0^2}{V_o^{2/3}}=\frac{\Delta}{p} = \left(\frac{p_*}{\Delta}\bu\right)^{1/n} .
\end{gather}
One has $\bu=\ell_0/V_o^{1/3}$ if $p_*=\Delta$ and
\begin{gather}\label{impqs}
n=\frac12 ,
\end{gather}
a choice corresponding to the improved quantization scheme~\cite{acs,aps3,APSV}. The set~$\{|v\rangle\}$ becomes the eigenstate basis of the volume operator, $v\propto p^{3/2}=V$. As the Universe expands, the comoving area gap shrinks to zero and the geometry is better and better described by classical general relativity, while near the big bang quantum ef\/fects become important.

Originally, the variables $p$ and $c$ were used instead of $v$ and $b$, corresponding to $\bu=1$ ($n=0$). In this ``old quantization scheme'', the states $|v\rangle=|\mu\rangle$ coincide with the basis eigenstates of the momentum operator~$\hat p$, with eigenvalues~$v\propto p$~\cite{abl,boj7}. This case leads to severe restrictions of the matter sector if the wavefunctions solving the Hamiltonian constraint are required to be normalizable and to reproduce the classical limit at large scales~\cite{NeSa}. Also for such reason, the improved quantization scheme seems to be the most natural and, as we have already seen, the most reasonable in a purely homogeneous context. However, later motivations lead us to keep~$n$,~$p_*$ and the other free parameters of the model as general as possible. In this case, $p_*$ is some physical squared length determined by the theory which may dif\/fer from the mass gap~$\Delta$.

\subsection{Ef\/fective equations of motion}

A semiclassical state $|\Psi_{\rm sc}\rangle$ is peaked around some point~$(v,b)$ in the classical phase space.
One can compute the expectation value of the Hamiltonian constraint operator thereon, using an appropriate inner product.
Accordingly, for the gravitational part of the Hamiltonian operator~\eqref{cig} we approximate its expectation value as
$\langle\Psi_{\rm sc}|\widehat{\sin \bu c}  \hat A  \widehat{\sin \bu
c}|\Psi_{\rm sc}\rangle \approx A_v\sin^2 \bu c$, and we may write (e.g.,~\cite{lqc1,Din08,Tav08})
\begin{gather}\label{Hg}
\langle\Psi_{\rm sc}|\hat C|\Psi_{\rm sc}\rangle \approx -\frac{6}{\g^2} \a\sqrt{p} \frac{\sin^2 \bu c}{\bu^2}+2\k^2 p^{3/2}\rho ,
\end{gather}
where the scalar f\/ield energy density is
\begin{gather}\label{rhola}
2\k^2\rho := \langle v|\hat C_{\rm mat}|\la\rangle = \frac{\nu P_\phi^2}{2p^{3/2}}+p^{3/2}V .
\end{gather}
We have two correction functions, $\a$ and $\nu$. The f\/irst is
\begin{gather}
\a=\frac{\s}{12q}v\left(\left|1+\frac1v\right|^\frac{6q}{\s}-\left|1-\frac1v\right|^\frac{6q}{\s}\right) ,\label{alpha}
\end{gather}
where
\begin{gather}\label{cmini}
\s=4(1+n) ,\qquad \frac13<q\leq 1 .
\end{gather}
The matter correction function is
\begin{gather}
\nu  :=  (\langle v|\widehat{v^{1-l}}\widehat{v^{l-1}}|v\rangle)^{\frac{6}{(1-l)\s}}  = \left[\frac{v}{2l}\left(\left|1+\frac1v\right|^l-\left|1-\frac1v\right|^l\right)\right]^{\tfrac{6}{(1-l)\s}},\label{nudef}
\end{gather}
where the ambiguity $l$ lies in the range \cite{boj12, QSD5}
\begin{gather}
\frac12\leq l <1 .
\end{gather}

When $\a=1$ and the matter sector is a massless free scalar f\/ield, equation~\eqref{Hg} is exact~\cite{bo10}. In general, however, the evolution of a f\/initely-spread semiclassical state will produce quantum f\/luctuations leading to additional corrections to equation~\eqref{Hg}~\cite{bo11, BHS}. Assuming that the semiclassical wave-packet of the Universe does not spread appreciably, we can stick with equation~\eqref{Hg} also in the presence of a nontrivial scalar potential. Then, the matter energy density~$\rho$ is given by equation~\eqref{rhola}.

The Hamilton equation of motion for the densitized triad gives the Hubble parameter
\begin{gather}\label{hub}
H = \a  \frac{\sin 2\bu c}{2\g a\bu} .
\end{gather}
In the classical limit, $c\to\g\dot a$ and the right-hand side tends to $\dot a/a$ for small $\bu c$. Combining equations~\eqref{Hg} and \eqref{hub}, one gets the Friedmann equation
\begin{gather}\label{frwlqc}
H^2=\frac{\k^2}{3} \rho\left(\a-\frac{\rho}{\rho_*}\right) ,
\end{gather}
where
\begin{gather}\label{rho*}
\rho_* \equiv \frac{3}{\g^2\k^2\bu^2 p} .
\end{gather}
The equation of motion of the scalar f\/ield is
\begin{gather*}
P_\phi =p^{3/2}\frac{\dot\phi}{N\nu} ,
\end{gather*}
while the equation for $\dot P_\phi$ leads to the ef\/fective Klein--Gordon equation
\begin{gather}\label{kglqc}
\ddot\phi+\left(3H-\frac{\dot\nu}{\nu}\right)\dot\phi+\nu U_{,\phi}=0 .
\end{gather}
As $\nu\geq 0$ has a maximum at $v=1$ and then decreases down to unity for large $v$, the friction term in equation~\eqref{kglqc} changes sign during the evolution of the universe, the f\/irst stage being of superacceleration.

Setting $\a=1=\nu$ in the equations of motion~\eqref{frwlqc} and~\eqref{kglqc}, one ignores inverse-volume corrections. On the other hand, in the limit $\sin 2\bu c\to 2\bu c$ one neglects holonomy corrections and the second term in equation~\eqref{frwlqc} is dropped.

The left-hand side of equation~\eqref{frwlqc} is positive def\/inite and, if $\rho>0$ ($\a>0$ if $n>-1$), the energy density is \emph{bounded from above}:
\begin{gather}\label{lowbo}
\rho\leq\a\rho_* .
\end{gather}
When $\rho_*\propto a^{2(2n-1)}$ varies with time, there is no constant absolute upper bound. This is avoided in the improved quantization~\eqref{impqs}, where the critical density is constant:
\[
\rho_* = \frac{3}{\g^2\k^2 p_*} .
\]
For the particular choice $p_*=\Delta$, the critical density is less than half the Planck density,
\begin{gather}\label{rho*im}
\rho_* = \frac{\sqrt{3}}{32\pi^2\g^3}\rho_\Pl\approx 0.41 \rho_\Pl ,
\end{gather}
where we used the value $\g\approx 0.238$ \cite{Mei04} from the computation of the entropy of non-rotating black-hole isolated horizons~\cite{ABCK,ABK,Mei04}. The numerical prefactor depends on equations~\eqref{minar2} and~\eqref{ineqal} and it could change in a more complete formulation of the model, but not in a way leading to qualitative dif\/ferences.

If the ambiguity $q$ is set equal to~1, $\a=1$ and the lower bound~\eqref{lowbo} is the f\/ixed constant~\eqref{rho*im}~\cite{aps3,sin06,SVV}. Thus, the avoidance of the big bang singularity in LQC is conf\/irmed at the kinematical level (via the spectrum of the inverse volume operator), by the full quantum Hamiltonian dynamics and through the ef\/fective dynamical equations. The big bang is replaced by a bounce at~$H=0$,
where the energy density is about half the Planck energy.

These results are encouraging but, of course, insuf\/f\/icient to establish a solid solution of the big bang issue.
First, the minisuperspace quantization is a toy model. Second, even choosing the improved quantization scheme, the critical density
$\a\rho_*$ may be non-constant if $q\neq 1$, in which case one might think that the neat bounce interpretation is lost.
Nevertheless, we have seen how loop quantum gravity could resolve the big bang singularity on one hand, and how far we are from a complete
understanding of the full theory from the other.


\subsection{Inverse-volume corrections in minisuperspace models}

We discuss now the correction functions $\a$ and $\nu$ from the point of view of their asymptotic limits, later stressing an interpretational issue.

On a semiclassical state the eigenvalues of $\widehat{|v|^{l-1}}$ are approximated by the classical variable~$v^{l-1}$ itself.
Consistently, the classical limit corresponds to a large-volume approximation where $v\gg 1$, while in the near-Planck regime
(``small volumes''; the reason for quotation marks will be soon clear) $v\ll1$. Since the momentum operator is
$\hat v= 2\widehat{p/\bu}$, the total $p$-dependence of~$v$ is ef\/fectively
\[
v =\frac{12\sqrt{3}}{\s}\frac{p_*}{\Delta}\left(\frac{p}{p_*}\right)^{\frac{\s}{4}} .
\]
``Near the Planck scale'' ($v\ll1$), the correction functions read
\begin{gather}
\a   \approx  v^{2-\frac{6q}{\s}}=: \a_1\dpl^{-q_\a},\label{qa}\\
\nu  \approx  v^{\frac{6(2-l)}{(1-l)\s}}=: \nu_1\dpl^{-q_\nu},\label{qn}
\end{gather}
where
\begin{gather*}
q_\a   =  1-\frac{3q}{\s} ,\qquad\a_1=\left(\frac{12\sqrt{3}}{\s}\frac{p_*}{\Delta}\right)^{2q_\a}  ,\qquad
q_\nu  = \frac{3(2-l)}{(1-l)\s} ,\qquad \nu_1=\left(\frac{12\sqrt{3}}{\s}\frac{p_*}{\Delta}\right)^{2q_\nu}
\end{gather*}
and
\begin{gather}\label{dp}
\dpl:= \left(\frac{p_*}{p}\right)^{\frac{\s}{2}}=\left(\frac{a_*}{a}\right)^\s .
\end{gather}

From the calculation leading to $\a$ and $\nu$, one can argue that the ``natural'' choice of the ambiguities $l$ and $q$ can be set at
the middle of their range:
\begin{gather}\label{natuiq}
l=\frac34 ,\qquad q=\frac12 .
\end{gather}
In the minisuperspace parametrization, the old quantization scheme corresponds to $\s=4$ and
\begin{gather*}
q_\a   =  \frac{5}{8} ,\qquad\a_1=3^{\frac{15}{8}}=O(10) ,\qquad
q_\nu  = \frac{15}{4} ,\qquad \nu_1=3^{\frac{45}{4}}=O(10^5) ,
\end{gather*}
while the improved scheme has
\[
\s=6
\]
and
\begin{gather*}
q_\a   =  \frac{3}{4} ,\qquad\a_1=2^{3/2}3^{3/4}\approx 6 ,\qquad
q_\nu  = \frac{5}{2} ,\qquad \nu_1=2^5 3^{5/2}\approx 500 .
\end{gather*}
In homogeneous models with $n=0$, the duration of this regime depends on the spin representation of the holonomies, small $j$ implying a very short super-inf\/lationary period and, actually, almost no intermediate stage between the discrete quantum regime and the continuum classical limit~\cite{boj12}. Since small-$j$ representations are theoretically favoured, this constitutes a problem. It will be relaxed in a dif\/ferent parametrization when inhomogeneities are taken into account.

In the quasi-classical limit (large volumes), equations~\eqref{alpha} and \eqref{nudef} can be approximated as
\begin{gather}
\a  \approx  1+\a_0\dpl ,\label{ca}\\
\nu \approx  1+\nu_0\dpl ,\label{cn}
\end{gather}
where
\begin{gather*}
\a_0   =  \frac{(3q-\s)(6q-\s)}{6^4}\left(\frac{\Delta}{p_*}\right)^2 ,\qquad
\nu_0  =  \frac{\s(2-l)}{6^3}\left(\frac{\Delta}{p_*}\right)^2 .
\end{gather*}
For the natural choice \eqref{natuiq}, the old and improved quantization schemes in minisuperspace parametrization correspond, respectively, to
\[
\s=4 ,\qquad \a_0=\tfrac{5}{2^5 3^4}\approx 0.002 ,\qquad\nu_0=\tfrac{5}{6^3}\approx 0.02 ,
\]
and
\[
\s=6 ,\qquad \a_0=\tfrac{1}{96}\approx 0.01 ,\qquad\nu_0=\tfrac{5}{144}\approx 0.03 .
\]
Taking $q=1$ instead, one gets a negative $\a_0=-1/648$ for $\s=4$ and $\a_0=0$ for $\s=6$.

Although one can resort to dif\/ferent quantization schemes, equations~\eqref{ca}, \eqref{qa}, \eqref{cn} and~\eqref{qn} maintain the same structure, where the coef\/f\/icients $\s$, $q_\a$, and $q_\nu$ are robust in the choice of the parameters, inasmuch as their order of magnitude does not change appreciably~\cite{BHKS0}. All these parameters can be set to their ``natural'' values, which are dictated by the form of the Hamiltonian or other considerations.

Now we examine an interpretational issue, already mentioned earlier, related to any para\-met\-ri\-za\-tion in pure minisuperspace. On an ideal FRW background, open and f\/lat universes have inf\/inite spatial volume and the super-Hamiltonian constraint is formally ill
def\/ined because it entails a~divergent integration of a spatially constant quantity over a comoving spatial slice $\Sigma$,
\[
\int_\Sigma \rmd^3 x=+\infty .
\]
To make the integral f\/inite, it is customary to def\/ine the constraint
on a freely chosen f\/inite region of size $V=a^3 V_o$, where $V_o$ is the corresponding comoving volume:
\[
\int_\Sigma \rmd^3 x\to\int_{\Sigma(V_o)} \rmd^3 x=V_o<+\infty .
\]
The volume appears in the correction function \eqref{dp} as $\dpl\sim
a^{-\s}\sim V^{-\s/3}$. To make $\dpl$ adimensional, one can use
the Planck length $l_\Pl$ to write
\begin{gather*}
\dpl \sim
\left(\frac{l_\Pl^3}{V_o}\right)^{\frac{\s}{3}} a^{-\s}.
\end{gather*}
Physically, the parameter $\s$ is related to how the number of plaquettes of an underlying discrete state
changes with respect to the volume as the universe expands. The latter is a phenomenological prescription for the area of holonomy plaquettes,
but ideally it should be an input from the full theory \cite{bo609}. For phenomenology
at the current level of precision, the most signif\/icant parameter among $\{\a_0,\nu_0,\s\}$ is~$\s$, which is not as much af\/fected by dif\/ferent
choices of the minisuperspace scheme.

Since $\delta_{\rm Pl}$ is $V_o$-dependent, inverse-volume corrections cannot strictly be made sense of in a~pure minisuperspace treatment.
To cast the problem in other words, the conformal invariance of the scale factor $a$ in a non-closed universe make statements such as $a_*/a\ll 1$
independent of any physical length scale. One could interpret $V_o$ as a regulator and send $V_o\to\infty$ at the end of calculations, so that in the
quasi-classical limit there are no inverse-volume corrections at all. However, the full theory does contain these corrections, and one should explain
why they do not appear in a cosmological setting. At best, this highlights some tension in the theoretical construction of the homogeneous LQC ef\/fective
dynamics. To get a clearer picture, we should include inhomogeneities already at the fundamental level. The study of midisuperspace models, mentioned
in the previous part, is a step in that direction.

\subsection{Models with $\textsc{k}\neq 0$ and $\Lambda\neq 0$}

The f\/lat ef\/fective dynamical model has been extended to cases with curvature and a cosmological constant.

For a closed universe, $\textsc{k}=1$, there is no f\/iducial volume problem, as mentioned in Section \ref{klam}, and inverse-volume corrections are meaningful also in a pure homogeneous and isotropic setting. The cyclic bounces appearing in the dynamics of the dif\/ference evolution equation \cite{APSV} exist also at the ef\/fective level \cite{BoT,LMNT,MHS}; in particular, the big crunch of classical closed universes can be avoided \cite{SiT}. The bounce persists in an open universe, $\textsc{k}=-1$ \cite{van06}. In general, all past and future strong curvature singularities are resolved in $\textsc{k} = \pm1$ isotropic models; for the closed model, weak singularities in the past evolution may also be resolved \cite{ViS}.

There is evidence that a cosmological constant, if suitably tuned, does not spoil the singularity resolution. When $\Lambda>0$ and $\textsc{k}=1$ \cite{MHS}, the bounce is preserved if the cosmological constant is suf\/f\/iciently small. Above a certain critical value, however, periodic oscillations take place. When $\Lambda<0$, recollapse of the universe is possible, even cyclically \cite{BeP, BoT}. Whatever the sign of the cosmological constant, the ef\/fective Friedmann equation is equation~\eqref{frwlqc}, with the critical density $\rho_*$ shifted by a constant, $\Lambda$-dependent term.

\section{Inhomogeneous models}\label{latti}

So far we have not given any motivation for taking $\bu\propto p^{-n}$. This is the next subject and it resides in a framework which does not enjoy the symmetries of a purely FRW background.

\subsection{Lattice ref\/inement}

In loop quantum gravity, the classical continuum of general relativity is replaced by the appearance of discrete
spatial structures. It is often expected that the scale of the discreteness is determined by the Planck length $l_\Pl$, but if discreteness is fundamental, its scale must be set by the dynamical parameters of some underlying state. Such states are spin networks, graphs in an embedding space whose edges $e$ are labeled by spin quantum numbers $j_e$. The quantum number determines the area of an elementary plaquette intersecting only one edge $e$, given by ${\cal
A}=\gamma l_\Pl^2 \sqrt{j_e(j_e+1)}$. The geometrical size of the plaquette changes only when the latter intersects another edge, thus increasing in quantum jumps. The scale is determined by the Planck length for dimensional reasons, but the actual size is given by the spin quantum number. Its values in a specif\/ic physical situation have to be derived from the LQG dynamical equations, a task which remains extremely dif\/f\/icult. However, given the form in which
$j_e$ appears in the dynamical equations, its implications for physics can be understood in certain phenomenological situations, such as cosmological scenarios. Then, instead of using the spin labels $j_e$, it is useful to refer to an elementary quantum-gravity length scale $L$, which needs not be exactly the Planck length.

The scale $L$ naturally arises if translation invariance is broken, e.g., by clustering matter or inhomogeneous perturbations. The comoving volume $V_o$ of the system can be discretized as a~lattice whose $\cN$ cells or patches are nearly isotropic, have characteristic comoving size $\ell_0^3$, and correspond to the vertices of the spin network associated with $V_o$. The proper size of a cell is
\begin{gather}\label{Nf}
L^3:=a^3\ell_0=\frac{V}{\cN} .
\end{gather}
To calculate the curvature at the lattice sites within $V_o$, we need to specify closed holonomy paths around such points. A generic holonomy plaquette is given by the composition of elementary holonomies over individual plaquettes. Therefore we set the length of the elementary holonomy to be that of the characteristic lattice cell. In other words, the elementary loops of comoving size~$\ell_0$ we have talked about until now def\/ine the cells' walls, while in a pure FRW background there is only one cell of volume $V_o$ (the number $\cN$ is arbitrary). We naturally identify the previously ad-hoc function $\bu(p)$ as the ratio of the cell-to-lattice size, under the requirement that the lattice be \emph{refined} in time:
\begin{gather}\label{N13}
\bu={\cal N}^{-1/3} .
\end{gather}
The patch size $\ell_0^3$ is independent of the size of the f\/iducial region, since both $V_o$ and ${\cal N}$ scale in the same way when the size of the region is changed. Physical predictions should not feature the region one chooses unless one is specif\/ically asking region-dependent questions (such as: What is the number of vertices in a given volume?). This addresses the issue of conformal invariance brief\/ly mentioned above in minisuperspace. In the presence of inhomogeneities there is no conformal freedom and, on the other hand, f\/luxes are determined by the inhomogeneous spin-network quantum state of the full theory associated with a given patch~\cite{boj11}. This implies that to change the f\/iducial volume~$a^3V_o$ would change the number of vertices of the underlying physical state. Therefore, there is no scaling ambiguity in the equations of motion~\cite{boj11,BH2}, although the physical observables will depend on the choice of spin-network state.

The spin-network state described by the lattice can be (and usually is) excited by the action of the Hamiltonian operator on the spin vertices, increasing their number and changing their edge labels \cite{RS3, inv-vol-lqg2}. This process has not yet been established univocally in the full theory, so it is convenient to parametrize the number of vertices as in equation~\eqref{Nf} \cite{bo609}, where the length $L(t)$ is state dependent and, by assumption, coordinate independent; its time dependence is inherited from the state itself. As the kinematical Hilbert space is usually factorized into gravitational and matter sectors, the problem here emerges of how to def\/ine a natural clock when matter does not enter in the def\/inition of a (purely geometrical) spin network. This issue will require a much deeper understanding of the theory. So, as unsatisfactory as equation~\eqref{Nf} may be, we take it as a phenomenological ingredient in the present formulation of inhomogeneous LQC.

The general form \eqref{bu2} of $\bu(p)$ is obtained if $L(t)$ scales as
\begin{gather*}
L\sim a^{3(1-2n)} .
\end{gather*}
Homogeneous models adopting equation~\eqref{buimp} feature holonomies which depend on triad variables; in other words, curvature components are constrained by the area operator although this does not appear in the full constraint. On the other hand, in inhomogeneous models the dependence of the parameter $\bu$ on $p$ is implemented at state (rather than operatorial) level, in closer conformity with the full theory \cite{bo609}.

As a side remark, the patches of volume $L^3$ f\/ind a most natural classical analogue in inhomogeneous cosmologies, in particular within the separate universe picture \cite{WMLL}. For quantum corrections, the regions of size $L^3$ are provided by an underlying discrete state and thus correspond to quantum degrees of freedom absent classically. However, the discrete nature of the state implies that inhomogeneities are unavoidable and no perfectly homogeneous geometry can exist. Given these inhomogeneities and their scale provided by the state, one can reinterpret them in a classical context, making use of the separate universe picture. There, the volume~$V$ can be regarded as a region of the universe where inhomogeneities are non-zero but small. This region is coarse grained into smaller regions of volume~$L^3$, each centered at some point~${\bf x}$, wherein the universe is~FRW and described by a ``local'' scale factor $a(t,{\bf x})=a_{\bf x}(t)$. The dif\/ference between scale factors separated by the typical perturbation wavelength $|{\bf x}'-{\bf x}|\sim \lambda\ll V^{1/3}$ def\/ines a spatial gradient interpreted as a metric perturbation. In a perfectly homogeneous context, $L^3\sim V$ and there is no sensible notion of cell subdivision of $V$; this is tantamount to stating that only the f\/iducial volume will enter the quantum corrections and the observables, $\cN=\cN_0$. On the other hand, in an inhomogeneous universe the quantity $L^3$ carries a time dependence which, in turn, translates into a momentum dependence. The details of the cell subdivision (number of cells per unit volume) are intimately related to the structure of the small perturbations and their spectrum. Thus, lattice ref\/inement is better suitable in the cosmological perturbation analysis. As long as perturbations are linear and almost scale invariant, the size of the volume within which the study is conducted is totally irrelevant.

\subsubsection{Critical density and quantum corrections}

From equations~\eqref{rho*}, \eqref{N13} and \eqref{Nf}, the critical density is
\begin{gather}\label{rhol}
\rho_* =\frac{3}{\g^2\k^2} \left(\frac{\cN}{V}\right)^{2/3}= \frac{3}{\g^2\k^2 L^2}.
\end{gather}
In all quantization schemes but the improved one ($n=1/2$), $\rho_*$ is not constant and depends on the dynamical patch size $L$. In any case, the critical density is a number density which depends neither on the size of the f\/iducial volume nor on coordinates, so it is physically well def\/ined even outside the improved quantization scheme.

Similar considerations hold for the quantum correction~$\dpl$. In a purely homogeneous universe, the only way to write down equation~\eqref{dp} is $\dpl\propto (l_\Pl/V^{1/3})^\s$, which is volume dependent. On the other hand, in the lattice interpretation
\begin{gather}\label{main}
\dpl=\left(\frac{\ell_\Pl}{L}\right)^{\tilde\s} ,
\end{gather}
and the same quantity is determined by the inhomogeneous state through the patch size $L$. Notice that $\tilde\s>0$ is not the parameter $\s$ determined by equation~\eqref{cmini}; $n=1/2$ will not imply $\tilde\s=6$. The inverse-volume corrections~\eqref{main} do not depend on holonomies due to the use of Thiemann's trick (such as equation~\eqref{eq:identidad-lig})~\cite{BCT2}. Another reason to understand this fact is that~$L^2$ is nothing but the expectation value of the f\/lux operator $\hat{F}_S=\int_S \rmd^2y\,E^a_in_a$ (through a surface $S$ with co-normal~$n_a$) on a semiclassical state~\cite{BCT2}. In inverse-volume as well as holonomy corrections, one refers to elementary building blocks of a discrete state, respectively, the plaquette areas and the edge lengths. A pure minisuperspace quantization makes use of macroscopic parameters such as the volume of some f\/iducial region, and f\/luxes are calculated on comoving areas $\sim V_o^{2/3}$.  On the other hand, in the lattice-ref\/inement formulation of loop quantum cosmology one uses the microscopic volume of a cell, and f\/luxes are def\/ined on comoving areas $\sim \ell_0^2$. This leads to equation~\eqref{main}, with some phenomenological parameter $\tilde\s$.

Intuitively, holonomy corrections become large when the Hubble scale $H^{-1}= a/\dot{a}\sim \g L$ is of the size of the discreteness scale, an extreme regime in cosmology. In terms of the classical energy density $\rho=3H^2/\k^2$, holonomy corrections can be quantif\/ied by the parameter
\[
\delta_{\rm hol}:= \frac{\rho}{\rho_*}= (\g HL)^2 .
\]
These are small when $\delta_{\rm hol}\ll 1$. In order to compare inverse-volume with holonomy corrections, we notice that
\begin{gather}\label{dpl}
\dpl=\big(\g l_\Pl H \delta_{\rm hol}^{-1/2}\big)^{\tilde\s} .
\end{gather}
For a universe of causal size $H^{-1}\sim l_\Pl$, inverse-volume corrections are considerable and behave very dif\/ferently from what is normally expected for quantum gravity. For small densities, holonomy corrections are small, but inverse-volume corrections may still be large because they are magnif\/ied by an inverse power of $\delta_{\rm hol}$. As the energy density decreases in an expanding universe, holonomy corrections fall
to small values, while inverse-volume corrections increase. For instance, in an inf\/lationary regime with a
typical energy scale of $\rho\sim 10^{-10}\rho_{\rm Pl}$, we can use equation~(\ref{dpl}) with $\tilde\s=4$ to write $\delta_{\rm hol} \sim 10^{-9}/\sqrt{\dpl}$. Small holonomy corrections of size $\delta_{\rm hol}<10^{-6}$ then require inverse-volume correction larger than $\dpl>10^{-6}$. This interplay of holonomy and inverse-volume corrections can make loop quantum cosmology testable, because it leaves only a f\/inite window for consistent parameter values, rather
than just providing Planckian upper bounds. It also shows that inverse-volume corrections become dominant for suf\/f\/iciently small densities (eventually, of course, they are suppressed as the densities further decrease).

\subsubsection{Lattice parametrization}

The lattice ref\/inement picture allows us to reinterpret minisuperspace quantization schemes in a~dif\/ferent language.
Equation \eqref{main} replaces the total lattice f\/iducial volume $V$ as the ``patch'' (i.e., cell) volume $L^3$ \cite{bo08}. This means that one makes the formal replacement $V\to V/\cN$ everywhere in minisuperspace expressions, which can be also justif\/ied as follows. At the kinematical level, internal time is taken at a f\/ixed value but the geometry still varies on the whole phase space. In this setting, we must keep $\cN$ f\/ixed to some constant $\cN_0$ while formulating the constraint as a~composite operator. Since the vertex density does not depend on the choice of f\/iducial volume, it is physically reasonable to expect the $\cN_0$ factor to be hidden in the kinematical quantity~$a_*$ (or~$p_*$). The net result is the Hamiltonian constraint operator of the previous sections.

However, when one solves the constraint or uses it for ef\/fective equations, one has to bring in the dynamical nature of $\cN$ from an underlying full state. This is the motivation for promo\-ting~$\cN$ to a time-dependent quantity. For some stretches of time, one can choose to use the scale factor~$a$ as the time variable and represent ${\cal N}(a)$ as a power law (equation~\eqref{N13}),
\begin{gather}\label{m}
\cN=\cN_0a^{6n} .
\end{gather}
Overall, quantum corrections are of the form \eqref{main},
\begin{gather}\label{13}
\dpl= \left(\ell_\Pl^3\frac{{\cal N}}{V}\right)^\frac{\tilde \s}{3}= \left(\ell_\Pl^3\frac{\cN_0}{V_o}\right)^{\frac{\tilde \s}{3}} a^{(2n-1)\tilde \s} ,
\end{gather}
where $\tilde \s>0$. This equation cannot be obtained in a pure minisuperspace setting.

The parameter $a$ plays two roles, one as a dynamical geometric quantity and the other as internal time. While writing down the semiclassical Hamiltonian with inverse-volume (and holonomy) corrections, one is at a non-dynamical
quantum-geometric level. Then, internal time is taken at a f\/ixed value but the geometry still varies on the whole phase space. In this setting, we must keep~$\cN$ f\/ixed while formulating the constraint as a composite operator. The net result is the Hamiltonian constraint operator of the basic formulation of loop quantum cosmology \cite{boj2,boj7} not taking into account any ref\/inement, corresponding to $n=0$ and $\tilde \s=\s$. On the other hand, equation~\eqref{m} captures operator as well as state properties of the ef\/fective dynamics. The parametrization of $\cN$ as a power law of the scale factor is simply a way to encode the qualitative (yet robust) phenomenology of the theory. The general viewpoint is similar to mean-f\/ield approximations which model ef\/fects of underlying degrees of freedom by a single, physically motivated function.

Comparing with the earlier minisuperspace parameterization, equation~\eqref{13} gives $\s=(1-2n)\tilde \s$. Since $\p\cN/\p V\geq 0$, one has $n\geq 0$: the number of vertices $\cN$ must not decrease with the volume, and it is constant for $n=0$. Also, $\ell_0\sim a^{1-2n}$ is the geometry as determined by the state; in a discrete geometrical setting, this has a lower non-zero bound which requires $n\leq 1/2$. In particular, for $n=1/2$ we have a constant patch volume as in the improved minisuperspace quantization scheme \cite{aps3}. In contrast with the minisuperspace parametrization \eqref{cmini}, in the ef\/fective parametrization of equation~\eqref{13} we have $\s=0$ for the improved quantization scheme $n=1/2$.
The range of $n$ is then
\begin{gather*}
0<n\leq \frac12 .
\end{gather*}
The critical density $\rho_*\propto a^{6(2n-1)}$, equation~\eqref{rhol}, is still constant for $n=1/2$.

The exponent $\tilde \s$ in equation~\eqref{m} can be taken as a small positive integer. In fact, the correction function $\dpl$ depends on f\/lux values, corresponding to $p$ for the isotropic background. Since $p$ changes sign under orientation reversal but the operators are parity invariant, only even powers of $p$ can appear, giving $\tilde \s=4$ as the smallest value. Therefore we set $\tilde \s\geq 4$.

To summarize, $\s$ is a time-independent parameter given by the quasi-classical theory and with range
\begin{gather*}
\s\geq 0 .
\end{gather*}
$\s$ may be dif\/ferent in $\a$ and $\nu$ for an inhomogeneous model, but we assume that the background equations~\eqref{ca} and~\eqref{cn} are valid also in the perturbed case. The coef\/f\/icients $\a_0$ and $\nu_0$ become arbitrary but positive parameters. In fact, from the explicit calculations of inverse-volume operators and their spectra in exactly isotropic models and for regular lattice states in the presence of inhomogeneities~\cite{boj6,boj8, BHKS0}, correction functions implementing inverse-volume corrections approach the classical value always from above. This implies that
\[
\a_0\geq 0 ,\qquad \nu_0\geq 0 .
\]
The lattice parametrization replaces the one for homogeneous LQC. In fact, strictly speaking, the use of one parametrization instead of the other is not a matter of choice. A perfectly homogeneous FRW background is an idealization of reality which, in most applications, turns out to be untenable. The study of cosmological perturbations with inverse volume corrections~\cite{BC,BCT1,BCT2,BH1, BH2,BHKS,BHKS2,CH} is an example in this respect. In that case, therefore, the lattice ref\/inement parametrization is not only useful, but also required for consistency. Ef\/fective linearized equations in the presence of holonomy corrections are under development, but we do not have complete control over them yet.
For vector and tensor modes a class of consistent constraints with a closed algebra is known~\cite{BH1,BH2,MCBG}, and therefore inspections of cosmological holonomy ef\/fects have been analyzed for gravitational waves~\cite{CMNS2,GB1,GB3,Mie1,GB2}. On the other hand, anomaly cancellation in the scalar sector has been worked out only recently~\cite{CaBa,CMBG, ed-scalar}.

We conclude with some comments on the superinf\/lationary phase of loop quantum cosmo\-lo\-gy. In the near-Planckian regime, the small-$j$ problem in the homogeneous parametrization is reinterpreted and relaxed in terms of the lattice embedding. The volume spectrum depends on the quadratic Casimir in $j$ representation: $ \bu^{-n}\sim V^{2/3}\sim \sqrt{C_2(j)}\sim j$. A higher-$j$ ef\/fect can be obtained as a ref\/inement of the lattice (smaller $\bu$) \cite{BCK}, thus allowing for long enough superacceleration. A change in $\bu(p)$ can be achieved by varying the comoving volume $V_o$. This is an arbitrary operation in pure FRW, while in inhomogeneous models $\bu$ is a physical quantity related to the number of vertices of the underlying reduced spin-network state. As long as a calculation of this ef\/fect from the full theory is lacking, we will not be able to predict the duration of the small-volume regime. More importantly, the closure of the constrained algebra is a wide open issue outside the approximation \eqref{ca}, \eqref{cn}, and early works on superinf\/lationary cosmological perturbations \cite{boj12,CaC,CMNS0,hos04,ShH, TSM} have not received a rigorous conf\/irmation.

\section{Conclusions}\label{concl}

The kinematical structure of LQG is well def\/ined although there are some technical dif\/f\/iculties in the construction of the physical theory.
One way of increasing our understanding of loop quantization is to apply it to simple systems. Here we look at applications to cosmological systems,
also with the hope of making progress in developing a realistic theory of quantum cosmology.

Owing to the presence of an underlying full theory, this procedure has been much more successful than the earlier WDW quantization.
Trying to follow the steps of LQG, using the fact that the underlying geometry is discrete and making physically well-motivated {\em Ans\"atze} about
the size of the f\/iducial cells, we have obtained a quantization scheme which ensures the resolution of the classical singularity as well as the
correct semiclassical limit. This is a generic feature of LQC of all minisuperspace models. We have described two examples
of minisuperspace LQC in detail:
\begin{itemize}\itemsep=0pt
\item The f\/lat FRW model is the simplest and the most rigorously studied model in LQC. We have used it to explain the kinematic structure of LQC, and to
describe the construction of the physical Hilbert space and of observables. These provide an evolution picture of the universe with respect to a massless
scalar f\/ield playing the role of a clock variable, which in turn serves to illustrate the mechanism of singularity resolution.
\item Bianchi I model serves as an example for the complications which arise in LQC, from the correct choice of the quantization scheme to
the construction of the physical Hilbert space. In this model the evolution picture is not complete yet as long as a basis of states diagonalizing the
Hamiltonian constraint remains unknown.
The Bianchi~I model also plays a crucial role in the hybrid quantization of Gowdy $T^3$ model.
\end{itemize}

Once we have got some handle on the construction of quantum theories of minisuperspace models, we need to consider models with f\/ield theory degrees
of freedom, i.e., midisuperspace models. They can serve as good toy models for testing the f\/ield-theoretical features of the full theory. Also a study of the fate of the classical singularity in these models is needed to ensure that the singularity resolution mechanism in LQC is generic and not an
artifact of the symmetry reduction. However, only one model, the polarized Gowdy $T^3$ model, has been studied so far in LQC. We have described and
compared two approaches, the hybrid quantization and the polymer quantization procedures. The hybrid quantization scheme,
although very successful, quantizes a part of the geometry in the Fock way, while the polymer quantization scheme is still incomplete. We feel the
need to study a complete loop quantization of this and other midisuperspace models for a better understanding of LQG/LQC.

The path we have followed in the f\/irst two parts is to bring more and more complicated systems under the ambit of loop quantization. We have started with f\/lat FRW, which is homogeneous and isotropic. Then we have removed the isotropy condition and studied Bianchi~I. Finally, we have tried to lift even the homogeneity condition (although retaining some other symmetries) and studied the Gowdy $T^3$ model.

Instead of looking at various models, one can take another path and look at the generic features of the theory and try to incorporate these as corrections to the classical equations. This ef\/fective equation approach, presented in Part~\ref{part3}, is a setting where the ef\/fects
of LQC can be incorporated in a way suitable for phenomenology and observations. We have described the progress made in both homogeneous and inhomogeneous contexts. In the latter, the main complication is the closure of the ef\/fective constraint algebra, whose study has been completed for inverse-volume corrections but is still ongoing when holonomy corrections are switched on. While verifying this property, we have to ensure that we retain the features which resolve the singularity while giving the correct large-scale behaviour. Once this is satisf\/ied, we can use this framework to make predictions and constrain the parameters of the model, at least in the semiclassical regime.

A number of issues are under active investigation in Loop quantization of cosmological models. Here we mention some interesting problems in the areas of LQC covered in this review. At the homogeneous level, in solvable LQC we are still trying to understand the nature of the bounce replacing the classical singularity and what are the characteristics of the universe at late times after the bounce. This has been extensively explored in the simple models where an evolution picture has been developed, but it remains a challenge to control the complete evolution even in FRW models with massive scalar f\/ields and in the Bianchi I models. Other Bianchi cosmologies need to be analyzed in greater detail because the non-trivial topologies and properties of the connections introduce additional subtleties in the quantization. Finding methods for solving these issues may indicate ways to address their counterparts in the full theory. Also, it would be important to understand whether some of the features of classical evolution (for example, the chaotic approach to singularity in Bianchi IX models) are retained after Loop quantization. Although we now have a good understanding of the big-bang type singularity from the LQC perspective, the nature and the possible resolutions of big-rip type singularities require further investigation.

Here we concentrated on the Hamiltonian formalism and on cosmological models constructed via the imposition of quantum constraints. Quantum cosmology can be also def\/ined via dif\/ferent sum-over-histories approaches. One is to recast the dynamics of LQC in a form resembling the spin-foam formulation of the full theory, and express it in terms of a ``vertex expansion'' \cite{ACH2,ACH1,CGO1,HRVW,HMQ}. This corresponds to a minisuperspace symmetry reduction of the path integral of the full theory. Another possibility, also dubbed ``spin-foam cosmology'' and attempting to get in closer contact with the full theory, is to evaluate spin-foam amplitudes on graphs interpreted as homogeneous and isotropic geometries \cite{BTRV,BRV,Hel11,LiMa,Rok10,Vid11,Vid10}. A third line of research, inspired by group f\/ield theory, consists in writing down a minisuperspace version of a~scalar f\/ield theory action where wavefunctions are promoted to f\/ields and the Hamiltonian plays the role of kinetic operator \cite{CGO2, GiO}. Interaction terms, possibly nonlocal, are responsible for topology changes and can give ef\/fective contributions to the Hamiltonian constraint in a linear mean-f\/ield approximation. All these approaches are at their f\/irst stages of development.

Much work needs to be done in midisuperspace models where only one cosmological example has been studied so far. Although the hybrid quantization
scheme has made signif\/icant progress, one should still fully justify the approach and check whether similar methods are viable for the other Gowdy models. Again, that may depend on the development of the other Bianchi setups. The polymer quantization scheme provides a genuine toy model suitable for trying to solve the problems facing the full theory, and for this reason it should be further ref\/ined. We also need to look
at other midisuperspace scenarios, particularly 1-Killing vector models which can be mapped to $2+1$ gravity with scalar f\/ields. Finally, it is not
clear how to obtain minisuperspace LQC as a sector of LQG and at what level we expect such an embedding, and whether it is possible at all. However, a similar exercise can be attempted in LQC by checking whether one can embed minisuperspace models into midisuperspace models.

The power and importance of the ef\/fective-dynamics technique have been slowly gaining momentum. Ef\/fective dynamics and ef\/fective quantum constraints provide the means to ask questions such as whether the LQC corrections are consistent with the observed universe and whether the underlying discreteness coming from LQC has any appreciable ef\/fect on the inf\/lationary imprint. All these challenges are currently being undertaken. Confrontation of LQC with experiments is one of the major open problems to be addressed. Detailed investigations on linear cosmological perturbations showed that the size of quantum corrections could be consi\-de\-rab\-ly larger than what previously expected. Both theoretical considerations and observations of the primordial spectra in the microwave sky can set limits on the ambiguity parameters in the model and highlight how the parameter range can change upon performing a symmetry enhancement from pure FRW to perturbed FRW backgrounds. Many of these results have been achieved when taking only inverse-volume corrections into account, but control over holonomy corrections is now suf\/f\/icient for completing the picture at the ef\/fective-constraints level (see Section~\ref{latti} and references therein).

The purpose of this review was to show the dif\/ferent facets of LQC and to describe the work done in various directions. These indicate that, as we move towards studying the LQC of models with lesser degrees of symmetry, a more holistic understanding of all these developments is needed. This is essential for further improving our hold on this quantum theory of gravity.

\subsection*{Acknowledgements}
K.B.\ would like to thank NSFC (Grant No.\ 10975017), China Postdoctoral Science Foundation (Grant No.\ 20100480223)
and the Fundamental Research Funds for the Central Universities for f\/inancial support.
M.M.B.\ is partially supported by the Spanish MICINN Projects No.\ FIS2008-06078-C03-03 and No.\ FIS2011-30145-C03-02.

\LastPageEnding

\end{document}